\documentclass[10pt]{article}
\usepackage{epsfig}
\usepackage{axodraw}
\usepackage{epsfig}
\usepackage{graphicx}
\usepackage{rotate}
\usepackage{latexsym}
\usepackage{amssymb}
\usepackage{fancybox}      
\newcommand\pubnumber{}
\newcommand\pubdate{\today}
\newcommand\hepnumber{hep-ph/0603121}

\def\csumc{Dipartimento di Fisica Teorica, Universit\`a di Torino, Italy\\
INFN, Sezione di Torino, Italy}
\def\support{\footnote{Work supported by MIUR under contract
2001023713$\_$006.}}
\def\Title#1{\begin{center} {\Large\bf #1 } \end{center}}
\newcommand{\Authors}[2]{\large\begin{center}{ \sc #1 \hspace{0.1cm} {\rm and}
\hspace{0.1cm} #2} \end{center}}

\def\Address#1{\begin{center}{ \it #1} \end{center}}

\newcommand\pubblock{\rightline{\begin{tabular}{l} \pubnumber\\
         \pubdate\\ \hepnumber \end{tabular}}}
\newenvironment{Abstract}{\begin{quotation}  }{\end{quotation}}

\def\Acknowledgments{\bigskip  \begin{center}
          \large\bf Acknowledgments\end{center}}
\def\email#1{\footnote{#1}}
\makeatletter
\def\section{\@startsection{section}{0}{\z@}{5.5ex plus .5ex minus
 1.5ex}{2.3ex plus .2ex}{\large\bf}}
\def\subsection{\@startsection{subsection}{1}{\z@}{3.5ex plus .5ex minus
 1.5ex}{1.3ex plus .2ex}{\normalsize\bf}}
\def\subsubsection{\@startsection{subsubsection}{2}{\z@}{-3.5ex plus
-1ex minus  -.2ex}{2.3ex plus .2ex}{\normalsize\sl}}

\renewcommand{\@makecaption}[2]{%
   \vskip 10pt
   \setbox\@tempboxa\hbox{\small #1: #2}
   \ifdim \wd\@tempboxa >\hsize     % IF longer than one line:
       \small #1: #2\par          %   THEN set as ordinary paragraph.
     \else                        %   ELSE  center.
       \hbox to\hsize{\hfil\box\@tempboxa\hfil}
   \fi}
 \def\citenum#1{{\def\@cite##1##2{##1}\cite{#1}}}
\def\citea#1{\@cite{#1}{}}
\newcount\@tempcntc
\def\@citex[#1]#2{\if@filesw\immediate\write\@auxout{\string\citation{#2}}\fi
  \@tempcnta\z@\@tempcntb\m@ne\def\@citea{}\@cite{\@for\@citeb:=#2\do
    {\@ifundefined
       {b@\@citeb}{\@citeo\@tempcntb\m@ne\@citea\def\@citea{,}{\bf }\@warning
       {Citation `\@citeb' on page \thepage \space undefined}}%
    {\setbox\z@\hbox{\global\@tempcntc0\csname b@\@citeb\endcsname\relax}%
     \ifnum\@tempcntc=\z@ \@citeo\@tempcntb\m@ne
       \@citea\def\@citea{,}\hbox{\csname b@\@citeb\endcsname}%
     \else
      \advance\@tempcntb\@ne
      \ifnum\@tempcntb=\@tempcntc
      \else\advance\@tempcntb\m@ne\@citeo
      \@tempcnta\@tempcntc\@tempcntb\@tempcntc\fi\fi}}\@citeo}{#1}}
\def\@citeo{\ifnum\@tempcnta>\@tempcntb\else\@citea\def\@citea{,}%
  \ifnum\@tempcnta=\@tempcntb\the\@tempcnta\else
  {\advance\@tempcnta\@ne\ifnum\@tempcnta=\@tempcntb \else\def\@citea{--}\fi
    \advance\@tempcnta\m@ne\the\@tempcnta\@citea\the\@tempcntb}\fi\fi}
\makeatother
\newcommand{\nl}{\nonumber\\}
\newcommand{\nn}{\nonumber}

\newcommand{\lpar}{\left(}                            % bracketing
\newcommand{\rpar}{\right)}

\newcommand{\bq}{\begin{equation}}                    % equationing
\newcommand{\eq}{\end{equation}}
\newcommand{\bqa}{\arraycolsep 0.14em\begin{eqnarray}}
\newcommand{\eqa}{\end{eqnarray}}
\newcommand{\ba}[1]{\begin{array}{#1}}
\newcommand{\ea}{\end{array}}
\newcommand{\ben}{\begin{enumerate}}
\newcommand{\een}{\end{enumerate}}
\newcommand{\bei}{\begin{itemize}}
\newcommand{\eei}{\end{itemize}}
\newcommand{\eqn}[1]{Eq.(\ref{#1})}
\newcommand{\eqns}[2]{Eqs.(\ref{#1})--(\ref{#2})}

\newcommand{\tabn}[1]{Tab.~\ref{#1}}

\newcommand{\fig}[1]{Fig.~\ref{#1}}
\newcommand{\figs}[2]{Figs.~\ref{#1}--\ref{#2}}
\newcommand{\sect}[1]{Section~\ref{#1}}
\newcommand{\sects}[2]{Section~\ref{#1} and \ref{#2}}

\newcommand{\appendx}[1]{Appendix~\ref{#1}}

\newcommand{\GeV}{\mathrm{GeV}}
\newcommand{\MeV}{\mathrm{MeV}}
\newcommand{\eV}{\mathrm{eV}}

\def\Re{\mathop{\operator@font Re}\nolimits}
\def\Im{\mathop{\operator@font Im}\nolimits}
\newcommand{\ord}[1]{{\cal O}\lpar#1\rpar}

\newcommand{\ph}{\gamma}

\newcommand{\wb}{W}

\newcommand{\barf}{\overline f}

\newcommand{\mw}{M_{_W}}

\newcommand{\mz}{M_{_Z}}

\newcommand{\mh}{M_{_H}}

\newcommand{\mt}{m_t}

\newcommand{\mzs}{M^2_{_Z}}

\newcommand{\spro}[2]{{#1}\cdot{#2}}
\newcommand{\li}[2]{\mathrm{Li}_{#1}\lpar\displaystyle{#2}\rpar} % polylog
\newcommand{\egam}[1]{\Gamma\lpar#1\rpar}               % Euler's Gamma
\newcommand{\intmomi}[2]{\int\,d^{#1}#2}

\newcommand{\intfx}[1]{\int_{\scriptstyle 0}^{\scriptstyle 1}\,d#1}
\newcommand{\intfxy}[2]{\int_{\scriptstyle 0}^{\scriptstyle 1}\,d#1\,
                        \int_{\scriptstyle 0}^{\scriptstyle #1}\,d#2}

\newcommand{\Beta}[2]{{\rm{B}}\lpar #1,#2\rpar}

\newcommand{\ointfxy}[2]{\int_{\scriptstyle 0}^{\scriptstyle 1}\,d#1\,d#2}
\newcommand{\cff}[7]{C_{#1}\lpar #2,#3,#4;#5,#6,#7\rpar}    
 
\newcommand{\ep}{\epsilon}
\newcommand{\tHss}{\mu^2}
\newcommand{\Reb}{{\rm{Re}}}
\newcommand{\Imb}{{\rm{Im}}}
\newcommand{\upar}[1]{u}

\newcommand{\ssA}{{\scriptscriptstyle{A}}}

\newcommand{\ssC}{{\scriptscriptstyle{C}}}
\newcommand{\ssD}{{\scriptscriptstyle{D}}}
\newcommand{\ssE}{{\scriptscriptstyle{E}}}

\newcommand{\ssG}{{\scriptscriptstyle{G}}}
\newcommand{\ssH}{{\scriptscriptstyle{H}}}
\newcommand{\ssI}{{\scriptscriptstyle{I}}}

\newcommand{\ssK}{{\scriptscriptstyle{K}}}

\newcommand{\ssM}{{\scriptscriptstyle{M}}}
\newcommand{\ssN}{{\scriptscriptstyle{N}}}

\newcommand{\ssR}{{\scriptscriptstyle{R}}}
\newcommand{\ssS}{{\scriptscriptstyle{S}}}

\newcommand{\ssV}{{\scriptscriptstyle{V}}}
\newcommand{\ssW}{{\scriptscriptstyle{W}}}
\newcommand{\ssX}{{\scriptscriptstyle{X}}}

\newcommand{\bqas}{\begin{eqnarray*}}
\newcommand{\eqas}{\end{eqnarray*}}
\def\app#1#2 {{\it Acta. Phys. Pol.} {\bf#1},#2}
\def\cpc#1#2 {{\it Computer Phys. Comm.} {\bf#1},#2}
\def\np#1#2 {{\it Nucl. Phys.} {\bf#1},#2}
\def\pl#1#2 {{\it Phys. Lett.} {\bf#1},#2}
\def\prep#1#2 {{\it Phys. Rep.} {\bf#1},#2}
\def\prev#1#2 {{\it Phys. Rev.} {\bf#1},#2}
\def\prl#1#2 {{\it Phys. Rev. Lett.} {\bf#1},#2}
\def\zp#1#2 {{\it Zeit. Phys.} {\bf#1},#2}
\def\sptp#1#2 {{\it Suppl. Prog. Theor. Phys.} {\bf#1},#2}
\def\mpl#1#2 {{\it Modern Phys. Lett.} {\bf#1},#2}
\def\jetp#1#2 {{\it Sov. Phys. JETP} {\bf#1},#2}
\def\fpj#1#2 {{\it Fortschr. Phys.} {\bf#1},#2}
\def\afp#1#2 {{\it Acta.Phys. Polon.} {\bf#1},#2}
\def\err#1#2 {{\it Erratum} {\bf#1},#2}
\def\ijmp#1#2 {{\it Int. J. Mod. Phys} {\bf#1},#2}
\def\nc#1#2 {{\it Nuovo Cimento} {\bf#1},#2}
\def\ap#1#2 {{\it Ann. Phys.} {\bf#1},#2}
\def\cmp#1#2 {{\it Comm. Math. Phys.} {\bf#1},#2}
\def\el#1#2 {{\it Europhys. Lett.} {\bf#1},#2}
\def\hpa#1#2 {{\it Helv. Phys. Acta} {\bf#1},#2}
\def\yf#1#2 {{\it Yad. Fiz.} {\bf#1},#2}
\def\nim#1#2 {{\it Nucl. Instrum. Meth.} {\bf#1},#2}
\def\spz#1#2 {{\it Sov. Pisma Zhetf} {\bf#1},#2}
\def\jetpl#1#2 {{\it JETP Lett.} {\bf#1},#2}
\def\sjnp#1#2 {{\it Sov. J. Nucl. Phys.} {\bf#1},#2}
\def\ptp#1#2 {{\it Progr. Theor. Phys. (Kyoto)} {\bf#1},#2}
\def\rmp#1#2  {{\it Rev. Mod. Phys.} {\bf#1},#2}
\def\zhetf#1#2 {{\it ZhETF} {\bf#1},#2}
\def\prs#1#2 {{\it Proc. Roy. Soc.} {\bf#1},#2}
\def\phys#1#2 {{\it Physica} {\bf#1},#2}

\newcommand{\egams}[1]{\Gamma^2\lpar#1\rpar}               % Euler's Gamma

\newcommand{\intfxx}[2]{\int_{\scriptstyle 0}^{\scriptstyle 1}\,d#1\,
                        \int_{\scriptstyle 0}^{\scriptstyle 1}\,d#2}
\def\bfi{\begin{figure}}
\def\efi{\end{figure}}

\newcommand{\hyper}[4]{{}_2F_1(#1\,,\,#2\,;\,#3\,;\,#4)}
\newcommand{\hyperf}[1]{{}_2F_1\left(#1\right)}

\newcommand{\dsimp}[1]{\int\,dS_{#1}}
\newcommand{\dssimp}[1]{\int\!dS_{#1}}
\newcommand{\dcub}[1]{\int\,dC_{#1}}
\newcommand{\dscub}[1]{\int\!dC_{#1}}

\newcommand{\aba}{\ssE}

\newcommand{\aca}{\ssI}
\newcommand{\acan}[1]{#1;\ssI}
\newcommand{\ada}{\ssM}
\newcommand{\adan}[1]{#1;\ssM}
\newcommand{\bba}{\ssG}

\newcommand{\bca}{\ssK}
\newcommand{\bcan}[1]{#1;\ssK}
\newcommand{\bbb}{\ssH}

\newcommand{\bX}{{\overline X}}

\newcommand{\bchi}{{\overline \chi}}
\newcommand{\bbeta}{{\overline \beta}}
\newcommand{\bxi}{{\overline \xi}}
\newcommand{\chiu}[1]{\chi_{_{#1}}}

\newcommand{\LB}{{\cal L}oop{\cal B}ack}
\newcommand{\GS}{{\cal G}raph{\cal S}hot}
\newcommand{\intsx}[1]{\int_{\scriptstyle 0}^{\scriptstyle 1}\!\!\!d#1}
\newcommand{\intsxy}[2]{\int_{\scriptstyle 0}^{\scriptstyle 1}\!\!\!d#1
                        \int_{\scriptstyle 0}^{\scriptstyle #1}\!\!\!d#2}
\newcommand{\intsxx}[2]{\int_{\scriptstyle 0}^{\scriptstyle 1}\!\!\!d#1
                        \int_{\scriptstyle 0}^{\scriptstyle 1}\!\!\!d#2}

\newcommand{\Bbt}{B_{_{\scriptstyle\chi}}}
\newcommand{\Xbt}{X_{_{\scriptstyle\chi}}}
\newcommand{\bXbt}{\bX_{_{\scriptstyle\chi}}}
\newcommand{\bmid}{\Bigr|}
\newcommand{\spliti}[6]{\int_{\scriptstyle #1,#2}^{\scriptstyle #2,#3}\,d#4\,#5_{\scriptstyle 1\oplus 2\,;\,#6}}
\newcommand{\sspliti}[6]{\int_{\scriptstyle #1,#2}^{\scriptstyle #2,#3}\!\!\!\!\!d#4\,#5_{\scriptstyle 1\oplus 2\,;\,#6}}
\newcommand{\triagi}[3]{\int_{\scriptstyle ( #1\,,\,#2\,,\,#3)}}

\begin{document}
\begin{titlepage}
\pubblock
\vfill
\def\thefootnote{\fnsymbol{footnote}}
\Title{Two-Loop Vertices in Quantum Field Theory:\\[5mm]
Infrared and Collinear Divergent Configurations\support}
\vfill
\Authors{Giampiero Passarino\email{giampiero@to.infn.it}}
{Sandro Uccirati\email{uccirati@to.infn.it}}
\Address{\csumc}
\vfill
\begin{Abstract}
\noindent 
A comprehensive study is performed of two-loop Feynman diagrams with 
three external legs which, due to the exchange of massless gauge-bosons, 
give raise to infrared and collinear divergencies. Their relevance in 
assembling realistic computations of next-to-next-to-leading corrections 
to physical observables is emphasised. A classification of infrared singular
configurations, based on solutions of Landau equations, is introduced.
Algorithms for the numerical evaluation of the residues of the infrared poles 
and of the infrared finite parts of diagrams are introduced and discussed 
within the scheme of dimensional regularization. Integral representations of 
Feynman diagrams which form a generalization of Nielsen - Goncharov 
polylogarithms are introduced and their numerical evaluation discussed.
Numerical results are shown for all different families of multi-scale, 
two-loop, three-point infrared divergent diagrams and successful comparisons 
with analytical results, whenever available, are performed. Part of these
results has already been included in a recent evaluation of electroweak 
pseudo-observables at the two-loop level. 
\end{Abstract}
\vfill
\begin{center}
\end{center}
\end{titlepage}
\def\thefootnote{\arabic{footnote}}
\setcounter{footnote}{0}
%--
\small
\thispagestyle{empty}
\tableofcontents
\setcounter{page}{1}
\normalsize
%--
\clearpage
%--
\section{Introduction \label{intro}}
%--
This paper belongs to a series devoted to numerical evaluation of the
multi-loop, multi-leg Feynman diagrams that appear in any renormalizable
quantum field theory. In~\cite{Passarino:2001wv} (hereafter I) the general 
strategy has been designed and in~\cite{Passarino:2001jd} (hereafter II) a 
complete list of results has been derived for two-loop functions with two
external legs, including their infrared divergent on-shell derivatives. 
Results for one-loop multi-leg diagrams have been shown 
in~\cite{Ferroglia:2002mz} and additional material can be found 
in\cite{Ferroglia:2002yr}. Two-loop three-point functions for infrared
convergent configurations have been considered in~\cite{Ferroglia:2003yj} 
(hereafter III), two-loop tensor integrals in~\cite{Actis:2004bp}.

Many mass scales appear in the computation of physical observables within the 
Standard Model, generating serious difficulties for the familiar 
analytical approach. Our purpose is to overcome these problems 
through a numerical approach. The application of our techniques has 
recently contributed to the evaluation of the two-loop fermionic correction to 
the effective electroweak mixing angle and of the full Higgs-mass dependence 
of the bosonic ones~\cite{Hollik:2005va}.

The approach described in~\cite{Passarino:2001wv} is primarily intended for 
evaluation of multi-loop diagrams with internal massive lines. However, QED 
and QCD are integral part of any realistic calculation and they lead to 
infrared singularities. Therefore, any method aimed to a numerical evaluation 
of diagrams must be able to handle the infrared problem and infrared/collinear 
configurations should be treatable within the same class of algorithms used 
for the non-infrared cases or within some simple extension of the latter. 

For one-loop diagrams we have seen that our methods allow us to extract the 
infrared pole in dimensional regularization with a residue and a finite part 
that can be treated numerically~\cite{Ferroglia:2002mz}. The procedure has 
been extended in II to cover the on-shell derivative of two-point
functions which are needed in the treatment of external legs.

It is the purpose of this paper to extend the study of infrared 
divergencies to two-loop three-point functions. All diagrams are computed 
within the scheme of dimensional regularization~\cite{Gastmans:1975sr} with 
space-time dimensionality $n = 4 - \ep$. Each loop in a diagram contributes 
at most one soft (zero gauge-boson mass) and one collinear (for zero fermion 
mass) $1/\ep$ term but the global order of the pole at $\ep = 0$ can be 
greater than two due to simultaneous occurrence of ultraviolet poles which 
are removed by the introduction of counter-terms.

To accomplish our goals we need an automatized procedure for handling infrared 
(and collinear) configurations: Landau equations~\cite{Landau:1959fi} represent
the proper tool since a necessary condition for the presence of infrared 
divergencies is that the Landau equations are fulfilled. 
Therefore, for each topology we build individual diagrams by filling all the 
lines with the line content of the theory, disregarding those configurations 
with vertex content not allowed by the theory itself. The generated result is 
examined and Landau equations studied for those diagrams that contain massless
gauge-boson: if they are fulfilled then we have an infrared divergent 
configuration.
The residue of the infrared pole(s) and the corresponding infrared finite 
part are then computed numerically.

This part of the procedure is relatively easy while the difficult task is 
connected to the numerical evaluation of residues and of finite parts. 
They will be given in terms of multi-dimensional integrals over Feynman 
parameters with integrands that are not positive defined and, according to our 
strategy, their evaluation requires introduction of smoothness algorithms.

Smoothness requires that, after suitable manipulations, the kernel in the 
integral representation and its first $N$ derivatives be continuous functions 
and, ideally, $N$ should be as large as possible. However, in most of the 
cases we will be satisfied with absolute convergence, e.g. logarithmic 
singularities of the kernel. This is particularly true when the large number 
of terms required by obtaining continuous derivatives of higher order leads to 
large numerical cancellations.

There is a general approach for extracting infrared poles which goes under the
name of sector decomposition~\cite{Binoth:2000ps}. 
We have examined this technique which, despite its great intrinsic 
possibilities, has its own problems: to name one it has been applied (so far)
mainly to unphysical kinematics where infrared residues and finite parts
are given in terms of positive definite integrands, i.e. it will not work 
properly around thresholds where the Feynman integrands are known to change 
their sign and imaginary parts show up. For recent developments see, however,
ref.~\cite{Binoth:2004jv}.

In our experience the form of the integrand, after many iterations of the 
sector decomposition technique, is such that one can hardly imagine to design 
adequate smoothness algorithms. For this reason we have, quite often, 
privileged algorithms that keep under control the smoothness of the Feynman
integrand at each step of the extraction of the infrared singularities. Usage 
of the whole machinery of hypergeometric functions has shown particularly 
useful in this respect.

One may wonder why to devote additional efforts to the problem of computing
infrared divergent diagrams, given the spectacular success of 
analytical evaluation in QED/QCD: here we refer, in particular to the  
results by~\cite{Davydychev:2003mv}, by~\cite{Davydychev:2002hy} and
by~\cite{Bonciani:2003hc} but also to~\cite{Birthwright:2004kk}.

The actual reason for pursuing this line of research is that QED and QCD are 
embedded in a more general theory, e.g. the standard model of 
fundamental interactions; from this point of view their handling is much more 
complicated. 
For instance there will be more than one mass scale for infrared divergent 
configurations, like in the decay of charged gauge-bosons and, with few 
exceptions, the analytical approach works only for very few scales or in the 
approximation where the scales themselves are arranged according to some fixed 
hierarchy, $m \ll M$ etc.

We are not claiming that a purely numerical approach is the final solution,
rather one should carefully mix (semi) analytical extraction of dominant 
corrections (e.g. leading and sub-leading collinear logarithms) with numerical 
evaluation of sub-dominant, process-dependent, terms; the latter should be 
transformed in a way that allows for a safe, stable, integration where 
apparent singularities of the integrand are absent or limited to a minimum 
amount. 
Whenever a cancellation of dominant terms is foreseen we have to organize the 
calculation in such a way that these terms drop out before any numerical 
integration is attempted. From this point of view the technique of reduction 
of an arbitrary diagram to generalized scalar integrals is not always the best 
choice; master integrals quite often are individually more divergent (e.g. in 
the collinear limit) than the complete answer. Our technique does not 
grant any privilege to master integrals -- from a computational point of 
view -- and, therefore, seems more appropriate in handling the problem.

The outline of the paper will be as follows: in \sect{conve} we define our 
conventions. In \sect{IRtools} we review some of the tools that have
been introduced to study infrared divergencies in quantum field theory.
The connection between infrared divergent configurations and Landau equations 
is described in \sect{IRSLE},
in \sects{SD}{ISSD} we present the procedure of sector decomposition while 
an alternative technique for extracting ultraviolet (if any) and infrared 
poles, based on properties of the hypergeometric function is given in 
\sect{IRHF}. In \sect{TBMBT} we present a discussion of threshold
singularities. Starting with \sect{IDTLV} we present our results for
all configurations, from \sect{vaba} to \sect{vbbb}.
In \sect{IRLclass} we present an explicit example of our procedure for
classifying infrared divergent diagrams.
Numerical results are summarized in \sect{numres}.
In Appendix we give a collection of technical details.
%--
\section{Notations and Conventions \label{conve}}
%--
Our conventions for dealing with arbitrary two-loop diagrams have been
introduced in Sect. 2 of II. Conventions that are specific for three-point 
functions have been introduced in Sect.~2 of III; also the various families of 
two-loop vertex diagrams have been classified in III but, for the reader's 
convenience, they are repeated in \figs{TLvertaba}{TLvertbbb}.
%--
\subsection{Integrals and integration measures}
%--
In particular, to keep our results as compact as possible, we introduce the 
following notations where $x_0 = y_0 = 1$:
%--
\bqa
\dsimp{n}(\{x\})\,f(x_1,\cdots,x_n) &\equiv& 
\prod_{i=1}^{n}\,\int_0^{x_{i-1}}\,dx_i\,f(x_1,\cdots,x_n),
\nl
\dcub{n}(\{x\})\,f(x_1,\cdots,x_n) &\equiv& \int_0^1\,\prod_{i=1}^{n}\,dx_i\,
\,f(x_1,\cdots,x_n),
%\nl
%\dcubs{\{x\}}{\{y\}}\,f(x_1,\cdots,x_{n_1},y_1,\cdots,y_{n_2}) &\equiv& 
%\int_0^1\,\prod_{i=1}^{n_1}\,dx_i\,
%\prod_{j=1}^{n_2}\,\int_0^{y_{j-1}}\,dy_j\,
%f(x_1,\cdots,x_{n_1},y_1,\cdots,y_{n_2}).
%\nl
\eqa
%--
Also, the so-called $'+'$-distribution will be extensively used, e.g.\
%--
\bqa
\dcub{n}(\{z\})\,\intfx{x}\,\frac{f(x,\{z\})}{x}\bmid_+ &=&
\dcub{n}(\{z\})\,\intfx{x}\,\frac{f(x,\{z\}) - f(0,\{z\})}{x},
\nl
\dcub{n}(\{z\})\,\intfx{x}\,\frac{f(x,\{z\})}{x-1}\bmid_+ &=&
\dcub{n}(\{z\})\,\intfx{x}\,\frac{f(x,\{z\}) - f(1,\{z\})}{x-1},
\nl
\dcub{n}(\{z\})\,\intfx{x}\,\frac{f(x,\{z\})\,\ln^n x}{x}\bmid_+ &=&
\dcub{n}(\{z\})\,\intfx{x}\,\frac{\Bigl[f(x,\{z\}) - 
f(0,\{z\})\Bigr]\,\ln^n x}{x}.
\label{plusdist}
\eqa
%--
The last relation in \eqn{plusdist} is used for evaluating integrals
of the following type:
%--
\bqa
\intfx{x}\,\frac{f(x)}{x^{1-\ep}} &=& \frac{f(0)}{\ep} +
\intfx{x}\,\frac{f(x)}{x}\bmid_+ +
\ep\,\intfx{x}\,\frac{f(x)\,\ln x}{x}\bmid_+ + \ord{\ep^2}.
\label{poleext}
\eqa
%--
Since we will have to split integrals during the evaluation of diagrams we
have introduced a special notation:
%--
\bq
\spliti{a}{c}{b}{x}{F}{\ssA}(x)= \int_a^c\,dx\,F_{1\,;\,\ssA}(x) +
\int_c^b\,dx\,F_{2\,;\,\ssA}(x).
\eq
%--
In other cases we have to integrate over a triangle, for which we introduce
the special notation ($\bX = 1 - X$)
%--
\bq
\triagi{0}{\bX}{x_1}\,dx_2 dx_3 =
\int_{\scriptstyle 0}^{\scriptstyle \bX\,x_1}\,dx_2\,
\int_{\scriptstyle 0}^{\scriptstyle x_2/\bX}\,dx_3 +
\int_{\scriptstyle \bX\,x_1}^{\scriptstyle x_1}\,dx_2\,
\int_{\scriptstyle 0}^{\scriptstyle (x_1-x_2)/X}\,dx_3,
\eq
%--
\subsection{Alphameric classification of Feynman diagrams\label{alphaC}}
%--
In our conventions any scalar two-loop diagram is identified by a capital 
letter ($S, V$ etc, for self-energies, vertices etc) indicating the number of 
external legs and by a triplet of numbers $(\alpha, \beta$ and $\gamma$) 
giving the number of internal lines (in the $q_1, q_2$ and $q_1 - q_2$ loops 
respectively). There is a compact way of representing this triplet: assume 
that $\gamma \ne 0$, i.e. that we are dealing with non-factorizable diagrams, 
then we introduce
$\kappa = \gamma_{\rm max}\,\Bigl[ \alpha_{\rm max}\,(\beta - 1) +
\alpha - 1\Bigr] + \gamma$
for each diagram. Furthermore, we can associate a letter of the alphabet 
to each $\kappa$: for $G = V$ we have $\alpha_{\rm max} = 2$ and
$\gamma_{\rm max} = 2$, therefore, the following correspondence holds:
%--
\bq
121 \to E, \quad 131 \to I, \quad 141 \to M, \quad
221 \to G, \quad 231 \to K, \quad 222 \to H.
\eq
%-- 
For $G = S$ we have $\alpha_{\rm max} = 2$ and $\gamma_{\rm max} = 1$,
therefore
%--
\bq
111 \to A, \quad 121 \to C, \quad 131 \to E, \quad 221 \to D.
\eq
%--
This classification is extensively used throughout the paper where we omit
the suffix $0$ for scalar diagrams.
%--
\subsection{Basic quadratic forms}\label{app:def}
%--
An $x$-dependent mass is always defined as
%--
\bq
m^2_x= \frac{m^2_1}{x} + \frac{m^2_2}{1-x}.
\label{xdepmass}
\eq
%--
In the following we introduce some quadratic forms that are widely used 
throughout the paper. 
%--
\bqa
\chi(x\,;\,P^2\,;\,m\,,\,M) &\equiv&
-P^2\,x^2 + (P^2-m^2+M^2)\,x + m^2 = -P^2\,(x-\Xbt)^2 + \Bbt,
\label{chidef}
\eqa
%--
where we define
%--
\bq
\Bbt = \frac{1}{4\,P^2}\,\lambda(-P^2,m^2,M^2), \quad
\Xbt = \frac{P^2-m^2+M^2}{2\,P^2}, \quad \bXbt = 1 - \Xbt.
\label{btfactors}
\eq
%--
Here $\lambda(x,y,z)= x^2+y^2+z^2-2\,(x y + x z + y z)$ is the usual K\"allen
lambda - function. If no ambiguity will arise we will simply write
$\chi(x)$. Furthermore, we introduce
%--
\bq
\bchi(x\,;\,P^2\,;\,m\,,\,M) \equiv
\chi(1-x\,;\,P^2\,;\,m\,,\,M)= -P^2\,x^2 + (P^2-M^2+m^2)\,x + M^2 = 
-P^2\,(x-\bXbt)^2 + \Bbt,
\label{chibdef}
\eq
%--
\bq
\beta(x,y\,;\,P^2\,;\,m\,,\,M) \equiv 
-P^2\,x^2 + (P^2-m^2+M^2)\,x\,y + m^2\,y^2 =
-P^2\,(x-\Xbt\,y)^2 + \Bbt\,y^2,
\label{betadef}
\eq
%--
\bq
\bbeta(x,y\,;\,P^2\,;\,m\,,\,M) \equiv 
-P^2\,x^2 + (P^2-M^2+m^2)\,x\,y + M^2\,y^2 = 
-P^2\,(x-\bXbt\,y)^2 + \Bbt\,y^2.
\label{betabdef}
\eq
%--
Once again, we will drop irrelevant arguments if no ambiguity may arise.
These quadratic forms will play a major role in the evaluation of the 
diagrams considered in this paper.
%--
\section{Tools for virtual infrared divergencies \label{IRtools}}
%--
Before starting a comprehensive study of two-loop, infrared divergent, 
vertices we collect in this Section a set of tools which are relevant for 
the general analysis of infrared divergencies in a spontaneously broken 
quantum field theory. 

First, we recall the classification of infrared configurations based on the 
study of Landau equations; we then move to a short review of the techniques 
which go under the general name of sector decomposition; the use of these 
techniques for handling integrable singularities in the evaluation of Feynman
diagrams is also discussed. Alternative techniques based on a representation 
of infrared divergent diagrams through hypergeometric functions will be 
illustrated by means of simple examples. Mellin - Barnes techniques will be 
introduced to study infrared configurations around their normal thresholds and
to extract collinear limits. Finally, we will present some of our new 
techniques which represent an extension of the Bernstein - Sato functional 
relations~\cite{BS}.
%--
\subsection{Infrared singularities and Landau equations \label{IRSLE}}
%--
As explained in the Introduction, the classification of infrared 
divergent Feynman diagrams is most conveniently based on the use of Landau 
equations~\cite{Landau:1959fi}. The whole procedure is better illustrated in 
terms of a simple scalar one-loop triangle. 
%--
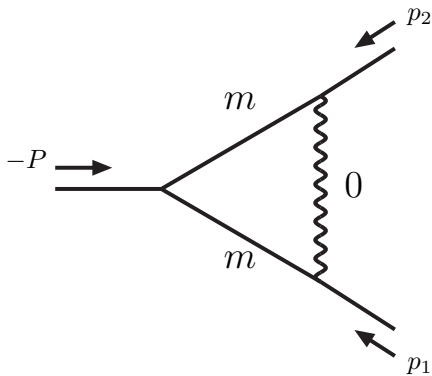
\begin{figure}[ht]
\begin{center}
\begin{picture}(150,75)(0,0)
 \SetWidth{1.5}
 \Line(0,0)(40,0)        \LongArrow(0,8)(20,8)        \Text(-11,7)[cb]{$-P$}
 \Line(128,-53)(100,-35) \LongArrow(128,-63)(114,-54) \Text(138,-70)[cb]{$p_1$}
 \Line(128,53)(100,35)   \LongArrow(128,63)(114,54)   \Text(138,62)[cb]{$p_2$}
 \Line(100,35)(40,0)                  \Text(70,30)[cb]{\Large $m$}
 \Line(100,-35)(40,0)                 \Text(70,-30)[cb]{\Large $m$}
 \Photon(100,-35)(100,35){2}{10}      \Text(113,-3)[cb]{\Large $0$}
\end{picture}
\end{center}
\vspace{2cm}
\caption[]{The scalar one-loop, three-point Green function with one massless 
internal line. All momenta are flowing inwards.\label{fig:threepoint}}
\label{oneloopV}
\end{figure}
%--
Consider the following Feynman parametric representation of the diagram:
%--
\bqa
\cff{0}{p^2_1}{p^2_2}{P^2}{m}{0}{m} &=& \frac{\mu^{\ep}}{i\,\pi^2}\,
\intfxy{x}{y}\,V^{-1-\ep/2}_{\ssC}(x,y),
\nl
V_{\ssC}(x,y)= -p^2_2\,x^2 - p^2_1\,y^2 -2\,\spro{p_1}{p_2}\,x\,y &+&
(p^2_2 - m^2)\,x - (p^2_2-P^2-m^2)\,y + m^2.
\label{nconv}
\eqa
%--
To regularize infrared divergencies we use the scheme of dimensional 
regularization, with the space-time dimensionality $n = 4- \ep$ and where $\mu$ 
is the 't Hooft unit of mass. In principle one should carefully distinguish 
between ultraviolet and infrared regulators ($\ep > 0$ or $\ep < 0$) but for a 
comprehensive discussion we refer to~\cite{Bardin:1999ak}.

It is important to recall that a necessary condition for the presence of
infrared divergencies is that the Landau equations are fulfilled. A proper 
solution of the set of Landau equations for the one-loop vertex of
\fig{oneloopV} requires that
%--
\bq
P^2 = -\,\frac{1}{2\,m^2_2}\,\Bigl\{ 2\,m^2_2\,(m^2_1+m^2_3) -
(m^2_1+m^2_2+p^2_1)\,(m^2_3+m^2_2+p^2_2) \pm
\Bigl[ \lambda(-p^2_1,m^2_1,m^2_2)\lambda(-p^2_2,m^2_3,m^2_2)\Bigr]^{1/2}
\Bigl\},
\label{exaathr}
\eq
%--
which, for $m_2 = 0$ and $m_1 = m_3 = m$ is satisfied if $p^2_1 = p^2_2 =
-m^2$ and $P^2$ is unconstrained~\footnote{In our metric spacelike $p$ 
implies positive $p^2$. Further $p_4 = i\,p_0$ with $p_0$ real for a physical 
four-momentum.}. The corresponding $V_{\ssC}$ reads
%--
\bq
V_{\ssC}(x,y) = m^2\,(1-x)^2 + m^2\,y^2 + (P^2 + 2\,m^2)\,(1-x)\,y,
\eq
%--
showing that $V_{\ssC} = 0$ for $x = 1$ and $y = 0$. The parametrization in 
\eqn{nconv} is certainly not the most convenient one to evaluate the
residue of infrared pole but this choice has been made deliberately, having 
in mind the more complex two-loop cases.

In any software package designed to compute (automatically) physical 
observables the classification of the infrared cases must be a built-in 
procedure.
The study of the Landau equations for a given family of diagrams is the most 
elegant way to classify its infrared divergent configurations. Again we refer 
to the one-loop vertex for illustration. The Landau equations are
%--
\bq
\alpha_1\,(q^2 + m^2_1) = 0, \quad
\alpha_2\,((q+p_1)^2 + m^2_2) = 0, \quad
\alpha_3\,((q+p_1+p_2)^2 + m^2_3) = 0,
\label{exalanone}
\eq
%--
\bq
\alpha_1\,q_{\mu} + \alpha_2\,(q+p_1)_{\mu} + 
\alpha_3\,(q+p_1+p_2)_{\mu} = 0.
\label{exalan}
\eq
%--
A solution of the system with $\alpha_i \ne 0, \forall i$ gives the leading 
singularity. 

Let us multiply \eqn{exalan} by $q$, by $p_1$ and by $p_2$ to obtain 
an homogeneous system of three equations where we use $q^2 = - 
m^2_1$ etc, from \eqn{exalanone}.
A necessary and sufficient condition to have a proper solution, i.e. not all 
the $\alpha_i = 0$, requires the determinant of coefficients to be zero,
thus fixing a relation between internal and external masses.
Any configuration that satisfy this constraint is a Landau singularity for 
the diagram which, however, does not necessarily imply that the diagram itself 
diverges at that configuration.
For the $C_0$ function this corresponds to the well-known anomalous threshold.

Consider the following case: $p^2_{1,2}= - m^2, m_{1,3} = m$ and $m_2 = 0$. 
It is easily found that the anomalous threshold condition is satisfied, 
therefore the configuration is a Landau singularity; however, the question is
of which kind. Let us insert the above values into the homogeneous 
system, what we obtain is
%--
\bq
m^2\,\alpha_1 + (\frac{1}{2}\,P^2 + m^2)\,\alpha_2 = 0,
\quad
(\frac{1}{2}\,P^2 + m^2)\,\alpha_1 + m^2\,\alpha_3 = 0.
\eq
%--
First of all we observe that it is not either $\alpha_1 = 0$ or $q^2 = -m^2_1$,
etc; it can be both. Secondly, our configuration, where $P^2 = (p_1+p_2)^2$ is
unconstrained, is a singularity. Finally these is a special case of the general
configuration discussed ($P^2$ free) which is even more singular giving, in the
infrared case, the true leading Landau singularity. To have $\alpha_{1,3} 
\neq 0$ we must require $P^2 = - 4\,m^2$ which gives, in the annihilation
channel, the well-known threshold singularity on top of the infrared one. 
This condition emerges also from the following argument: inserting
$p^2_{1,2} = - m^2$ in the anomalous threshold condition one obtains
$m^2_2\,P^2 = 0$, and $m^2_2\,( m^2_2 - P^2 - 4\,m^2) = 0$,
%--
%\bq
%\eq
%--
corresponding to the two signs of \eqn{exaathr}. In a certain sense the
constraint $p^2 = -\,4\,m^2$ is buried inside the anomalous threshold 
condition. To summarize: that all the propagators in a diagram are on-shell
and that the consistency relation is satisfied does not necessarily imply that
all $\alpha_i$ are different from zero: the infrared case is a clear example.
Note the presence of a potential singularity also at $P^2 = 0$ which, however,
is not physical, i.e. is not lying on the physical Riemann sheet. The latter 
fact can be seen by inspecting the explicit analytical result,
%--
\bq
C^{\rm IR}_0 = \frac{2}{\beta\,P^2}\,\ln\frac{\beta+1}{\beta-1}\,
\frac{1}{\ep} + \,\,\mbox{IR finite}, \quad \beta^2 = 1 + 4\,\frac{m^2}{P^2}.
\eq
%--
Our strategy for the general classification of infrared divergent 
configurations will be on a diagram-by-diagram basis; we assume a certain 
number of zero internal masses with at least one unconstrained external 
momentum. Then we fix the remaining parameters to satisfy the consistency 
relation for the Landau equations. Finally, we return to the original set 
of Landau equations and look for additional constraints that are necessary 
in order to generate the true leading singularity. The presence of a 
threshold-like singularity on top of the infrared poles is the sign that, 
after extracting these poles, we still have complications for the residual 
integrations that cannot be solved with a naive use of the method to be 
described in the next subsection.
%--
\subsection{Sector decomposition \label{SD}}
%--
Once a diagram has been identified as infrared divergent we must proceed
with the extraction of the infrared pole(s) and with the evaluation of the
infrared residue(s) and of the finite part.

To continue our discussion we reconsider \eqn{nconv} and change variables, 
$x = 1 - x'$ and $y = (1-x')\,y'$. Next we split the integration by using 
$1 = \theta(x-y)+\theta(y-x)$, remap the two integrals into $[0,1]^2$ and 
obtain
%--
%\bq
\bqa
%C^{\ssI\ssR}_0 = \frac{\mu^{\ep}}{i\,\pi^2}\,\,\intfxx{x}{y}\,\Bigl[
C^{\ssI\ssR}_0 = \frac{\mu^{\ep}}{i\,\pi^2}\,\,\intfxx{x}{y}\,&\Bigl[&
( x^{-1-\ep} - x^{-\ep})\,\chi_x^{-1-\ep/2} +
( y^{-1-\ep} - x\,y^{-\ep})\,\chi_y^{-1-\ep/2}\Bigr],
%\eq
%--
%\bqa
\nl
\chi_x &=&
m^2\,\Bigl[ (1+y)^2 + x\,y\,(x\,y - 2 - 2\,y) \Bigr] +
P^2\,y\,(1 - x),
\nl
\chi_y &=&
m^2\,\Bigl[ (1+x)^2 + x\,y\,(x\,y - 2 - 2\,x) \Bigr ] +
P^2\,x\,(1 - x\,y).
\eqa
%--
The infrared pole can now be extracted, giving
$C^{\ssI\ssR}_0 = P_{\ssI\ssR} + F_{\rm fin}$,
%--
%\bqa
\bq
%P_{\ssI\ssR} &=& 2\,\frac{\mu^{\ep}}{i\,\pi^2}\,\ointfxy{x}{y}\,x^{-1-\ep}\,
P_{\ssI\ssR} = 2\,\frac{\mu^{\ep}}{i\,\pi^2}\,\ointfxy{x}{y}\,x^{-1-\ep}\,
%\Bigl[ m^2\,y^2 + (P^2+2\,m^2)\,y + m^2\Bigr]^{-1-\ep/2} 
\Bigl[ m^2\,(1+y)^2 + P^2\,y \Bigr]^{-1-\ep/2} 
%\nl
%{}&=& -\frac{2}{\ep}\,\frac{\mu^{\ep}}{i\,\pi^2}\,\intfx{y}\,
= -\frac{2}{\ep}\,\frac{\mu^{\ep}}{i\,\pi^2}\,\intfx{y}\,
%\Bigl[ m^2\,y^2 + (P^2+2\,m^2)\,y + m^2\Bigr]^{-1-\ep/2}.
\Bigl[ m^2\,(1+y)^2 + P^2\,y \Bigl]^{-1-\ep/2}.
\label{exasm}
\eq
%\eqa
%--
This procedure of extracting infrared poles, known as sector 
decomposition~\cite{Binoth:2000ps}, cannot be applied in its naive version to 
thee-point functions.
The reason is that the residue of the infrared pole(s) is now given by an
integral where the integrand is not positive defined over the parametric
hyper-cube and, therefore, we cannot perform a straightforward numerical
integration, at least if we want to avoid brute force and time consuming 
methods. 
Note that in our procedure we will use the van der Bij - Veltman
parametrization~\cite{vanderBij:1983bw} of two-loop integrals and not the 
more familiar Cvitanovic - Kinoshita one~\cite{Cvitanovic:1974uf}.

To summarize the complete algorithm of sector decomposition, 
we perform the following steps (for each family of diagrams):
At first, infrared and eventually collinear divergent configurations are 
selected by using the corresponding set of Landau equations which have been 
derived, diagram-by-diagram in III. Examples for two-loop vertices are given 
in \figs{fig131}{fig222}. For a similar classification in QED with a massive 
regulator we refer to~\cite{Cvitanovic:1974sv}.
sector decomposition is applied to a graph $G$ leading to an expression
%--
\bq
G = \sum_{l=0}^{L}\,\frac{g_l(\ep)}{\ep^l} = \sum_{l=0}^{L}\,\frac{G_l}{\ep^l},
\qquad
G_l = \frac{1}{B^{\ssG}_l}\,\int_{\ssS}\,dx\,{\cal G}_l(x),
\label{generalclass}
\eq
%--
since each loop contributes at most one soft and collinear $1/\ep^2$ term
($\ep$ being the dimensional regulator) the highest value for $L$ in a 
two-loop diagram is four; higher values in the Laurent expansion are only 
possible if ultraviolet divergencies are also present. In those cases the 
proper procedure is: ultraviolet poles are removed by local counter-terms, 
analytical continuation in $\ep$ is performed and infrared poles are extracted.
Smoothness algorithms are derived for each component $G_l$, as indicated in 
\eqn{generalclass},
%--
%\bq
%G_l = \frac{1}{B^{\ssG}_l}\,\int_{\ssS}\,dx\,{\cal G}_l(x),
%\label{generalclass}
%\eq
%--
where $x$ is a vector of Feynman parameters, $S$ is some simplex, ${\cal G}_l$
is an integrable function (in the limit $\delta \to 0$) and $B^{\ssG}_l$ is
a function of masses and external momenta whose zeros correspond to true
singularities of $G$, if any.
Smoothness requires that the kernel in \eqn{generalclass} and its first
$N$ derivatives be continuous functions and, ideally, $N$ should be as large
as possible. However, in most of the cases we will be satisfied with
absolute convergence, e.g. logarithmic singularities of the kernel. This is
particularly true around the zeros of $B^{\ssG}_l$ where the large number of
terms required by obtaining continuous derivatives of higher order leads to
large numerical cancellations.
%--
\subsection{Integrable singularities and sector decomposition \label{ISSD}}
%--
As it will become evident in the following sections, where we explicitly 
evaluate the diagrams, most of our integral representations will have
the form
%--
\bq
G = \int_{\ssS}\,dx\,\frac{1}{A(x)}\,\ln \Bigl[ 1 + \frac{A(x)}{B(x)}\Bigr],
\quad \hbox{or} \quad G(x) = \int_{\ssS}\,dx\,\frac{1}{A(x)}\,
\li{n}{\frac{A(x)}{B(x)}},
\label{maybesingular}
\eq
%--
where $x$ is a vector of Feynman parameters, $S$ is some simplex and
$A, B$ are multivariate quadratic forms. \eqn{maybesingular} generalizes 
the Nielsen-Goncharov family of polylogarithms~\cite{Kolbig:1983qt} based 
on monomials in one variable, see \appendx{app:Npoly}. Our integral 
representations are well-behaved around $A(x) \approx 0$ but numerical 
instabilities could arise when simultaneously $A(x) \approx B(x) \approx 0$ 
as it will always be the case for collinear singularities.
A nice solution to this problem~\cite{Passarino:2004kx} is to adopt a sector 
decomposition to factorize their common zero. Eventually, for some special 
configuration of internal masses and external invariants, this procedure 
will describe the correct behavior around a genuine singularity of $G$, 
e.g. $A$ and $B$ having a common zero of the same order. An example is 
provided by the following integral,
%--
\bq
J(a) = \ointfxy{x}{y}\,\frac{1}{x}\,\ln\Bigl[ 1 + \frac{x}{x+a\,y}\Bigr]
= \ointfxy{x}{y}\,\Bigl[ 
\ln\,\Bigl( 1 + \frac{1}{1+a\,y}\Bigr) +
\frac{1}{x}\,\ln\,\Bigl( 1 + \frac{x}{x+a}\Bigr)\Bigr].
\label{thisissing}
\eq
%--
After performing a sector decomposition in 
\eqn{thisissing} we obtain that $a = 0$ is indeed an end-point singularity of 
$J$. All our numerical results are based on additional sector decompositions 
of integrands of the form shown in \eqn{maybesingular}, although this is not 
explicitly indicated in the text.
%--
\subsection{Infrared power-counting \label{ipc}}
%--
Before continuing the discussion on the evaluation of infrared divergent 
configurations it is convenient to introduce the concept of infrared 
power-counting. Given any parametric representation of a Feynman diagram we 
perform transformation of variables such that $0 \le x_i \le 1$ with 
%--
\bqa
G &=& \dcub{N}\,\chiu{\ssN}^{-\mu},
\qquad
\chiu{\ssN} = \sum_{l_1\dots l_n}\,a_{l_1\dots l_n}\,\prod_{i=1}^n\,x^{l_i}_i
\theta(\sum_{i=1}^n\,l_i - L)
\eqa
%--
and $\chiu{\ssN} = 0$, for $x_i = 0, \,\forall i$, with $L \ge 1$. After 
introducing polar coordinates the integral can be written as
%--
\bq
G = \int\,d\Omega_{n-1}\,\int_0^{\scriptstyle R(\Omega)}\,dr\,
r^{n-\mu\,L-1}\,\chi^{-\mu}_{\ssN,{\rm red}}(r,\Omega),
\label{IRpc}
\eq
%--
with $\chi_{\ssN,{\rm red}}(0,\Omega) \ne 0$. Define $A,B$ such that
$n - \mu\,L - 1 = A - B\,\ep$. If $A \ge 0$ the integration does not lead 
to an $\ep$-pole and the corresponding diagram is infrared safe.
A typical example is represented by $V^{\aba}$ of \fig{TLvertaba} where,
although $\chiu{\aba} = 0$ at the hedge of the integration region, 
power-counting shows that no infrared pole arises.
%--
\subsection{Infrared singularities and hypergeometric functions \label{IRHF}}
%--
We have seen in the previous sections that a general method for extracting 
the infrared poles (within dimensional regularization) of Feynman diagrams is 
based on sector decomposition.
However, in most of the two-loop cases sector decomposition has drawbacks:
first of all the number of generated sectors tends to increase considerably 
and then the procedure creates new integrands with polynomials of very high 
degree, the higher the number of iterations the higher the degree. The 
consequence is that one cannot find an adequate smoothness algorithm to 
handle the final integration.
We give a second, alternative, procedure once again illustrated in terms of a 
$C_0$ function.

The polynomial $V_{\ssC}$ of \eqn{nconv} can be rewritten ($P^2 = - s,\,
p^2_i= -m^2$) as
%--
\bq
V_{\ssC}(x,y)= m^2\,x^2 + m^2\,y^2 + (s - 2\,m^2)\,x\,y - 
2\,m^2\,x - (s - 2\,m^2)\,y +m^2,
\eq
%--
and we perform the transformation $y= y' + \alpha\,x$ with 
$m^2\,\alpha^2 + (s - 2\,m^2)\,\alpha + m^2 = 0$. This transformation is 
designed to make $V$ linear in $x$ and, whenever possible we always seek for 
a transformation that makes the Feynman integrand linear in one of the 
variables. After some straightforward manipulation we obtain
$i\,\pi^2\,C_0 = \mu^{\ep}\,(C^1_0 + C^2_0)$ weher
%--
\bq
C^1_0 = (1 - \alpha)\,\intfxy{y}{x}\,V^{-1-\ep/2}_1(x,y),
\quad
C^2_0 = \alpha\,\intfxy{y}{x}\,V^{-1-\ep/2}_2(x,y),
\label{xint}
\eq
%--
with $V$-polynomials given by
%--
\bqa
V_1 &=& \Bigl[ s\,(1+\alpha)\,y - s +2\,(1-\alpha)\,m^2\Bigr]\,x -
        \alpha\,s\,y^2 - 2\,(1-\alpha)\,m^2\,y + m^2,
\nl
V_2 &=& \Bigl[ \alpha\,s + 2\,(1-\alpha)\,m^2\Bigr]\,x\,y -
\Bigr[ \alpha\,s - (2\,\alpha - 1)\,m^2\Bigr]\,y^2.
\label{thetwoV}
\eqa
%--       
In \eqn{xint} the $x$-integration has the general form
%--
\bqa
I_i(y) &=& \int_0^y\,dx \Bigl[ B_i(y) - A_i(y)\,x\Bigr]^{-1-\ep/2} = 
y\,B^{-1-\ep/2}_i\,\hyper{1+\ep/2}{1}{2}{\frac{A_i}{B_i}\,y}.
\qquad i= 1,2.
\eqa
%--
Using well-known properties of the hypergeometric function we obtain,
for $|\arg(-z)| < \pi$,
%--
\bqa
\hyper{1+\frac{\ep}{2}}{1}{2}{z} &=& \frac{2}{\ep}\,\Bigl[
-\,\hyper{1}{1+\frac{\ep}{2}}{1+\frac{\ep}{2}}{1-z}
+\,(1 - z)^{-\ep/2}\,\hyper{1}{1-\frac{\ep}{2}}{1-\frac{\ep}{2}}{1-z}\Bigr],
\eqa
%--
from which we derive
%--
\bq
I_i(y) = -\,\frac{2}{A_i\,\ep}\,B^{-\ep/2}_i\,\Bigl[ 1 - 
\Bigl( 1 - \frac{A_i}{B_i}\,y\Bigr)^{-\ep/2}\Bigr].
\eq
%--
Given the form of $A_1$ in \eqn{thetwoV} we can simply expand around $\ep = 0$
obtaining
%--
\bq
C^1_0 = ( \alpha - 1 )\,
\intfx{y}\,\frac{1}{A_1(y)}\,\ln\Bigl[ 1 - {\frac{A_1(y)}{B_1(y)}}\,y
\Bigr],
\eq
%--
which is well-behaved for $A_1(y)= 0$. However, for $i = 2$, we find
%--
\bq
A_2(y) = a(s,m^2)\,y, \qquad
B_2(y) =  b(s,m^2)\,y^2,
\eq
%--
\bq
a(s,m^2)= -\,\alpha\,s + 2\,(\alpha-1)\,m^2 = \frac{1}{4}\,
\beta\,(\beta+1)^2\,m^2,
\qquad
b(s,m^2)= -\,\alpha\,s + (2\,\alpha - 1)\,m^2 = 
\frac{1}{16}\,(\beta+1)^4\,m^2.
\eq
%--
Therefore, $C^2_0$ is infrared divergent and we get
%--
\bqa
C^2_0 &=& \frac{\alpha}{a(s,m^2)}\,b^{-\ep/2}(s,m^2)\,\intfx{y}\,y^{-1-\ep}\,
\ln \Bigl[ 1 - \frac{a(s,m^2)}{b(s,m^2)}\Bigr]\,\Bigl\{
1 - \frac{\ep}{4}\,\ln \Bigl[ 1 - \frac{a(s,m^2)}{b(s,m^2)}\Bigr]\Bigr\}
\nl
{}&=& 2\,\frac{\alpha}{a(s,m^2)}\,
\ln \frac{\beta-1}{\beta+1}\,\Bigl[
\frac{1}{\ep} - \frac{1}{2}\,
\ln \frac{\beta-1}{\beta+1} - \frac{1}{2}\,\ln\,b(s,m^2)\Bigr],
\eqa
%--
with $\beta^2 = 1 - 4\,m^2/s$.
The infrared pole has been isolated and infrared residue and finite part
are already in a form that allows for direct numerical integration.
For a general diagram they are rational functions of the residual Feynman 
parameters and we have been able to derive adequate smoothness algorithms
for their integration.
%--
\subsection{Threshold behavior and Mellin-Barnes transforms \label{TBMBT}}
%--
In computing infrared residues and finite parts of diagrams we face one
additional complication: in general we end up with integrands which 
are not positive definite. Therefore, around those configurations of the 
external parameters where threshold singularities occur the algorithm has to be
modified. The general idea is to isolate the singular behavior 
and to write, for the regular part, an expansion in some K\"allen function of 
the external parameters. We borrow the relevant technique from another problem,
the large energy expansion of Feynman diagrams which is best performed
by using Mellin-Barnes transforms~\cite{Roth:1996pd}.

The whole idea is better illustrated with a simple example. Consider the 
infrared residue of $C_0$:
%--
\bqa
P_{\ssI\ssR} &=& -\,\frac{2}{\ep}\,\frac{\mu^{\ep}}{i\,\pi^2}\,
B_{1+\ep/2} \lpar -m^2\,;\,m^2\,,\,P^2+4\,m^2\rpar, 
\nl
B_{\alpha}\lpar p^2\,;\,m_1^2\,,\,m_2^2\rpar &=& 
\intsx{x}\,\Bigl[ \chi(x) - i\,\delta\Bigr]^{-\alpha},
\qquad
\chi(x) = -p^2\,x^2 + (p^2+m_2^2-m_1^2)\,x + m_1^2,
\label{iexam}
\eqa
%--
and $\delta \to 0_+$. Let $x_{\pm}$ be the roots of $\chi-i\,\delta = 0$. If 
they are complex or real but external ($\not\in\,[0,1]$) then the numerical 
evaluation is straightforward; when they are real and internal 
$(\in [0,1]$) the integration contour can be distorted and the integral 
can be computed unless a pinch will occur in $[0,1]$~\cite{elop} (or an 
end-point singularity); this happens for $\lambda(-P^2,m^2,m^2) = 0$. 

The integral in \eqn{iexam} is simple enough to be computed analytically but
in our approach we will pretend to treat it numerically; distortion is 
performed, unless a pinch occurs. Then the question will arise of what to do 
around those parametric regions where $\lambda = 0$. Here we describe our 
solution for extracting the leading and sub-leading behavior of $P_{\ssI\ssR}$ 
around $\lambda = 0$. First we rewrite $\chi$ as
%--
\bq
\chi(x) = m^2\,(x - x_0)^2 - \frac{\lambda}{4\,m^2} - i\,\delta, 
\qquad\qquad
x_0 = \frac{s}{2\,m^2} - 1,
\eq
%--
and consider the case $P^2 = -s$ with $ s \ge 0$. Further, we assume that 
$0 \le x_0 \le 1$ and obtain
%--
\bq
B_{\alpha} = 
\frac{m^{-2\,\alpha}}{2\,\pi\,i}\,\intsx{x}\,
\int_{-i\,\infty}^{+i\,\infty}\!\!\!\!\!\!\!\!ds\,\,
B(s,\alpha-s)\,\rho^{\alpha-s}\,Q^{-s}, 
\qquad\qquad
\rho^{-1}= -\,\frac{\lambda}{4\,m^4} - i\,\delta,
\qquad
Q = (x - x_0)^2 - i\,\delta.
\label{vstrip}
\eq
%--
where $B$ denotes the Euler beta-function. \eqn{vstrip} is valid in the 
vertical strip $0 < \Reb\,s < \alpha$. We choose $\alpha < 1/2$, require 
$0 < \Reb\,s < 1/2$ and perform analytical continuation to obtain
%--
\bq
\intfx{x}\,Q^{-s} = \sum_{X=x_0}^{1-x_0}\,\frac{X^{1-2\,s}}{1-2\,s}.
\eq
%--
Since we are interested in the limit $|\rho| \to \infty$, the $s$-integral
will be closed over the right-hand complex half-plane at infinity, with
simple poles at $s = 1/2$ and $s = \alpha + k\; k \ge 0$. In this way we 
obtain
%--
\bq
B_{\alpha} = 
B(\frac{1}{2},\alpha-\frac{1}{2})\,m^{-2\,\alpha}\,
\Bigl( -\,\frac{\lambda}{4\,m^4} - i\,\delta\Bigr)^{1/2-\alpha} 
+ 
\quad 
\mbox{sub-leading}, 
\qquad  
s \to 4\,m^2.
\eq
%--
The result can be easily generalized to the case of unequal masses. In 
this case we have also have a pseudo-threshold, $s = (m_1-m_2)^2$, which is not
a singularity on the first Riemann shhet since $x_0 \not\in [0,1]$.
%--
\subsection{Old and new Bernstein - Sato - Tkachov functional 
relations\label{newBT}}
%--
For all one-loop multi-leg diagrams we have developed computational
techniques based on the proposal introduced in I. Most of two-loop 
infrared convergent diagrams can be computed by following the same strategy. 
However, some extension has to be introduced to deal with the general 
case.
To derive new algorithms, we recall the definition of Bernstein - 
Sato polynomials~\cite{BS}: if $V(x)$ is a polynomial in several variables 
then there is a non-zero polynomial $b(\mu)$ and a differential operator 
${\cal P}(\mu)$ with polynomial coefficients such that
%--
\bq
{\cal P}(\mu)\,V^{\mu+1}(x) = b(\mu)\,V^{\mu}(x).
\eq
%--
The Bernstein-Sato polynomial is the monic polynomial of smallest degree 
amongst such $b(\mu)$.
If $V(x)$ is a non-negative polynomial then $V^{\mu}(x)$, initially defined 
for $\mu$ with non-negative real part, can be analytically continued to a 
meromorphic distribution-valued function of $\mu$ by repeatedly using the 
functional equation
%--
\bq
V^{\mu}(x) = \frac{1}{b(\mu)}\,{\cal P}(\mu)\,V^{\mu+1}(x).
\eq
%--
The Bernstein - Sato - Tkachov theorem~\cite{Tkachov:1997wh} tells us that 
for any finite set of polynomials $V_i(x)$, where $x = \,\lpar x_1,\dots, 
x_{\ssN}\rpar$ is a vector of Feynman parameters, there exists an identity of 
the following form (hereafter a BST identity):
%--
\bq
{\cal P}\,\lpar x,\partial\rpar \prod_i\,V_i^{\mu_i+1}(x) =
B_{\ssV}\,\prod_i\,V_i^{\mu_i}(x).
\label{functr}
\eq
%--
where ${\cal P}$ is a polynomial of $x$ and $\partial_{i} = \partial/\partial
x_i$; $B_{\ssV}$ and all coefficients of ${\cal P}$ are polynomials of 
$\mu_i$ and of the coefficients of $V_i(x)$.
Furthermore, if the polynomial $V$ is of second degree we have a master 
formula, due to F.~V.~Tkachov~\cite{Tkachov:1997wh}. We write the polynomial 
as $V(x) = x^t\,H\,x + 2\,K^t\,x + L$, where $x^t=(x_1,...,x_n)$, $H$ is an 
$n \times n$ matrix, $K$ is an $n$ vector. The solution to the problem of 
determining the polynomial ${\cal P}$ is as follows:
%--
\bq
{\cal P}= 1 - \frac{(x-X_{\ssV})^t\,\partial_x}{2\,(\mu+1)}, \qquad
B_{\ssV} = L - K^t\,H^{-1}\,K, \qquad X_{\ssV}= - H^{-1}\,K.
\eq
%--
Therefore we have:
%--
\bq
V^{\mu}(x)= \frac{1}{B_{\ssV}}\,
\Big[ 1 - \frac{(x-X_{\ssV})^t\,\partial_x}{2\,(\mu+1)}\Big]\,V^{\mu+1}(x),
%\quad \mbox{for} \quad \mu \neq -\,1;
\qquad
V^{-1}(x)=  \frac{1}{B_{\ssV}}\,
\Big[1 - \frac{1}{2}\,(x-X_{\ssV})^t\,\partial_x\,\ln V(x)\Big].
\label{BT}
\eq
%--
The list of BST relations must be extended to cover infrared singular cases. 
It often happens that $V(x)$ is not complete and so other BST relations, 
originally defined in \eqn{functr}, are needed.
A typical example is when $V$ is linear in one variable
$V(x,y)= h\,(x-x_0)^2 + c\,y + b$. For this polynomial the $H$ matrix is 
singular and it can be easily shown that the following relations hold:
%--
\bq
V^{\mu}= 
\frac{1}{b}\,
\bigg[
1 - \frac{(x-x_0)\,\partial_x  + 2\,y\,\partial_y}{2\,(\mu+1)}\,
\bigg]\,
V^{\mu+1},
\qquad\qquad
V^{\mu}= 
\frac{1}{c}\,\frac{1}{\mu+1}\,\partial_y\,V^{\mu+1}.
\eq
%--
A better way to proceed without loss of generality is to introduce a
${\cal P}_0$ and a ${\cal P}_1$ with the property that
%--
\bq
{\cal D}_{\pm} = {\cal P}_0 \pm {\cal P}_1^t\,\partial_x, \qquad
{\cal D}_+\,V(x) = B_{\ssV}.
%\Big[ {\cal P}_0 + {\cal P}_1^t\,\partial_x \Big]\,V(x) = B_{\ssV}.
\label{P0P1}
\eq
%--
BST relations can be written as
%--
\bq
V^{\mu}(x)= \frac{1}{B_{\ssV}}\,
\Big[ {\cal P}_0 + \frac{1}{\mu+1}\,{\cal P}_1^t\,\partial_x \Big]\,
V^{\mu+1}(x),
\quad \mbox{for} \quad \mu \neq -\,1;
\qquad\quad
V^{-1}(x)= 
\frac{1}{B_{\ssV}}\,
\Big[ {\cal P}_0 + {\cal P}_1^t\,\partial_x\,\ln V(x) \Big].
\label{BTnew}
\eq
%--
A first extension of \eqn{BT} is given by the following example:
%--
\bq
V^{-1}(x)\,\ln^n V(x)= \frac{1}{B_{\ssV}}\,\Big[
{\cal P}_0\, + 
\frac{1}{n+1}\,{\cal P}_1^t\,\partial_x\,\ln V(x)\Big]\,\ln^n V(x).
\label{BTlogn}
\eq
%--
\eqn{BTlogn} can be easily generalised to arbitrary powers of $V$:
%--
\bq
V^{\mu}(x)\,\ln^n V(x)= \frac{1}{B_{\ssV}}\,
\bigg\{
{\cal P}_0\,\ln^n V(x)
+ \sum_{k=0}^{n}\,\frac{n!}{(n-k)!}\,\frac{(-1)^k}{(\mu+1)^{k+1}}\,
  {\cal P}_1^t\,\partial_x\,\ln^{n-k} V(x) \bigg\}\,V^{\mu+1}(x).
\label{BTmlogn}
\eq
%--
As a next step we seek for similar relations, holding for Nielsen 
polylogarithms~\cite{Kolbig:1983qt} and for the hypergeometric 
function~\cite{ellip}. These new relations can be obtained with the help
of the following formulae:
%--
\bq
\frac{d}{dx}\,\li{n+1}{x} = \frac{1}{x}\,\li{n}{x},
\qquad
\frac{d}{dx}\,\hyper{a-1}{b}{c}{x} = 
\frac{a-1}{x}\,\Big[\hyper{a}{b}{c}{x} - \hyper{a-1}{b}{c}{x}
\Big].
\eq
%--
Given a generic polynomial $A(x)$, it is easily verified that
%--
\bq
\frac{1}{V}\,\li{n}{\frac{A}{V}}= 
\frac{1}{B_{\ssV}}\,
\Bigg[ \frac{1}{A}\,\li{n}{\frac{A}{V}}
{\cal D}_+\,A
- {\cal P}_1^t\,\partial_x\,\li{n+1}{\frac{A}{V}}
\Bigg],
\label{BTlin}
\eq
%--
\bqa
V^{-a}\,\hyper{a}{b}{c}{\frac{A}{V}} 
&=& 
\frac{1}{B_{\ssV}}\,
\bigg\{
\frac{V^{1-a}}{A}\,
\Big[
  \hyper{a}{b}{c}{\frac{A}{V}}\,{\cal D}_+\,A
- \hyper{a-1}{b}{c}{\frac{A}{V}}\,{\cal P}_1^t\,\partial_x\,A
\Big]
\nl
{}&+& 
  \frac{{\cal P}_1^t}{1-a}\,
  \partial_x\,\Big[ V^{1-a}\,\hyper{a-1}{b}{c}{\frac{A}{V}} \Big]
\bigg\}.
\label{BThyper}
\eqa
%--
In particular for $n=0,1$, we have
$\li{0}{x} = x/(1-x)$ and $ \li{1}{x} = -\,\ln(1-x)$, for which the 
corresponding relations are
%--
\bq
\frac{1}{V}\,\frac{A}{V-A}= \frac{1}{B_{\ssV}}\,
\Bigg[
%\frac{1}{V-A}\,\Big( {\cal P}_0 + {\cal P}_1^t\,\partial_x \Big)A
\frac{1}{V-A}\,{\cal D}_+\,A
+ {\cal P}_1^t\,\partial_x\,\ln\Big(1-\frac{A}{V}\Big)
\Bigg],
\label{BTli0}
\eq
%--
\bq
\frac{1}{V}\,\ln\Big(1-\frac{A}{V}\Big) = \frac{1}{B_{\ssV}}\,
\Bigg[ \frac{1}{A}\,\ln\Big(1-\frac{A}{V}\Big)\,
%\Big( {\cal P}_0 + {\cal P}_1^t\,\partial_x \Big)\,A
{\cal D}_+\,A
+ {\cal P}_1^t\,\partial_x\,\li{2}{\frac{A}{V}}
\Bigg].
\label{BTli1}
\eq
%--
When the BST factor, $B_{\ssV}$, vanishes we have:
%--
\bqa &{}&
\frac{1}{V-A}\,{\cal D}_+\,A=
%\frac{1}{V-A}\,\Big( {\cal P}_0 + {\cal P}_1^t\,\partial_x \Big)A=
-\,{\cal P}_1^t\,\partial_x\,\ln\Big(1-\frac{A}{V}\Big),
\qquad\qquad \mbox{for} \quad B_{\ssV} = 0.
\label{BT0li0}
\eqa
%--
Another set of (new) relations is obtained starting from known properties of
polylogarithms,
%--
\bq
\frac{d}{dx}\,\frac{\li{n+1}{x}}{x}= 
\frac{1}{x^2}\,\Bigl[ \li{n}{x} - \li{n+1}{x} \Bigr].
\eq
%--
For $x = A/V$, when $n = 1$, we easily obtain:
%--
\bq
\frac{A}{V^2}\,\ln\Big(1-\frac{V}{A}\Big) = \frac{1}{B_{\ssV}}\,
\Bigg[
%A\,\Big( {\cal P}_0 - {\cal P}_1^t\,\partial_x \Big)
A\,{\cal D}_-\,
\frac{1}{V}\,\ln\Big(1-\frac{V}{A}\Big)\,
+ {\cal P}_1^t\,\partial_x\,\ln\Big(1-\frac{A}{V}\Big)
\Bigg].
\label{BT2li1}
\eq
%--
Using \eqn{BT2li1} for $n = 2$, we obtain:
%--
\bq
\frac{A}{V^2}\,\li{2}{\frac{V}{A}} = \frac{1}{B_{\ssV}}\,
\Bigg\{
A\,{\cal D}_-\,
\frac{1}{V}\,\li{2}{\frac{V}{A}}
+ {\cal P}_1^t\,\partial_x\,
  \bigg[   \frac{A}{V}\,\ln\Big(1-\frac{V}{A}\Big)
         - \ln\Big(1-\frac{A}{V}\Big)
  \bigg]
\Bigg\}
\label{BT2li2}
\eq
%--
The same procedure can be iterated to get a similar relation for 
$A/V^2\,\li{n}{V/A}$.

One can derive other relations of this type but here we have restricted our
attention to those which are actually used in this paper.
Their application in computing Feynman integrals is based on the  
integration by parts of the terms containing ${\cal P}_1^t\,\partial_x$ 
(when it is not applied just on $A$). 

After applying BST relations we end up with integrands which show 
a {\em less} divergent behaviour with respect to the original one; in most 
cases the integrand is smooth enough to allow for a stable numerical 
integration. When this is not the case we reiterate the use of BST functional
relations; they are followed by a second integration by parts, etc, etc.

A typical example of smoothness algorithm corresponds to apply a BST 
functional relation to \eqn{exasm}. In this case we obtain
%--
\bqa
i\,\,\pi^2\,m^2\,P_{\ssI\ssR} &=& - \frac{2}{\ep}\,
\lpar\frac{\mu}{m}\rpar^{\ep}\,\intfx{y}\,v^{-1-\ep/2}(y),
\qquad
v(y) = y^2 + (z+2)\,y + 1, \quad z = \frac{P^2}{m^2};
\eqa
%--
\bqa
\intfx{y}\,v^{-1-\ep/2}(y) &=& -\,\frac{4}{z\,(z + 4)}\,
\Bigl[ 1 -\frac{1}{4}\,(z + 4)\,\ln\,(z + 4)
+  \frac{1}{2}\,\intfx{y}\,\ln\,v(y) + \ord{\ep}\Bigr],
\eqa
%--
showing the additional threshold singularity at $P^2 = -4\,m^2$. Note the
absence of a singularity at $P^2 = 0$ on the first Riemann sheet.
%--
\section{Derivatives of two-loop self-energies and infrared poles.
\label{dsabaip}}
%--
In II we have defined the on-shell derivative of a two-point function,
where possibly some of the internal masses are zero, as the 
$\partial/\partial p^2$ derivative evaluated at the mass shell of one of the 
non-zero internal masses. These derivatives are used to construct wave
function renormalization factors and a complete list of results has been
shown in II. 

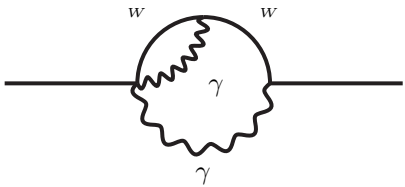
\begin{figure}[th]
\vspace{0.5cm}
\[
  \vcenter{\hbox{
  \begin{picture}(150,0)(0,0)
  \SetWidth{1.5}
  \Line(0,0)(50,0)
  \CArc(75,0)(25,0,90)
  \PhotonArc(75,0)(25,-180,0){2}{7}
  \CArc(75,0)(25,90,180)
  \PhotonArc(50,25)(25,-90,0){2}{7}
  \Line(100,0)(150,0)
  \Text(50,25)[cb]{$\ssW$}
  \Text(100,25)[cb]{$\ssW$}
  \Text(75,-38)[cb]{$\ph$}
  \Text(80,-5)[cb]{$\ph$}
  \end{picture}}}
\]
\vspace{0.5cm}
\caption[]{A two-loop diagram contribution to the $\wb$-boson self-energy.}
\label{wwf}
\end{figure}

Here, we briefly review the subject; consider the on-shell derivative of 
$S^{\ssC}$ (for the simplest topology $S^{\ssA}$ the on-shell 
derivative is infrared finite). Once again, a necessary condition for the 
presence of infrared divergencies is that the Landau equations are fulfilled: 
for $S^{\ssC}$, we see that 
$s = (m_1 + m_2 \pm m_4)^2$ and $m^2_3 = (m_1 + m_2)^2$ are satisfied by 
$m_2 = m_4 = 0$, $ m_1 = m_3 = m$ and $s = m^2$. 
It is easily seen that $S^{\ssC}$ is not infrared divergent but 
its derivative with respect to $p^2$ shows an infrared pole when 
computed on-shell.
To continue our discussion we consider the case $m_2 = m_4 = 0$ and 
$m_1 = m_3 = m$, a typical example of which is shown in \fig{wwf}.
%--
%--
After a straightforward calculation we obtain
%--
\bq
S^{\ssC}_p = \frac{1}{m^2}\,\Bigl(\frac{\tHss}{\pi\,m^2}\Bigr)^{\ep}\,
\Bigl[ -\,\frac{2}{\ep^2} - 2\,(2 - \gamma)\,\frac{1}{\ep} - 7 +
\gamma\,(4 -\gamma) - \frac{1}{2}\,\zeta(2) + \ord{\ep}\Bigr].
\label{irder}
\eq
%--
Results for $S^{\ssE}_p \equiv S^{131}_p$ have been derived in Sect 7.4 of II;
results for $S^{\ssD}_p \equiv S^{221}_p$ have been derived in Sect 7.9 of II;
$\gamma$ is the Euler constant and $\zeta$ the Riemann zeta function.

%--
\section{Infrared divergent two-loop vertices \label{IDTLV}}
%--
Starting with this section we present a detailed discussion, diagram - by - 
diagram, of two-loop infrared divergent vertices. 
Once again, evaluating a specific diagram means to derive a set of algebraic
manipulations that return a multi-dimensional integral representation of
the diagram for which numerical methods can be safely applied. For infrared 
configurations this means evaluating both the residues of the infrared poles 
and the finte parts. Special limits always require additional refinements
of the procedure. The key ingredients in our derivation are: BST functional
relations, the whole machinery of hypergeometric functions and Mellin-Barnes
transforms.

Several diagrams belong to the general $G^{1 \ssN 1}$ family which is 
specified in terms of a set of momenta $k_i$ which are linear combinations 
of the external momenta $p_j$; $k_i = p_1 + \dots + p_i$. Our parametrization 
is as follows:
%--
\bqa
G^{1 \ssN 1} = - \left( \frac{\mu^2}{\pi} \right)^{\ep}\,\Gamma(N-2+\ep)
  \int_0^1\,dx && \,\dsimp{N}(y,u_1,\cdots,u_{\ssN-1})\, 
\Bigl[ x(1-x) \Bigr]^{-\ep/2}\,(1-y)^{\ep/2-1}\,\chi^{2-\ssN-\ep}_{1 \ssN 1},
\nl
\chiu{1 \ssN 1} &=& u^t {\cal H} u + 2 {\cal K}^t u + 
( m_x^2 - m_3^2 )\,(1-y) + m_3^2,
\label{normalform}
\eqa
%--
where we have introduced the following quantities:
%--
\bq
{\cal H}_{ij} = - \spro{p_i}{p_j}, \quad
{\cal K}_i = \frac{1}{2}\,( k^2_i - k^2_{i-1} + m^2_{i+3} - m^2_{i+2} ), \quad
i,j = 1 , \dots , N - 1
\label{defHK}
\eq
%--
and where the $x$-dependent mass is defined in \eqn{xdepmass}.
%--
Three out of six of the non-trivial two-loop vertex families can be described
in terms of the representation of \eqn{normalform}.
%--
\subsection{The $V^{\aba}$ diagram \label{vaba}}
%--
The $V^{\aba}$ diagram of \fig{TLvertaba} is representable as
%--
\bq
\pi^4\,V^{\aba} = \mu^{2\ep}\,\int\,d^nq_1 d^nq_2\,
\frac{1}{[1]_{\aba}[2]_{\aba}[3]_{\aba}[4]_{\aba}},
\eq
%--
\bqa
[1]_{\aba} &=& q^2_1+m^2_1, \quad [2]_{\aba} = (q_1-q_2)^2+m^2_2,
\quad
[3]_{\aba} = (q_2-p_2)^2+m^2_3, \quad [4]_{\aba} = (q_2-P)^2+m^2_4.
\eqa
%--
Although $V^{\aba}$ does not exhibit infrared poles it is simple enough to 
illustrate our procedure. 
%--
\begin{figure}[ht]
\begin{center}
\begin{picture}(150,75)(0,0)
 \SetWidth{1.5}
 \Line(0,0)(40,0)         \LongArrow(0,8)(20,8)          \Text(-11,7)[cb]{$-P$}
 \Line(128,-53)(100,-35)  \LongArrow(128,-63)(114,-54)   \Text(138,-70)[cb]{$p_1$}
 \Line(128,53)(100,35)    \LongArrow(128,63)(114,54)     \Text(138,62)[cb]{$p_2$}
 \CArc(100,-35)(70,90,150)     \Text(65,30)[cb]{$1$}
 \CArc(40,70)(70,270,330)      \Text(78,0)[cb]{$2$}
 \Line(100,-35)(100,35)        \Text(107,-3)[cb]{$3$}
 \Line(100,-35)(40,0)          \Text(70,-30)[cb]{$4$}
\end{picture}
\end{center}
\vspace{2cm}
\caption[]{The irreducible two-loop vertex diagrams $V^{\aba}$. External 
momenta are flowing inwards.} 
\label{TLvertaba}
\end{figure}
%--
A necessary condition for the presence of infrared singularities is that the 
corresponding Landau equations are fulfilled. For $V^{\aba}$ this requires 
the condition $A\,P^4 + 2\,B\,P^2 + C = 0$, with coefficients
%--
\bqa
A &=& m^2_3,
\qquad
B =  -\Bigl[p^2_2+(m_1+m_2)^2-m^2_3\Bigr]\,
       (p^2_1-m^2_3+m^2_4)-2\,m^2_3\,(p^2_1+p^2_2),
\nl
C &=& \frac{1}{4\,m^2_3}\,\Bigl[ B^2 - 
     \lambda(-p^2_1,m^2_3,m^2_4)\,\lambda(-p^2_2,(m_1+m_2)^2,m^2_3)\Bigr].
\eqa
%--
Let us concentrate on infrared divergencies due to photons. Then any tri-linear
vertex has at most one photon line and any quadri-linear one has at most two
photon lines. We are looking for solutions of the Landau equations where
two external momenta are on some mass-shell and the third one is unconstrained.
The procedure will be as follows: 
%--

\vspace{0.1cm}\noindent
Let $m_1 = 0$ and $P^2$ be a free parameter. We immediately find 
that Landau equations admit a solution with $\alpha_i \ne 0, \forall i$ if 
$m_3 = 0, \; p^2_1 = -\,m^2_4, \; p^2_2 = -\,m^2_2$.

\noindent
Let $m_1 = 0$ and $p^2_1$ be free, we find
$m_2 = 0, \; P^2 = -\,m^2_4, \; p^2_2 = -\,m^2_3$.

\noindent
Let $m_1 = 0$ and $p^2_2$ be free, we find
$m_4 = 0, \; P^2 = -\,m^2_2, \; p^2_1 = -\,m^2_3$.
%--

\vspace{0.1cm}\noindent
Another possibility is to start with $m_3 = 0$. We proceed as follows:

\vspace{0.1cm}\noindent
Let $m_3 = 0$ and $P^2$ be free. We obtain
$p^2_1 = -\,m^2_4, \; p^2_2 = -\,(m_1+m_2)^2$,
which is of no interest since we do not expect a theory with such a peculiar
relation among masses.

\noindent
Let $m_3 = 0$ and $p^2_1$ be free. We obtain
$m_1 + m_2 = 0$, which is of no interest because it requires three photon 
lines in the same vertex. 

\noindent
Let $m_3 = 0$ and $p^2_2$ be free. We obtain
$m_4 = 0, \; p^2_1 = 0, \; P^2 = -(m_1+m_2)^2$, again of no interest.
%--

\vspace{0.1cm}\noindent
Because of the symmetry of the diagram the $m_2 = 0$ and $m_4 =0$ cases are
already covered. Consider $m_1 = m_3 = 0$ and $p^2_1 = -\,m^2_4, 
p^2_2 = -\,m^2_2$. Using \eqn{normalform} we obtain
%--
\bqa
\chiu{\aba}(x,y,z) &=& -m^2_2\,\frac{1-y}{1-x} - m^2_2\,(1-y)^2 -
m^2_4\,z^2 - (P^2 + m^2_2 + m^2_4)\,(1-y)\,z + m^2_2\,(1-y),
\eqa
%--
showing that $\chiu{\aba} = 0$ for the specified configuration at $y = 1$ and
$z = 0$. However, $\chiu{\aba}$ appears with exponent $-\ep$ in the parametric
representation for $V^{\aba}$ so that no infrared pole will show up.
To continue our analysis we consider the following cases:
%--

\vspace{0.1cm}\noindent
$m_1 = m_3 = 0$ and $P^2$ free. We obtain
$p^2_1 = -m^2_4, \; p^2_2 = -m^2_2$.

\noindent
$m_1 = m_3 = 0$ and $p^2_1$ or $p^2_2$ free. We obtain that 
$m_2 = 0$ and the case is of no interest because it requires three photon 
lines in a vertex.

\vspace{0.1cm}\noindent
The last case is $m_1 = m_3 = m_4 = 0$ which requires either $p^2_1 = 0$ or
$m_2 = 0$, i.e. more than two photon lines in a vertex. 
%--
\subsection{The $V^{\aca}$ diagram \label{vaca}}
%--
The $V^{\aca}$ family of diagrams, shown in \fig{TLvertaca}, is representable 
as
%--
\bqa
\pi^4\,V^{\aca} &=& \mu^{2\ep}\,\int\,d^nq_1 d^nq_2\,
\frac{1}{[1]_{\aca}[2]_{\aca}[3]_{\aca}[4]_{\aca}[5]_{\aca}},
\eqa
%--
with propagators
%--
\bqas
[1]_{\aca} &=& q^2_1+m^2_1, \qquad [2]_{\aca} = (q_1-q_2)^2+m^2_2,
\qquad
[3]_{\aca} = q^2_2+m^2_3,
\eqas
%--
\bqa
[4]_{\aca} &=& (q_2+p_1)^2+m^2_4, \qquad [5]_{\aca} = (q_2+P)^2+m^2_5.
\eqa
%--
\begin{figure}[ht]
\begin{center}
\begin{picture}(150,75)(0,0)
 \SetWidth{1.5}
 \Line(0,0)(42,0)         \LongArrow(0,8)(20,8)          \Text(-11,7)[cb]{$-P$}
 \Line(128,-53)(100,-35)  \LongArrow(128,-63)(114,-54)   \Text(138,-70)[cb]{$p_1$}
 \Line(128,53)(100,35)    \LongArrow(128,63)(114,54)     \Text(138,62)[cb]{$p_2$}
 \CArc(55,-9)(15,0,360)         \Text(75,-3)[cb]{$1$}\Text(38,-27)[cb]{$2$}
 \Line(100,-35)(67,-15.75)      \Text(80,-37)[cb]{$3$}
 \Line(100,-35)(100,35)         \Text(107,-3)[cb]{$4$}
 \Line(100,35)(45,3)            \Text(68,23)[cb]{$5$}
\end{picture}
\end{center}
\vspace{2cm}
\caption[]{The irreducible two-loop vertex diagrams $V^{\aca}$. External 
momenta are flowing inwards.} 
\label{TLvertaca}
\end{figure}
%--
Referring to \sect{IRSLE} we see that the leading Landau singularity for 
$V^{\aca}$ is equivalent to the sub-leading one, with $\alpha_1 = 
\alpha_2 = 0$, when the peculiar condition $m_3 = m_1+m_2$ is added. 
Therefore we discuss first those infrared configurations that correspond 
to sub-leading Landau singularities, and postpone the special case  
$m_3 = m_1 + m_2$.

If $\alpha_3 = 0$ the reduced diagram, i.e. the one where the third 
propagator is shrunk to a point, corresponds to a $V^{\aba}$ topology which 
is free from infrared poles, as explicitly shown in \sect{vaba}.

If $\alpha_1 = \alpha_2 = 0$ the reduced diagram is a one loop three-point 
function and the classification of the infrared singularities is simpler. We 
obtain:
$1) \; m_3 = 0, \; P^2 = -m^2_5, \; p^2_1 = - m^2_4$,
or
$2) \; m_4 = 0, \; p^2_1 = - m^2_3, \; p^2_2 = - m^2_5$,
or
$3) \; m_5 = 0, \; P^2 = - m^2_3, \; p^2_2 = - m^2_4$.
%--
They correspond to the configurations shown in \fig{fig131} (the first and
the third are actually the same).
%--
\begin{figure}[ht]
\begin{center}
\begin{picture}(150,75)(0,0)
 \SetWidth{1.5}
 \Line(0,0)(40,0)                     \Text(0,5)[cb]{$-P$}
 \DashLine(128,-53)(100,-35){2}       \Text(138,-65)[cb]{$p_1$}
 \DashLine(128,53)(100,35){5}         \Text(138,57)[cb]{$p_2$}
 \DashLine(100,35)(40,0){5}           \Text(65,25)[cb]{$M$}
 \DashLine(100,-35)(40,0){2}          \Text(65,-33)[cb]{$m$}
 \Photon(100,-3)(100,35){2}{5}
 \CArc(100,-18)(15,0,360)             \Text(80,-9)[cb]{$1$}\Text(123,-22)[cb]{$2$}
 \Text(15,50)[cb]{\Large $V^{\aca}_a$}
 \end{picture}
\qquad \qquad \qquad \qquad
\begin{picture}(150,75)(0,0)
 \SetWidth{1.5}
 \Line(0,0)(42,0)                     \Text(0,5)[cb]{$-P$}
 \DashLine(128,-53)(100,-35){2}       \Text(138,-65)[cb]{$p_1$}
 \DashLine(128,53)(100,35){5}         \Text(138,57)[cb]{$p_2$}
 \DashLine(100,35)(70,17.5){5}        \Text(75,30)[cb]{$M$}
 \CArc(55,9)(15,0,360)                \Text(38,20)[cb]{$1$}\Text(74,-6)[cb]{$2$}
 \DashLine(100,-35)(45,-3){2}         \Text(65,-33)[cb]{$m$}
 \Photon(100,-35)(100,35){2}{10}
 \Text(15,50)[cb]{\Large $V^{\aca}_b$}
\end{picture}
\end{center}
\vspace{2cm}
\caption[]{The $V^{\aca}$ infrared configurations. The photon line represents 
a general massless particle while the dashed and the continuous lines 
represent different massive particles. The mass of the two particles in the 
bubble are $m_1$ and $m_2$.}
\label{fig131}
\end{figure}
%--
\subsubsection{Extraction of UV and IR poles \label{extUVIRp}}
%--
For both configurations $1) \equiv 3)$ and $2)$ we perform the transformation 
$y= 1-y'$; the diagram reads as follows:
%--
\bq
V^{\aca}_i = - \lpar \frac{\mu^2}{\pi} \rpar^{\ep}\,\egam{1+\ep}\,
\intfx{x}\,\dsimp{2}(\{z\})\,\int_0^{1-z_1}\,dy\,
[x(1-x)]^{-\ep/2}\,y^{\ep/2-1}\,(a_{\acan{i}}\,y + Z_{\acan{i}})^{-1-\ep}
\label{innermost}
\eq
%--
where $i = a,b$, and $a_{\acan{a}} = m_x^2$, $a_{\acan{b}} = m_x^2 - M^2$, with
$m_x^2$ defined in \eqn{xdepmass}, and
%--
\bq
Z_{\acan{a}} = \beta\lpar z_2\,,\,z_1\,;\,P^2\,;\,m^2\,,\,M^2\rpar
\qquad
Z_{\acan{b}} = \beta\lpar 1 - z_1\,,\, 1-z_2\,;\,P^2\,;\,m^2\,M^2\rpar
\eq
%--
and $\beta$ defined in \eqn{betadef}. $V^{\aca}$ is also ultraviolet 
divergent and we look first for a procedure that extracts the ultraviolet 
pole. Consider the innermost integral appearing in \eqn{innermost}
%--
\bq
{\cal Y}_{\acan{i}} = Z_{\acan{i}}^{-1-\ep}\,\int_0^{1-z_1}\,dy\,y^{\ep/2-1}\,
(1 + \frac{a_{\acan{i}}}{Z_{\acan{i}}}\,y)^{-1-\ep}.
\eq
%--
It is convenient to evaluate this integral in terms of an hypergeometric
function~\cite{ellip}: 
%--
\bq
{\cal Y}_{\acan{i}} = \frac{2}{\ep}\,Z_{\acan{i}}^{-1-\ep}\,z^{\ep/2}_1\,
\hyper{1+\ep}{\frac{\ep}{2}}{1+\frac{\ep}{2}}{-\,\frac{1}{\zeta_{\acan{i}}}},
\qquad
\zeta_{\acan{i}} = \frac{Z_{\acan{i}}}{a_{\acan{i}}\,(1-z_1)}.
\label{hereishyper}
\eq
%--
Using a well-known property of hypergeometric functions (see 
appendix~\ref{app:hyp}) we obtain
%--
\bq
{\cal Y}_{\acan{i}} = 
\frac{2}{\ep}\,\frac{\egams{1+\ep/2}}{\egam{1+\ep}}\,
a_{\acan{i}}^{-\ep/2}\,Z_{\acan{i}}^{-1-\ep/2}\,- 
\frac{2}{2+\ep}\,a_{\acan{i}}^{-1-\ep}\,(1-z_1)^{-1-\ep/2}\,
\hyper{1+\ep}{1+\frac{\ep}{2}}{2+\frac{\ep}{2}}{-\,\zeta_{\acan{i}}},
\label{secfinite}
\eq
%--
with the result that the ultraviolet pole has been extracted. The second 
term in \eqn{secfinite} is finite, so we can set $\ep = 0$, obtaining for 
$V^{\aca}_i$:
%--
\bq
V^{\aca}_i = 
- \lpar \frac{\mu^2}{\pi} \rpar^{\ep}\,\frac{2}{\ep}\,\egams{1+\frac{\ep}{2}}\,
  {\cal X}_{\acan{i}}\,{\cal Z}_{\aca}
+ \intsx{x}\,\dsimp{2}(\{z\})\,
  \frac{\ln ( 1 + \zeta_{\acan{i}} )}{Z_{\acan{i}}}\, + \ord{\ep},
\label{v131ir}
\eq
%--
\bq
{\cal X}_{\acan{i}} = \intfx{x}\,[x(1-x)]^{-\ep/2}\,a_{\acan{i}}^{-\ep/2},
\qquad
{\cal Z}_{\aca} = 
\dsimp{2}(\{z\})\,Z_{\acan{i}}^{-1-\ep/2}.
\eq
%--
In the last expression we have dropped out the index $i$ for 
${\cal Z}_{\aca}$, since both cases produce the same result (it is easily 
seen by transforming $z_1 \rightarrow 1-z_1 $, $z_2 \rightarrow 1-z_2 $ and 
$z_1 \leftrightarrow z_2 $, for the case b).

The second term in \eqn{v131ir} is well-behaved when integrated over $x$, 
$z_1$ and $z_2$ and additional manipulations are not needed. As far as the 
first term is concerned we can write, after some straightforward manipulation, 
%--
\bq
{\cal X}_{\acan{i}} = 
1 - \frac{\ep}{2}\,\intfx{x} \ln V_{\acan{i}}
+ \frac{\ep^2}{8}\,\intfx{x} \ln^2 V_{\acan{i}}
\label{expandXaca}
\eq
%--
where the two new quadratic forms are
%--
\bq
V_{\acan{a}} = m_1^2\,(1-x) + m_2^2\,x,
\qquad
V_{\acan{b}} = \chi\lpar x\,;\, -M^2\,;\,m_1^2\,,\,m_2^2\rpar.
\label{quadaca}
\eq
%--
For ${\cal Z}_{\aca}$ we map the integration region into the square $[0,1]^2$ 
obtaining:
%--
\bq
{\cal Z}_{\aca} = 
\!\dssimp{2}(\{z\})\,
\beta^{-1-\ep/2}\lpar z_2,z_1;P^2;m^2,M^2\rpar
%\nl
%{}&=& 
=
\!\dscub{2}(\{z\})\,z_1^{-1-\ep}\,\chi^{-1-\ep/2}(z_2) =
- \frac{1}{\ep}\intsx{z}\,\chi^{-1-\ep/2}(z).
\label{hereMB}
\eq
%--
where $\beta$ has been defined in \eqn{betadef}.
For the $z_2$-integral we use one BST iteration, integrate by parts and 
expand in $\ep$ obtaining:
%--
\bq
{\cal Z}_{\aca} = 
%{\cal R}^{\aca}_{-1}\,\ep^{-1} + {\cal R}^{\aca}_0 + {\cal R}^{\aca}_1\,\ep,
{\cal R}^{\aca}_1\,\ep^{-1} + {\cal R}^{\aca}_2 + {\cal R}^{\aca}_3\,\ep,
\eq
%--
where the coefficients in the $\ep$ expansion are
%--
\bq
{\cal R}^{\aca}_i =
\frac{(-1)^n}{2^n\,n!\,\Bbt}
\bigg\{
\intsx{z}\,\ln^{n-1}\!\chi(z)\,\Big[ \ln\chi(z) + 2\,n \Big]
- L^n_{\chi},
\bigg\}
\qquad \qquad 
L^n_{\chi} = \bXbt\,\ln^n\,\chi(1) + \Xbt\,\ln^n\,\chi(0).
\label{Racan}
\eq
%\bqa
%{\cal R}^{\aca}_{-1} &=& - \frac{1}{2\,\Bbt}\Big[ 
%\intfx{z}\,\ln\chi(z) - L^1_{\chi} + 2 \Big]
%\nl
%{\cal R}^{\aca}_{0} &=& \frac{1}{8\,\Bbt}\Big\{ 
%\intfx{z}\,\ln\chi(z)\,\Bigl[ \ln\chi(z) + 4 \Bigr] 
%- L^2_{\chi} \Big\},
%\quad L^n_{\chi} = \bXbt\,\ln^n\,\chi(1) + \Xbt\,\ln^n\,\chi(0),
%\nl
%{\cal R}^{\aca}_{1} &=&  - \frac{1}{48\,\Bbt}\,\Bigl\{ 
%\intfx{z}\,\ln^2\,\chi(z)\,\Bigl[ \ln\,\chi(z) + 6 \Bigr] 
%- L^3_{\chi} \Bigr\}
%\label{Racan}
%\eqa
%--
The BST factor $\Bbt$ and co-factor $\Xbt$ are collected in \eqn{btfactors}.
Of course, the BST method fails when masses and external momenta are such that 
$\Bbt$ is very small. The general solution to this problem is based on the 
method of Mellin-Barnes transforms. For this particular diagram, however, the 
situation is very easy. Referring to \eqn{hereMB} we have integrals of the 
form
%--
\bq
{\cal Z}_{\aca} = - \frac{1}{\ep}\,H_{\aca}, \qquad
H_{\aca} = \intfx{z}\, \chiu{\aca}^{-1-\ep/2}(z),
\quad
\chiu{\aca}(z) \equiv \chi \lpar z\,;\,p^2\,;\, m^2_a\,,\,m^2_b\rpar
\label{abpve}
\eq
%--
which we want to evaluate in the limit $\Bbt \to 0$ ($\chi$ is defined in 
\eqn{chidef}). The integral in \eqn{abpve} is a $B_{\alpha}$ function of 
\eqn{iexam} and can be evaluated for an arbitrary values of
the exponent $\alpha$. Setting $\alpha = 1+\ep/2$ in \eqn{iexam} we obtain
%--
\bqa
B_{1+\ep/2} \lpar p^2\,;\,m^2_a\,,\,m^2_b\rpar 
&=& 
\bigg[
B \left( \frac{1}{2},\frac{1+\ep}{2} \right)\,\rho^{(1+\ep)/2}
- \sum_{X=a}^{1-a}\,\frac{X^{-1-\ep}}{1+\ep}
+ {\cal O}\big( \rho^{-1} \big)
\bigg]\,s^{-1-\ep/2}.
\nl
{}&=& 
\Bigg\{
\pi\rho^{1/2} - \sum_{X=a}^{1-a} \frac{1}{X} - 
\frac{\ep}{2}\,
\bigg[
\pi\rho^{1/2}\,\ln\frac{4s}{\rho} - 
\sum_{X=a}^{1-a} \frac{\ln X^2\,s + 2}{X}
+ {\cal O}\big( \rho^{-1} \big)
\bigg]
\Bigg\}\,s^{-1},
\eqa
%--
where $B$ is the Euler beta function and where we have introduced
%--
\bq 
p^2 =  - s, \qquad \rho = -\frac{4 s^2}{\lambda- i\,\delta}, \qquad
a = \frac{s + m_a^2 - m_b^2}{2\,s}, \qquad
\lambda= \lambda(s\,,\,m_a^2\,,\,m_b^2),
\label{a1}
\eq
%--
where $\lambda(x,y,z)$ is the K\"allen function and where we assume $s > 0$.
%--
Collecting all pieces together, we get:
%--
\bq
\ovalbox{\boldmath $V^{\aca}_i$} =
- \,\lpar \frac{\mu^2}{\pi} \rpar^{\ep}\,\egam{1+\ep}\,\Big(\, 
      \frac{1}{\ep^2}\,V^{\acan{i}}_{-2}
\,+\, \frac{1}{\ep}\,V^{\acan{i}}_{-1} 
\,+\, V^{\acan{i}}_0 \,\Big)
\eq
%--
\bqa
V^{\acan{i}}_{-2} =
2\,{\cal R}^{\aca}_1,
\qquad && \qquad
V^{\acan{i}}_{-1}=
2\,{\cal R}^{\aca}_2 - {\cal R}^{\aca}_1\,\intfx{x} \ln V_{\acan{i}},
\nl
V^{\acan{i}}_0 =
2\,{\cal R}^{\aca}_3
-\frac{\zeta(2)}{2}\,{\cal R}^{\aca}_1
&-& \intsx{x}\,\bigg[ 
  {\cal R}^{\aca}_2\,\ln V_{\acan{i}}
- \frac{{\cal R}^{\aca}_1}{4}\,\ln^2 V_{\acan{i}}
- \dssimp{2}(\{z\})\,\frac{\ln ( 1 + \zeta_{\acan{i}} )}{Z_{\acan{i}}}
\bigg].
\eqa
%--
\subsection{The $V^{\ada}$ diagram \label{vada}}
%--
The $V^{\ada}$ family of diagrams, given in \fig{TLvertada}, is representable 
as
%--
\bq
\pi^4\,V^{\ada} = \mu^{2\ep}\,\int\,d^nq_1 d^nq_2\,
\frac{1}{[1]_{\ada}[2]_{\ada}[3]_{\ada}[4]_{\ada}[5]_{\ada}[6]_{\ada}},
\eq
%--
\bqas
[1]_{\ada} &=& q_1^2 + m_1^2, \quad
[2]_{\ada} = (q_1-q_2)^2 + m_2^2, \quad
[3]_{\ada} = q_2^2 + m_3^2,
\eqas
%--
\bqa
[4]_{\ada} &=& (q_2+p_1)^2 + m_4^2, \quad
[5]_{\ada} = (q_2+P)^2 + m_5^2, \quad
[6]_{\ada} = q_2^2 + m_6^2.
\eqa
\label{defv141}
%--
\begin{figure}[ht]
\begin{center}
\begin{picture}(150,75)(0,0)
 \SetWidth{1.5}
 \Line(0,0)(40,0)         \LongArrow(0,8)(20,8)          \Text(-11,7)[cb]{$-P$}
 \Line(128,-53)(100,-35)  \LongArrow(128,-63)(114,-54)   \Text(138,-70)[cb]{$p_1$}
 \Line(128,53)(100,35)    \LongArrow(128,63)(114,54)     \Text(138,62)[cb]{$p_2$}
 \Line(57,-10)(40,0)            \Text(46,-16)[cb]{$6$}
 \CArc(70,-17.5)(15,0,360)      \Text(80,0)[cb]{$1$}\Text(58,-40)[cb]{$2$}
 \Line(100,-35)(83,-25)         \Text(89,-42)[cb]{$3$}
 \Line(100,-35)(100,35)         \Text(107,-3)[cb]{$4$}
 \Line(100,35)(40,0)            \Text(68,23)[cb]{$5$}
\end{picture}
\end{center}
\vspace{2cm}
\caption[]{The irreducible two-loop vertex diagrams $V^{\ada}$. 
External momenta are flowing inwards.} 
\label{TLvertada}
\end{figure}
%--
If $m_3 \ne m_6$ then $V^{\ada}$ is the difference of two $V^{\aca}$ diagrams,
%--
\bq
V^{\ada} = \frac{1}{m_6^2 - m_3^2}\,
\Bigl[
V^{\aca}(P^2;m_1,m_2,m_3,m_4,m_5) - V^{\aca}(P^2;m_1,m_2,m_6,m_4,m_5)
\Bigr].
\eq
%--
In general Landau equations are the same for $V^{\ada}$ and for $V^{\aca}$ so 
that the classification of potentially infrared singular $V^{\ada}$ vertices
follows closely the discussion presented in the previous section (see 
\fig{fig141}), the main difference being in the exponent of the integrand.
Configuration c) is a special case of configuration b) and will be treated 
in a second step.
%--
\begin{figure}[ht]
\begin{center}
\begin{picture}(140,75)(0,0)
 \SetWidth{1.5}
 \Line(0,0)(40,0)                     \Text(5,5)[cb]{$-P$}
 \DashLine(128,-53)(100,-35){2}       \Text(128,-65)[cb]{$p_1$}
 \DashLine(128,53)(100,35){5}         \Text(128,57)[cb]{$p_2$}
 \DashLine(100,35)(40,0){5}           \Text(65,25)[cb]{$M$}
 \DashLine(100,-35)(40,0){2}          \Text(65,-33)[cb]{$m$}
 \Photon(100,15)(100,35){2}{3}
 \Photon(100,-15)(100,-35){2}{3}
 \CArc(100,0)(15,0,360)               \Text(78,-3)[cb]{$1$}\Text(123,-3)[cb]{$2$}
 \Text(25,50)[cb]{\Large $V^{\ada}_a$}
\end{picture}
%\qquad \qquad \qquad \qquad
\qquad\qquad
\begin{picture}(140,75)(0,0)
 \SetWidth{1.5}
 \Line(0,0)(40,0)                     \Text(5,5)[cb]{$-P$}
 \DashLine(128,-53)(100,-35){2}       \Text(128,-65)[cb]{$p_1$}
 \DashLine(128,53)(100,35){5}         \Text(128,57)[cb]{$p_2$}
 \DashLine(100,35)(83,25){5}          \Text(90,36)[cb]{$M$}
 \DashLine(57,10)(40,0){5}            \Text(43,10)[cb]{$M$}
 \CArc(70,17.5)(15,0,360)             \Text(59,34)[cb]{$1$}\Text(82,-6)[cb]{$2$}
 \DashLine(100,-35)(40,0){2}          \Text(65,-33)[cb]{$m$}
 \Photon(100,-35)(100,35){2}{10}
 \Text(25,50)[cb]{\Large $V^{\ada}_b$}
\end{picture}
\qquad
\end{center}
\vspace{1.5cm}
\begin{center}
\begin{picture}(140,75)(0,0)
 \SetWidth{1.5}
 \Line(0,0)(40,0)                     \Text(5,5)[cb]{$-P$}
 \DashLine(128,-53)(100,-35){2}       \Text(128,-65)[cb]{$p_1$}
 \DashLine(128,53)(100,35){5}         \Text(128,57)[cb]{$p_2$}
 \DashLine(100,35)(83,25){5}          \Text(85,34)[cb]{$M$}
 \DashLine(57,10)(40,0){5}            \Text(46,12)[cb]{$M$}
 \DashCArc(70,17.5)(15,30,210){6}
 \PhotonArc(70,17.5)(15,-150,30){1}{10}
 \DashLine(100,-35)(40,0){2}          \Text(65,-33)[cb]{$m$}
 \Photon(100,-35)(100,35){2}{10}
 \Text(25,50)[cb]{\Large $V^{\ada}_c$}
\end{picture}
\end{center}
\vspace{2.cm}
\caption[]{The $V^{\ada}$ infrared configurations. The photon line represents 
a massless particle while the dashed and the solid lines represent different 
massive particles. Furthermore, the masses of the two particles in the bubble 
are $m_1$ and $m_2$.}
\label{fig141}
\end{figure}
%\vspace{1cm}
%--
\subsubsection{Evaluation of $V^{\ada}$, cases a) and b) \label{evaladaab}}
%--
For the first two configurations of \fig{fig141} we have
%--
\bq
V^{\ada}_i  =
- \lpar \frac{\mu^2}{\pi} \rpar^{\ep}\,\egam{2+\ep}\,
\intfx{x}\,\dsimp{2}(\{z\})\,\int_0^{1-z_1}\!\!\!dy\,
[x(1-x)]^{-\ep/2}\,y^{\ep/2-1}\,(1-z_1-y)\,
(a_{\adan{i}}\,y + Z_{\adan{i}})^{-2-\ep}
\label{innermostada}
\eq
%--
where $i = a,b$, $a_{\adan{a}} = m_x^2$, $a_{\adan{b}} = m_x^2 - M^2$, with
$m_x^2$ defined in \eqn{xdepmass} and with
%--
\bq
Z_{\adan{a}} = \beta\lpar z_2\,,\,z_1\,;\,P^2\,;\,m^2\,,\,M^2\rpar
\qquad
Z_{\adan{b}} = \beta\lpar 1-z_1\,,\,1-z_2\,;\,P^2\,;\,m^2\,,\,M^2\rpar
\label{defas}
\eq
%--
and $\beta$ defined in \eqn{betadef}. $V^{\ada}$ is also ultraviolet divergent 
and we need, once more, a procedure for extracting the ultraviolet 
pole. Consider the innermost integral in \eqn{innermostada}
%--
\bqa
V^{\ada}_i &=&
- \lpar \frac{\mu^2}{\pi} \rpar^{\ep}\,\egam{2+\ep}\,
\intfx{x}\,\dsimp{2}(\{z\})\,[x(1-x)]^{-\ep/2}\,{\cal Y}_{\adan{i}},
\nl
{\cal Y}_{\adan{i}} &=& Z_{\adan{i}}^{-2-\ep}\,\int_0^{1-z_1}\,dy\,
y^{\ep/2-1}\,(1 - z_1 - y)\,
(1 + \frac{a_{\adan{i}}}{Z_{\adan{i}}}\,y)^{-2-\ep}.
\label{generalada}
\eqa
%--
The integral of \eqn{generalada} is expressible in terms of hypergeometric 
functions,
%--
\bq
\int_0^b\,dy\,y^{s-1}\,(1 + \alpha\,y)^{-\nu} =
\frac{1}{s}\,b^s\,\hyper{\nu}{s}{s+1}{-b\alpha}, \quad
\Reb\,s > 0,\; |\arg(1 + \alpha\,b)| < \pi.
\eq
%--
The result is further transformed according to well-known properties (see 
appendix~\ref{app:hyp}) and we obtain that:
%--
\bq
{\cal Y}_{\adan{i}} = 
\frac{1}{\egam{2+\ep}}\,Z_{\adan{i}}^{-1-\ep/2}\,a_{\adan{i}}^{-\ep/2}\,
\Bigl[\,
\egam{2+\frac{\ep}{2}}\egam{\frac{\ep}{2}}(1-z_1)\,Z_{\adan{i}}^{-1}
\,-\, \egams{1+\frac{\ep}{2}}a_{\adan{i}}^{-1}
\,\Bigr] 
\,+\, \Phi_{\adan{i}},
\eq
%--
where the function $\Phi_{\adan{i}}$ is defined by the following relation:
%--
\bqa
\Phi_{\adan{i}} &=& 
a_{\adan{i}}^{-2-\ep}\,(1-z_1)^{-1-\ep/2}\,
\Bigl[ - \frac{2}{4+\ep}\,
\hyper{2+\ep}{2+\frac{\ep}{2}}{3+\frac{\ep}{2}}{-\,\zeta_{\adan{i}}}
\nl
{}&+& \frac{2}{2+\ep}\,
  \hyper{2+\ep}{1+\frac{\ep}{2}}{2+\frac{\ep}{2}}{-\,\zeta_{\adan{i}}}\Bigr]
= \frac{1}{a_{\adan{i}}\,Z_{\adan{i}}}\,\Bigl[ 
1 - \frac{\ln ( 1 + \zeta_{\adan{i}} )}{\zeta_{\adan{i}}}\,\Bigr],
\label{Phi}
\eqa
%--
where we used $\zeta_{\adan{i}} = Z_{\adan{i}}/a_{\adan{i}}(1-z_1)$.
%--
In \eqn{Phi} we have put $\ep = 0$ and the last result follows after 
using the well-known relations for the hypergeometric function collected
in appendix~\ref{app:hyp}.

After inserting these relations in \eqn{generalada} we obtain
%--
\bqa
V^{\ada}_i &=&
- \lpar \frac{\mu^2}{\pi} \rpar^{\ep}\!\egams{1+\frac{\ep}{2}}
\Bigl[
\Bigl( \frac{2}{\ep} + 1 \Bigr)\,{\cal X}^1_{\adan{i}}\,{\cal Z}^1_{\adan{i}}
\,-\, {\cal X}^2_{\adan{i}}\,{\cal Z}^2_{\ada}
\Bigr] 
\,-\, J_{\adan{i}}.
\label{lastfirst}
\eqa
%--
where we have introduced
%--
\bqa
{\cal X}^n_{\adan{i}} &=& 
\intfx{x}\,[x(1-x)]^{-\ep/2}\,a_{\adan{i}}^{1-n-\ep/2},
\quad
{\cal Z}^1_{\adan{i}} = \dsimp{2}(\{z\})\,(1-z_1)\,Z_{\adan{i}}^{-2-\ep/2},
\nl
{\cal Z}^2_{\ada} &=& \dsimp{2}(\{z\})\,Z_{\ada}^{-1-\ep/2},
\quad
J_{\adan{i}} = \dsimp{2}(\{z\})\,\intfx{x}\,\Phi_{\adan{i}},
\label{J}
\eqa
%--
Note that ${\cal X}^1_{\adan{i}}$ has been already computed while analyzing 
the $V^{\aca}$ diagram, \eqn{expandXaca} and \eqn{quadaca}; therefore
%--
\bq
{\cal X}^1_{\adan{i}} = {\cal X}_{\acan{i}}, \qquad
V_{\adan{a,b}} = V_{\acan{a,b}},
\label{Vada}
\eq
%--
For ${\cal X}^2_{\adan{i}}$ we distinguish between the two cases a) and b).
The computation of ${\cal X}^2_{\adan{a}}$ is straightforward,
since $a_{\adan{a}}$ is positive definite in $x \in [0,1]$, and gives
%--
\bq
{\cal X}^2_{\adan{i}} = 
{\cal X}^{2,0}_{\adan{i}} + {\cal X}^{2,1}_{\adan{i}}\,\ep,
\qquad
i= a,b;
\qquad\qquad\qquad
{\cal X}^{2,n}_{\adan{a}} = 
\frac{(-1)^n}{2^n\,n!}\,\intsx{x}\,x\,(1-x)\,V_{\adan{a}}^{-1}\,\ln^n V_{\adan{a}}
\eq
%\bq
%{\cal X}^2_{\adan{a}} = \intfx{x}\,x\,(1-x)\,V_{\adan{a}}^{-1}\,
%\Big[ 1 - \frac{\ep}{2}\,\ln V_{\adan{a}} \Big]
%\eq
%--
For ${\cal X}^2_{\adan{b}}$ instead we use the BST-method to increase the 
power of $V_{\adan{b}}$ and obtain:
%--
\bq
{\cal X}^{2,0}_{\adan{b}} =  
-\,\frac{1}{2B_{\ada}}\!\intsx{x}\,
\Big( \alpha_{\ada}\ln V_{\adan{b}} - \frac{1}{3} \Big),
\qquad\quad
{\cal X}^{2,1}_{\adan{b}} =  
-\,\frac{1}{2B_{\ada}}\!\intsx{x}\,\bigg\{
\ln V_{\adan{b}}\Big[ x(1-x) - \frac{\alpha_{\ada}}{4}\ln V_{\adan{b}}\Big]
\bigg\},
\eq
%\bq
%{\cal X}^2_{\adan{b}} =  -\,\frac{1}{2\,B_{\ada}}\,\intsx{x}\,\Bigg\{
%\alpha_{\ada}\,\ln V_{\adan{b}} - \frac{1}{3}
%+ \ep\,\ln V_{\adan{b}}\,\Bigl[
%x\,(1-x) - \frac{\alpha_{\ada}}{4}\,\ln V_{\adan{b}}\Bigr]\Bigg\},
%\eq
%--
where we use
%--
\bq
a_{\ada}= 3\,x^2 - 2\,(1+X_{\ada})\,x + X_{\ada},
\qquad\qquad
X_{\ada}= \frac{M^2+m_1^2-m_2^2}{2\,M^2}
\qquad\quad
B_{\ada} = - \,\frac{\lambda(M^2,m^2_1,m^2_2)}{4\,M^2}
\label{adaBTfac}
\eq
%--
The case $B_{\ada} = 0$ requires $M = m_1 \pm m_2$ which 
in all realistic situations means $m_1 (m_2)= 0$ and $m_2 (m_1)= M$.
This is equivalent to configuration c) which will be treated in the next 
subsection.

When examining ${\cal Z}^1_{\adan{i}}$ we transform 
$z_2 \rightarrow z_1\,z_2$ for $i=a$ and 
$z_1 \rightarrow 1-z_1,\; 
z_2 \rightarrow 1-z_2,\; 
z_1 \leftrightarrow z_2,\;
z_2 \rightarrow z_1\,z_2$ for $i=b$.
In both cases the $z_1$ integration can be performed, giving
%--
\bqa
{\cal Z}^1_{\adan{a}} &=&
\dcub{2}\,(1-z_1)\,z_1^{-3-\ep}\,\chi(z_2)^{-2-\ep/2} =
\frac{1}{(1+\ep)(2+\ep)}\intfx{z_2}\,\chi(z_2)^{-2-\ep/2}.
\nl
{\cal Z}_{\adan{b}}^1 &=&
\dcub{2}\,z_1^{-2-\ep}\,z_2\,\chi(z_2)^{-2-\ep/2} =
-\frac{1}{1+\ep}\intfx{z_2}\,z_2\,\chi(z_2)^{-2-\ep/2}.
\label{e20}
\eqa
%--
For the $z_2$-integration we use twice the BST-method 
followed by integration by parts and expand around $\ep = 0$ obtaining:
%--
\bq
{\cal Z}^1_{\adan{i}} = 
\frac{1}{4\,\Bbt}\,(\, {\cal R}_0^{\adan{i}} + {\cal R}_1^{\adan{i}}\,\ep \,)
\qquad\qquad
i= a,b
\eq
%--
\bqa
{\cal R}_0^{\adan{a}} &=& \frac{1}{2\,\Bbt}\,\Bigl[ 
S^{01}_{\chi} - \bXbt\,\ln M^2 - \Xbt\,\ln m^2 + 2 \Bigr]
+ \frac{\bXbt}{M^2} + \frac{\Xbt}{m^2}
\nl
{\cal R}_1^{\adan{a}} &=& -\,\frac{1}{8\,\Bbt}\, 
\lpar S^{02}_{\chi} + 8\,S^{01}_{\chi} + 8 \rpar 
+ \frac{1}{8}\,\bXbt\,\Bigl(\frac{\ln M^2}{\Bbt} - \frac{4}{M^2}\Bigr)\,
( \ln M^2 + 4 ) 
+ 
\frac{1}{8}\,\bXbt\,\Bigl(\frac{\ln m^2}{\Bbt} - \frac{4}{m^2}\Bigr)\,
( \ln m^2 + 4 )
\nl
{\cal R}_0^{\adan{b}} &=& 
-\,\frac{\Xbt}{\Bbt}\,
\lpar S^{01}_{\chi} - \bXbt\,\ln M^2 - \Xbt\,\ln m^2 + 2 \rpar 
- 2\,\frac{\bXbt}{M^2}
\nl
{\cal R}_1^{\adan{b}} &=& 
\frac{1}{4\,\Bbt}\,\Bigl[ 
\Xbt\,S^{02}_{\chi} - 8\,S^{11}_{\chi} + 14\,\Xbt\,S^{01}_{\chi} 
- \Xbt^2\,\ln m^2\,( \ln m^2 + 6 )
- \bXbt\,\ln M^2\,( \Xbt\,\ln M^2 + 6\,\Xbt - 4 )
\nl
{}&-& 4 + 12\,\Xbt \Bigr]
+ \frac{\bXbt}{M^2}\,( \ln M^2 + 3 ),
\qquad
S^{nm}_{\chi} = \intfx{z}\,z^n\,\ln^m \chi(z).
\eqa
%--
%\bqa
%{\cal R}_0^{\adan{a}} &=& \frac{1}{2\,\Bbt}\,\Bigl[ 
%\intfx{z}\,\ln\chi(z) - \bXbt\,\ln M^2 - \Xbt\,\ln m^2 + 2 \Bigr]
%+ \frac{\bXbt}{M^2} + \frac{\Xbt}{m^2}
%\nl
%{\cal R}_1^{\adan{a}} &=& -\,\frac{1}{8\,\Bbt}\, 
%\Bigl\{ \intfx{z}\,\ln\chi(z)\,\Big[ \ln\chi(z) + 8 \Big] + 8\Bigr\}
%+ \frac{1}{8}\,\bXbt\,\Bigl(\frac{\ln M^2}{\Bbt} - \frac{4}{M^2}\Bigr)\,
%( \ln M^2 + 4 ) 
%\nl
%{}&+& 
%%+
%\frac{1}{8}\,\bXbt\,\Bigl(\frac{\ln m^2}{\Bbt} - \frac{4}{m^2}\Bigr)\,
%( \ln m^2 + 4 )
%\nl
%{\cal R}_0^{\adan{b}} &=& 
%-\,\frac{\Xbt}{\Bbt}\,
%\Bigl[ \intfx{z}\,\ln\chi(z) - \bXbt\,\ln M^2 - \Xbt\,\ln m^2 + 2 \Bigr] 
%- 2\,\frac{\bXbt}{M^2}
%\nl
%{\cal R}_1^{\adan{b}} &=& 
%\frac{1}{4\,\Bbt}\,\bigg\{ 
%\intfx{z}\,\ln\chi(z)\,\Bigl[ \Xbt\,\ln\chi(z) - 8\,z + 14\,\Xbt \Bigr]
%- \Xbt^2\,\ln m^2\,( \ln m^2 + 6 )
%\nl
%{}&-& \bXbt\,\ln M^2\,( \Xbt\,\ln M^2 + 6\,\Xbt - 4 )
% - 4 + 12\,\Xbt \bigg\}
%+ \frac{\bXbt}{M^2}\,( \ln M^2 + 3 ).
%\eqa
%--
For ${\cal Z}^2_{\ada}$ we can follow the derivation given for $V^{\aca}$:
%--
\bq
{\cal Z}^2_{\ada} = 
%{\cal R}^{\ada}_{-1}\,\ep^{-1} + {\cal R}^{\ada}_0 + {\cal R}^{\ada}_1\,\ep,
{\cal R}^{\ada}_1\,\ep^{-1} + {\cal R}^{\ada}_2,
\qquad\qquad {\cal R}^{\ada}_n = {\cal R}^{\aca}_n,
\eq
%--
with ${\cal R}^{\aca}_n$ given in \eqn{Racan}.
%--
%\bqa
%{\cal R}^{\ada}_{-1} &=& - \frac{1}{2\,\Bbt}\Big[ 
%\intfx{z}\,\ln\chi(z) - \bXbt\,\ln\chi(1) - \Xbt\,\ln\chi(0) + 2 \Big]
%\nl
%{\cal R}^{\ada}_{0} &=& \frac{1}{8\,\Bbt}\Big\{ 
%\intfx{z}\,\ln\chi(z)\,\Bigl[ \ln\chi(z) + 4 \Bigr] 
%- \bXbt\,\ln^2\,\chi(1) - \Xbt\,\ln^2\,\chi(0) \Big\}
%\nl
%{\cal R}^{\ada}_{1} &=& - \frac{1}{48\,\Bbt}\,\Bigl\{ 
%\intfx{z}\,\ln^2\,\chi(z)\,\Bigl[ \ln\,\chi(z) + 6 \Bigr] 
%- \bXbt\,\ln^3\,\chi(1) - \Xbt\,\ln^3\,\chi(0)
%\Bigr\}.
%\eqa
%--
One expression remains to be computed:
%--
\bq
J_{\adan{i}} = \dsimp{2}(\{z\})\,\intfx{x}\,
\frac{1}{a_{\adan{i}}\,Z_{\adan{i}}}\,\Bigl[ 
1 - \frac{\ln ( 1 + \zeta_{\adan{i}} )}{\zeta_{\adan{i}}}\,\Bigr].
\eq
%--
Here there is an additional problem, namely the integrand is well-behaved in 
the limit $Z_{\adan{i}} \to 0$ but not when $a_{\adan{i}} \to 0$.

For case a), $a_{\adan{a}}$ never vanishes in the integration interval and 
we can simply write:
%--
\bq
J_{\adan{a}} = \dcub{2}(\{z\})\,\intfx{x}\,
\frac{1}{m_x^2\,z_1\,\chi(z_2)}\,\Bigl[ 
1 - \frac{\ln ( 1 + \eta_{\adan{a}} )}{\eta_{\adan{a}}}\Bigr],
\qquad\qquad
\eta_{\adan{a}}= \frac{z_1^2\,\chi(z_2)}{m_x^2\,(1-z_1)}.
\eq
%--
On the contrary, to compute $J_{\adan{b}}$ is more convenient to reexamine
\eqn{Phi}. Setting $\ep = 0$ we have:
%--
\bqa
J_{\adan{b}} &=& -\,\dsimp{2}(\{z\})\,\intfx{x}\,
a_{\adan{b}}^{-2}\,(1-z_1)^{-1}\,
\Bigl[ \frac{1}{2}\,\hyper{2}{2}{3}{-\,\zeta_{\adan{b}}} 
- \hyper{2}{1}{2}{-\,\zeta_{\adan{b}}}\Bigr]
\eqa
%--
As the next step we use the BST relation of \eqn{BThyper} which, in the 
present case, reads:
%--
\bqa
V_{\adan{b}}^{-2}\,\hyper{2}{b}{c}{\frac{A_{\adan{b}}}{V_{\adan{b}}}} &=&
\frac{1}{B_{\ada}}\,\Bigg\{
V_{\adan{b}}^{-1}\,\hyper{2}{b}{c}{\frac{A_{\adan{b}}}{V_{\adan{b}}}}
+ \frac{x-X_{\ada}}{2}\,\partial_x\,  \Big[ 
%+ \frac{1}{2}\,(x-X_{\ada})\,\partial_x\,  \Big[ 
  V_{\adan{b}}^{-1}\,\hyper{1}{b}{c}{\frac{A_{\adan{b}}}{V_{\adan{b}}}}  \Big]
\nl
%{}&-& \frac{1}{2}\,(x-X_{\ada})\,
{}&-& \frac{x-X_{\ada}}{2}\,
  \frac{\partial_x\,A_{\adan{b}}}{A_{\adan{b}}}\,V_{\adan{b}}^{-1}\,  \Big[ 
  \hyper{2}{b}{c}{\frac{A_{\adan{b}}}{V_{\adan{b}}}}
   - \hyper{1}{b}{c}{\frac{A_{\adan{b}}}{V_{\adan{b}}}}  \Big]\Bigg\}.
\eqa
%--
where $B_{\ada}$ and $X_{\ada}$ are defined in \eqn{adaBTfac} and 
$A_{\adan{b}}= - x\,(1-x)\,Z_{\adan{b}}/(1-z_1)$. After integration by parts 
we get:
%--
\bqa
J_{\adan{b}} &=& \frac{1}{2\,B_{\ada}}\,\dsimp{2}(\{z\})\,\intfx{x}\,
\frac{1}{Z_{\adan{b}}}\Big\{3\,x\,(1-x)\,\Big[ 
1 - \frac{\ln ( 1 + \zeta_{\adan{i}} )}{\zeta{\adan{i}}}\,\Big]
+ a_{\ada}\,\ln ( 1 + \zeta_{\adan{i}} )\Big\}
\eqa
%--
where $a_{\ada}= 3\,x^2 - 2\,(1+X_{\ada})\,x + X_{\ada}$ as 
defined in \eqn{adaBTfac}.
To obtain the last formula we have used the explicit expression for 
the hypergeometric functions given in appendix~\ref{app:hyp}.

The case where masses and momenta are such that $\Bbt$ is vanishing small 
must be treated separately.
In the following we will show how Mellin-Barnes techniques can be used
in this kinematical situation (for the cases under discussion).
%--
For case a) the situation is particularly simple: from \eqn{e20}
we see that when $\Bbt$ is very small we have to evaluate 
%the integral
$B_{2+\ep/2}(P^2\,;\,m^2\,,\,M^2)$
%--
%\bq 
%\intfx{x}\,\chi^{-2-\ep/2}\lpar x\,;\,P^2\,;\,m^2\,,\,M^2\rpar
%\label{abpvetwo}
%\eq
%--
%with $\chi$ defined in \eqn{chidef}
, which is of the general form of 
\eqn{iexam}. With $\alpha = 2+\ep/2$ in \eqn{iexam} and discarding the 
terms with positive powers of $\lambda(-P^2,m^2,M^2)$ we have,
%--
\bq 
B_{2+\ep/2} = 
\bigg[
B \left( \frac{1}{2},\frac{3 + \ep}{2} \right)\,\rho)^{3/2+\ep/2} 
- \frac{1}{3+\ep}\,\left[ a^{-3 -\ep} + (1-a)^{-3-\ep}\right] 
+ {\cal O}\big( \rho^{-1} \big)
\bigg]\,s^{-2-\ep/2},
\label{gd}
\eq
%--
where we have introduced $P^2 =  - s\, , \, \rho = - (4 s^2)/\lambda$ and
$a = (s + m^2 - M^2)/(2\,s)$.
It suffices to expand the result of \eqn{gd} in powers of
$\ep$ and to retain the first two terms in the expansion:
%--
\bq
B_{2+\ep/2} = [ I^{(0)} + I^{(1)}\,\ep + \ord{\ep^2} + 
\ord{\rho^{-1}} ]\,s^{-2},
\eq
%--
%where the $I^{(i)}$ are given by
%--
\bq
I^{(0)} = 
\frac{\pi}{2}\,\rho^{3/2} 
- \frac{1}{3}\,\sum_{X=a}^{1-a} \frac{1}{X^3},
%- \frac{1}{3}\,\left[\frac{1}{(1-a)^3}+ \frac{1}{a^3} \right],
\qquad\qquad
I^{(1)} =  
\frac{\pi}{4}\,\rho^{3/2}\,\left( 1+ \ln\frac{\rho}{4\,s} \right) 
+ \frac{1}{18}\,\sum_{X=a}^{1-a} \frac{3\ln X^2\,s + 2}{X^3}.
%\frac{1}{9}\, \left[\frac{1}{(1-a)^3}+ \frac{1}{a^3} + 3\,
%\left(\frac{\ln(1-a)}{(1-a)^3}+ \frac{\ln a}{a^3} \right)
%\right]\, .
\label{exp}
\eq
%--
%
% The original text is at the and of the file
%
The case b) is quite similar; consider the integral
%--
\bq
Y_{\alpha} = 
\intfx{x}\, x \left[ \chi(x) - i\delta \right]^{-\alpha}
= \intfx{x}\, (x-a)\, \left[ \chi(x) - i\delta \right]^{-\alpha} + a\,B_{\alpha},
\label{yalpha}
\eq
%--
Since $\chi(x)= s\,[(x-a)^2 + \rho^{-1}]$, the integral in \eqn{yalpha}  
gives:
%--
\bq
Y_{\alpha} = 
\frac{M^{2(1-\alpha)}+m^{2(1-\alpha)}}{2\,s\,(1-\alpha)} + a\,B_{\alpha}.
\eq
%--
Setting $\alpha=2+\ep/2$ and expanding in $\ep$ we get:
%--
\bq
Y_{2+\ep/2} = 
-\,\frac{1}{2\,s}\,
\Bigg\{
\frac{1}{M^2} - \frac{1}{m^2}
- \frac{\ep}{2}\,\bigg[ \frac{\ln M^2+1}{M^2} - \frac{\ln m^2+1}{m^2} \bigg]
\Bigg\}
+ a\,B_{2+\ep/2},
\eq
%--
which conclude our discussion.
Collecting all pieces together, we get:
%--
\bq
\ovalbox{\boldmath $V^{\ada}_i$} =
- \,\lpar \frac{\mu^2}{\pi} \rpar^{\ep}\,\egam{1+\ep}\,
\Big(\, \frac{1}{\ep}\,V^{\adan{i}}_{-1} \,+\, V^{\adan{i}}_0 \,\Big),
\qquad\qquad i= a,b
\eq
%--
\bq
V^{\adan{i}}_{-1} =
  \frac{{\cal R}_0^{\adan{i}}}{2\,\Bbt} 
- {\cal X}_{\adan{i}}^{2,0}\,{\cal R}_1^{\ada},
\qquad\quad
V^{\adan{i}}_0 =
  \frac{{\cal R}_1^{\adan{i}}}{2\,\Bbt} 
- \frac{{\cal R}_0^{\adan{i}}}{4\,\Bbt}
  \Big( \intsx{x}\ln V_{\adan{a}} - 1 \Big)
- {\cal X}_{\adan{i}}^{2,0}\,{\cal R}_2^{\ada}
- {\cal X}_{\adan{i}}^{2,1}\,{\cal R}_1^{\ada}
+ J_{\adan{i}}.
\eq
%--
\subsubsection{Evaluation of $V^{\ada}_c$ \label{evaladac}}
%--
Configuration c) of \fig{fig141} is a special case of b).
For this case the polynomial $V_{\ada}$ (see \eqn{Vada}) takes the form
%--
\bq
V_{\adan{c}}(x) = x^2\,M^2 \qquad \hbox{giving} \qquad
a_{\adan{c}}(x) = \frac{x}{1-x}\,M^2.
\eq
%--
$V_{\adan{c}}(x)$ has a double zero for $x = 0$ and the corresponding 
BST factor is therefore zero.
As a consequence the diagram acquires an extra infrared pole, confirming 
the presence of a leading Landau singularity.
To evaluate this special case, we reconsider \eqn{innermostada}, which 
in the present configuration reads:
%--
\bq
V^{\ada}_c =
-\!\lpar \frac{\mu^2}{\pi} \rpar^{\ep}\!\!\egam{2+\ep}\!\!
\intsx{x}\!\dssimp{2}(\{z\})\!\!\int_0^{1-z_1}\!\!\!\!\!\!\!\!\!\!\!dy\,\,\,
[x\,\!(1-x)]^{-\ep/2}\,y^{\ep/2-1}(1-z_1-y)
\Big(\, \frac{x\,y\,M^2}{1-x} + Z_{\adan{c}} \Big)^{-2-\ep},
\eq
%--
\bq
Z_{\adan{c}} = \beta\lpar 1 - z_1\,,\,1- z_2\,;P^2\,;\,m^2\,,\,M^2\rpar,
\eq
%\bqa
%V^{\ada}_c &=&
%- \lpar \frac{\mu^2}{\pi} \rpar^{\ep}\egam{2+\ep}\,
%\intsx{x}\dsimp{2}(\{z\})\,\int_0^{1-z_1}\!\!\!\!\!\!dy\,\,
%[x(1-x)]^{-\ep/2}\,y^{\ep/2-1}\,(1-z_1-y)\nl
%{}&\times&
%\Big(\, \frac{x\,y\,M^2}{1-x} + Z_{\adan{c}} \,\Big)^{-2-\ep},
%\qquad
%Z_{\adan{c}} = \beta\lpar 1 - z_1\,,\,1- z_2\,;,P^2\,;\,m^2\,,\,M^2\rpar,
%\eqa
%--
with $\beta$ defined in \eqn{betadef}. We perform the following set of 
transformations:
$y = (1-z_1)\,y' \,,$ and
$z_1 = 1-z'_1 \,,\, 
z_2 = 1-z'_2 \,,\,$ followed by
$z'_1 = z''_1\,z'_2$,
obtaining:
%--
\bq
V^{\ada}_c = - \lpar \frac{\mu^2}{\pi} \rpar^{\ep}\,\egam{2+\ep}\,
\dcub{4}(x,y,\{z\})\,
\lpar\frac{z_1}{z_2}\,\frac{y}{x(1 - x)}\rpar^{\ep/2}\,\frac{1 - y}{y}\,z_1
\Big[\, \frac{x\,y\,z_1\,M^2}{1-x} + z_2\,\chi(z_1) \,\Big]^{-2-\ep},
\eq
%--
where $\chi(z_1) \equiv \chi\lpar z_1\,;\,P^2\,;\,m^2\,,\,M^2\rpar$ of 
\eqn{chidef}.
Integrating over $z_2$ gives an hypergeometric function and we use its 
properties (see appendix \ref{app:hyp}) to get:
%--
\bqa
V^{\ada}_c &=&
- \lpar \frac{\mu^2}{\pi} \rpar^{\ep}\,\frac{2}{2-\ep}\,\egam{2+\ep}\,
\dcub{3}(x,y,z)\,x^{-2-3\ep/2}\,(1-x)^{2+\ep/2}\,\,z^{-1-\ep/2}\,
\nl
{}&\times& y^{-3-\ep/2}\,(1-y)(M^2)^{-2-\ep}\,
\hyper{2+\ep}{1-\frac{\ep}{2}}{2-\frac{\ep}{2}}
      {-\,\frac{(1-x)\,\chi(z)}{x\,y\,z\,M^2}}
\nl
{}&=& - \,\lpar \frac{\mu^2}{\pi} \rpar^{\ep}\,\dcub{3}(x,y,z)\,
\bigg[\,\egam{1-\frac{\ep}{2}}\,\egam{1+\frac{3}{2}\,\ep}\,
x^{-1-2\ep}\,(1-x)^{1+\ep}\,z^{-\ep}\,
\nl
{}&\times& y^{-2-\ep}\,(1-y)\,(M^2)^{-1-3\ep/2}\,\chi^{-1+\ep/2}(z)\,\,
-\,\,\frac{2}{2+3\,\ep}\,\egam{2+\ep}\,
[x\,(1-x)]^{-\ep/2}\,z^{1+\ep/2}\,
\nl
{}&\times& y^{-1+\ep/2}\,(1-y)\,\chi^{-2-\ep}(z)\,
\hyper{2+\ep}{1+\frac{3}{2}\,\ep}{2+\frac{3}{2}\,\ep}
      {-\,\frac{x\,y\,z\,M^2}{(1-x)\,\chi(z)}}\,\bigg].
\label{firstterm}
\eqa
%--
In the first term of \eqn{firstterm} the integrations over $x$ and $y$ are 
trivial and generate the double pole in $\ep$.
The second term of \eqn{firstterm} shows a single pole, hidden in the 
$y$ integration, which can be extracted as in \eqn{poleext}.
We obtain:
%--
\bq
V^{\ada}_c =
- \,\lpar \frac{\mu^2}{\pi} \rpar^{\ep}\,
\Big(\, V^{\ada}_{c,0} \,+\, V^{\ada}_{c,1} \,+\, V^{\ada}_{c,2} \,\Big)
\eq
%--
\bqa
V^{\ada}_{c,0} &=&
\egam{1-\frac{\ep}{2}}\,\egam{1+\frac{3}{2}\,\ep}\,
B(-2\,\ep,2+\ep)\,B(-1-\ep,2)\,(M^2)^{-1-3\ep/2}\,
\intsx{z}\,z^{-\ep}\,\chi^{-1+\ep/2}(z),
\nl
V^{\ada}_{c,1} &=&
-\,\frac{2}{\ep}\,\frac{2}{2+3\,\ep}\,\egam{2+\ep}\,
B\Big( 1-\frac{\ep}{2},1-\frac{\ep}{2} \Big)\,
\intsx{z}\,z^{1+\ep/2}\,\chi^{-2-\ep}(z),
\nl
V^{\ada}_{c,2} &=&
\dcub{3}(x,y,z)\,\frac{z}{\chi^2(z)}\,
\frac{(1-x)\,\chi(z)+x\,z\,M^2}{(1-x)\,\chi(z)+x\,y\,z\,M^2}.
\eqa
%--
For $V^{\ada}_{c,0}$ we perform a Laurent expansion in $\ep$ which gives:
%--
\bq
V^{\ada}_{c,0}= - \,\frac{1}{2\,M^2}\,\egam{1+\ep}\,
\Big(\,  {\cal R}^{\adan{c}}_{-2}\,\ep^{-2}
+ {\cal R}^{\adan{c}}_{-1}\,\ep^{-1}
+ {\cal R}^{\adan{c}}_0\,\Big),
\eq
%--
\bqa
{\cal R}^{\adan{c}}_{-2} &=& J_0^0, 
\qquad\qquad
{\cal R}^{\adan{c}}_{-1} =
\Big(  1 - \frac{3}{2}\,\ln M^2 \Big)\,J_0^0
 + \frac{1}{2}\,J_0^1 - J_1^0,
\nl
{\cal R}^{\adan{c}}_0 &=&
  \Big( \frac{9}{8}\,\ln^2 M^2 - \frac{3}{2}\,\ln M^2 
  + \frac{11}{4}\,\zeta(2) + 1 \Big)\,J_0^0
+ \Big(  1 - \frac{3}{2}\,\ln M^2 \Big)\,
  \Big( \frac{1}{2}\,J_0^1 - J_1^0 \Big)\,
+ \frac{1}{8}\,J_0^2 - \frac{1}{2}\,J_1^1 + \frac{1}{2}\,J_2^0,
\nn
\eqa
%--
where $J_n^k$ is defined by:
%--
\bq
J_n^k= \intsx{z}\,\ln^n z\,\frac{\ln^k\chi(z)}{\chi(z)}
\label{defJnk}
\eq
%--
This class of integrals can be treated by using \eqn{BTlogn} which, in the 
present case, reads as follows:
%--
\bq
\chi^{-1}(z)\,\ln^k \chi(z)=
\frac{1}{\Bbt}\,
\Big[
\ln^k \chi(z) - \frac{z-\Xbt}{2\,(k+1)}\,\partial_z\,\ln^{k+1}\chi(z)
\Big].
\eq
%--
After integration by parts, we get:
%--
\bqa
J_n^0 &=& \frac{1}{2\,\Bbt}\,\Big[
\intsx{z}\,\ln^{n-1}z\,\Big( n\,\frac{z-\Xbt}{z} + \ln z \Big)\,
\ln\frac{\chi(z)}{\chi(0)} + 2\,(-1)^{n}\,\egam{n+1}\Big],
\qquad\quad n\ge 1,
\nl 
J_n^1 &=& \frac{1}{4\,\Bbt}\,\intsx{z}\,\Big\{
4\,\ln^n z\,\ln\chi(z)
+ \ln^{n-1}z\,\Big( n\,\frac{z-\Xbt}{z} + \ln z \Big)\,
\ln\frac{\chi(z)}{\chi(0)}\,[ \ln\chi(z) + \ln\chi(0) ]
\Big\}, \qquad\quad n \ge 1,
\nl 
J_0^{k-1} &=& \frac{1}{2\,\Bbt}\,\frac{1}{k}\,\Big\{
\intsx{z}\,\ln^{k-1}\chi(z)\,[ \ln\chi(z) + 2\,k ]
- L^k_{\chi}\Big\},
\label{evalJnk}
\eqa
%--
with $L^k_{\chi}$ given in \eqn{Racan}. The same strategy is adopted for 
the second integral, thus obtaining:
%--
\bq
V^{\ada}_{c,1}= \egam{1+\ep}\,
\Big(\, -\,2\,I_0^0\,\ep^{-1} - I_0^0 - I_1^0 + 2\,I_0^1 \,\Big)
\eq
%--
where $I_n^k$ is now defined by:
%--
\bq
I_n^k= \intsx{z}\,z\,\ln^n z\,\frac{\ln^k\chi(z)}{\chi^2(z)}
\label{defInk}
\eq
%--
This class of integrals can be treated by using \eqn{BTmlogn} which, in the 
present case, becomes:
%--
\bq
\chi^{-2}(z)\,\ln^k\chi(z)= \frac{1}{\Bbt}\, \Big[
\chi^{-1}(z)\,\ln^k \chi(z)
+ \frac{z-\Xbt}{2}\,\sum_{l=0}^{k}\,\frac{k!}{l!}\,
  \partial_z\,\chi^{-1}(z)\,\ln^{l} \chi(z)\Big].
\eq
%--
After integration by parts, we repeat the whole procedure for $I_n^k$;
the BST relation to be used in this second step is:
%--
\bq
\chi^{-1}(z)\,\ln^k\chi(z)= \frac{1}{\Bbt}\,\Big[
\ln^k \chi(z) - \frac{z-\Xbt}{2\,(k+1)}\,\partial_z\,\ln^{k+1}\chi(z)\Big].
\eq
%--
The second integration by parts gives:
%--
\bqa
I_0^k &=& \frac{1}{4\,\Bbt^2}\,\bigg\{\intsx{z}\,\bigg[
  \frac{\Xbt}{k+1}\,\ln^{k+1}\chi(z)
- (4\,z-5\,\Xbt)\,\ln^k\chi(z)
- (8\,z-5\,\Xbt)\,k!\,{\cal L}_{k-1}(z)\bigg]
\nl
{}&-& \bXbt\,\bigg[
  \frac{\Xbt}{k+1}\,\ln^{k+1}\chi(1)
- (2-\Xbt)\,k!\,{\cal L}_k(1) \bigg]
- k!\,\bXbt\,\Big[ \bXbt\,{\cal L}_{k+1}(0) + 2 \Big]
\bigg\}
 + \frac{k!}{2}\,\frac{\bXbt}{\Bbt}\,\frac{1 + {\cal L}_k(1)}{\chi(1)}
\nl
I_1^0 &=& \frac{1}{4\,\Bbt^2}\intsx{z}\,\bigg\{
  \Big( \Xbt\!\ln z - 2\,z + 3\,\Xbt \Big)\,\ln\chi(z)
- \frac{\Xbt^2}{z}\ln\frac{\chi(z)}{\chi(0)}
+ \bXbt^2\!\ln\chi(1)- \Xbt^2\!\ln\chi(0) - 1 \bigg\},
\label{evalInk}
\eqa
%--
%\bq
%{\cal L}_k(z) = \sum_{l=1}^{k}\,\frac{1}{l!}\,\ln^l\chi(z).
%\eq
%--
where $ {\cal L}_k(z) = \sum_{l=1}^{k}\,\ln^l\chi(z)/l! $.
Finally we evaluate $V^{\ada}_{c,2}$. First the integration over $y$ 
is performed:
%--
\bq
V^{\ada}_{c,2} = \intsxx{x}{z}\,
\bigg[ \frac{1-x}{x\,M^2\,\chi(z)} + \frac{z}{\chi^2(z)} \bigg]\,
\ln \Big( 1 + \frac{x\,z\,M^2}{(1-x)\,\chi(z)} \Big).
\label{ast}
\eq
%--
In the second term of \eqn{ast} we transform the argument of the logarithm to 
get
%--
%In the second term of \eqn{ast} we use:
%%--
%\bq
%\ln \Big( 1 + \frac{a-i\,\delta}{b-i\,\delta} \Big)=
%\ln(a-i\,\delta) - \ln(b-i\,\delta)
%+ \ln \Big( 1 + \frac{b-i\,\delta}{a-i\,\delta} \Big),
%\eq
%%--
%and get
%--
\bq
V^{\ada}_{c,2} = \intsxx{x}{z}\,\bigg[ 
\frac{1-x}{x\,M^2\,\chi(z)}\,
\ln \Big( 1 + \frac{x\,z\,M^2}{(1-x)\,\chi(z)} \Big)
+ \frac{z}{\chi^2(z)}\,
\ln \Big( 1 + \frac{(1-x)\,\chi(z)}{x\,z\,M^2} \Big)\bigg]
+ \ln M^2\,I_0^0 + I_1^0 - I_0^1,
\label{againft}
\eq
%--
where $I_n^k$ is given by \eqn{evalInk}. The first two terms in \eqn{againft} 
are treated according to \eqn{BTli1} and \eqn{BT2li1} which in, the present 
case, become
%--
\bqa
\frac{1}{\chi(z)}\,\ln\Big(1-\frac{a\,z}{\chi(z)}\Big) &=&
\frac{1}{\Bbt}\,\Bigg[
\frac{1}{2}\,\Big( 1 + \frac{\Xbt}{z} \Big)\,
\ln\Big(1-\frac{a\,z}{\chi(z)}\Big)\,
- \frac{z-\Xbt}{2}\,\partial_z\,\li{2}{\frac{a\,z}{\chi(z)}}\Bigg],
\\
\frac{z}{\chi^2(z)}\,\ln\Big(1-\frac{\chi(z)}{a\,z}\Big)
&=& \frac{1}{\Bbt}\,\Bigg[
z\,\Big( 1 + \frac{z-\Xbt}{2}\,\partial_z \Big)
\frac{1}{\chi(z)}\,\ln\Big(1-\frac{\chi(z)}{a\,z}\Big)\,
- \frac{z-\Xbt}{2\,a}\,\partial_z\,\ln\Big(1-\frac{a\,z}{\chi(z)}\Big)\Bigg].
\nn
\eqa
%--
where $a= - x\,M^2/(1-x)$. After integration by parts, we finally obtain:
%--
\bqa
V^{\ada}_{c,2} 
&=& 
\frac{1}{2\,\Bbt}\,\intsxx{x}{z}\,\Bigg\{
\frac{1-x}{x\,M^2}\,\bigg[ 
  \frac{\Xbt}{z}\,\ln \Big( 1 + \frac{x\,z\,M^2}{(1-x)\,\chi(z)} \Big)
+ \li{2}{-\frac{x\,z\,M^2}{(1-x)\,\chi(z)}}\bigg]
\nl
&{}&
+ \frac{\Xbt}{\chi(z)}\,\ln \Big( 1 + \frac{(1-x)\,\chi(z)}{x\,z\,M^2} \Big)
\Bigg\} + \frac{\bXbt}{M^2\,\Bbt}\,\zeta(3) + \ln M^2\,I_0^0 + I_1^0 - I_0^1
\eqa
%--
Collecting all pieces together, we get:
%--
\bq
\ovalbox{\boldmath $V^{\ada}_c$} =
- \,\lpar \frac{\mu^2}{\pi} \rpar^{\ep}\,\egam{1+\ep}\,\Big(\, 
      \frac{1}{\ep^2}\,V^{\adan{c}}_{-2}
\,+\, \frac{1}{\ep}\,V^{\adan{c}}_{-1} 
\,+\, V^{\adan{c}}_0 \,\Big)
\eq
%--
\bq
V^{\adan{c}}_{-2}=
- \frac{{\cal R}^{\adan{c}}_{-2}}{2\,M^2},
\qquad
V^{\adan{c}}_{-1}=
- \frac{{\cal R}^{\adan{c}}_{-1}}{2\,M^2} - 2\,I_0^0,
\qquad
V^{\adan{c}}_0=
- \frac{{\cal R}^{\adan{c}}_0}{2\,M^2} 
- I_0^0 -I_1^0 + 2\,I_0^1 + V^{\ada}_{c,2}\,.
\eq
%--
\subsection{The $V^{\bba}$ diagram  \label{vbba}}
%--
This family of diagrams, corresponding to \fig{TLvertbba}, can be written as
%--
\bq
\pi^4\,V^{\bba} =  \mu^{2\ep}\,\int\,d^nq_1 d^nq_2\,
\frac{1}{[1]_{\bba}[2]_{\bba}[3]_{\bba}[4]_{\bba}[5]_{\bba}},
\eq
%--
where we have introduced the following notation:
%--
\bqas
[1]_{\bba} = q^2_1 + m^2_1,  \quad
[2]_{\bba} = (q_1+p_1)^2+ m^2_2,  \quad
[3]_{\bba} = (q_1-q_2)^2 + m^2_3,  
\eqas
%--
\bq
[4]_{\bba} = (q_2+p_1)^2 + m^2_4,  \quad
[5]_{\bba} = (q_2+P)^2 + m^2_5.
\label{def221}
\eq
%--
%\begin{figure}[ht]
\begin{figure}[t]
\begin{center}
\begin{picture}(150,75)(0,0)
 \SetWidth{1.5}
 \Line(0,0)(40,0)         \LongArrow(0,8)(20,8)          \Text(-11,7)[cb]{$-P$}
 \Line(128,-53)(100,-35)  \LongArrow(128,-63)(114,-54)   \Text(138,-70)[cb]{$p_1$}
 \Line(128,53)(100,35)    \LongArrow(128,63)(114,54)     \Text(138,62)[cb]{$p_2$}
 \Line(100,-35)(40,0)           \Text(70,-33)[cb]{$1$}
 \Line(100,0)(100,35)           \Text(107,-21)[cb]{$2$}
 \Line(100,0)(40,0)             \Text(83,5)[cb]{$3$}
 \Line(100,-35)(100,0)          \Text(107,15)[cb]{$4$}
 \Line(100,35)(40,0)            \Text(70,25)[cb]{$5$}
\end{picture}
\end{center}
\vspace{2cm}
\caption[]{The irreducible two-loop vertex diagrams $V^{\bba}$. External 
momenta are flowing inwards.} 
\label{TLvertbba}
\end{figure}
%--
The leading Landau singularity corresponds to the following four solutions:
%--
\bqa
P^2 &=& - \frac{1}{4\,m_2^2\,m_4^2}\,
\Big[
  (m_2^2-m_3^2+m_4^2)\,(p_1^2+m_1^2-m_2^2)\,(-p_2^2+m_4^2-m_5^2)
- 4\,m_2^2\,m_4^2\,(p_1^2+p_2^2)
\\
{}&{}&
- \eta\,(p_2^2-m_4^2+m_5^2)\,(\lambda_{12})^{1/2}
+ \sigma\,(p_1^2+m_1^2-m_2^2)\,(\lambda_{13})^{1/2}
+ \,\sigma \eta\,(m_2^2-m_3^2+m_4^2)\,(\lambda_{23})^{1/2}\Big],
\nl
\lambda_{ij} &=& \lambda_i\,\lambda_j, \qquad
\lambda_1 = \lambda(m_2^2,m_3^2,m_4^2), \;
\lambda_2 = \lambda(-p_1^2,m_1^2,m_2^2), \;
\lambda_3 = \lambda(-p_2^2,m_4^2,m_5^2).
\eqa
%--
where $\eta\,\sigma = \pm 1$. 
From a study of these conditions it follows that $V^{\aba}$ is free from
QED - like infrared divergent configurations.
%--
\subsection{The $V^{\bca}$ diagram \label{vbca}}
%--
This family of diagrams, depicted in \fig{TLvertbca}, can be written as
%--
\bq
\pi^4\,V^{\bca} =  \mu^{2\ep}\,
\int\,d^nq_1 d^nq_2\,
\frac{1}{[1]_{\bca}[2]_{\bca}[3]_{\bca}[4]_{\bca}[5]_{\bca}[6]_{\bca}},
\label{defBCA}
\eq
%--
where we have introduced the following notation for propagators:
%--
\bqas
[1]_{\bca} &=& q^2_1 + m^2_1,  \quad
[2]_{\bca} =  (q_1+P)^2+ m^2_2,  \quad
[3]_{\bca} =  (q_1-q_2)^2 + m^2_3,  
\eqas
%--
\bqa
[4]_{\bca} &=& q^2_2 + m^2_4,  \quad
[5]_{\bca} =   (q_2+p_1)^2 + m^2_5,  \quad
[6]_{\bca} =   (q_2+P)^2 + m^2_6,
\label{defbca}
\eqa
%--
\begin{figure}[ht]
\begin{center}
\begin{picture}(150,75)(0,0)
 \SetWidth{1.5}
 \Line(0,0)(40,0)         \LongArrow(0,8)(20,8)          \Text(-11,7)[cb]{$-P$}
 \Line(128,-53)(100,-35)  \LongArrow(128,-63)(114,-54)   \Text(138,-70)[cb]{$p_1$}
 \Line(128,53)(100,35)    \LongArrow(128,63)(114,54)     \Text(138,62)[cb]{$p_2$}
 \Line(70,-17.5)(40,0)             \Text(53,-21)[cb]{$1$}
 \Line(70,17.5)(40,0)              \Text(53,14)[cb]{$2$}
 \Line(70,-17.5)(70,17.5)          \Text(77,-3)[cb]{$3$}
 \Line(100,-35)(70,-17.5)          \Text(82,-39)[cb]{$4$}
 \Line(100,-35)(100,35)            \Text(107,-3)[cb]{$5$}
 \Line(100,35)(70,17.5)            \Text(82,31)[cb]{$6$}
\end{picture}
\end{center}
\vspace{2cm}
\caption[]{The irreducible two-loop vertex diagrams $V^{\bca}$. 
External momenta are flowing inwards.} 
\label{TLvertbca}
\end{figure}
%--
The leading Landau singularity for $V^{\bca}$ is given in Eq.~(321) of III. 
Our strategy will be to search for solutions of the Landau equations where 
two external momenta are on some mass-shell and the third one is 
unconstrained. Suppose that $P^2$ is unconstrained, then we must
have $m_3 = 0$ and, moreover, the first condition of Eq.~(321) of III requires
$m_4 = m_1$ and $m_6 = m_2$. The second condition will be written as
%--
\bqa
4\,m^2_3\,m^2_4\,p^2_2 &=&
  (m^2_1-m^2_3-m^2_4)\,(m^2_2-m^2_3-m^2_6)\,(p^2_1+m^2_4+m^2_5)
- 4\,m^2_3\,m^2_4\,(m^2_5+m^2_6)
\nl
{}&+& \eta_1\,(m^2_1-m^2_3-m^2_4)\,
\lambda_{23}^{1/2}
- \eta_2\,(m^2_2-m^2_3-m^2_6)\,\lambda_{13}^{1/2} +
\eta_3\,(p^2_1+m^2_4+m^2_5)\,\lambda_{12}^{1/2},
\label{seqm}
\eqa
%--
where $\eta_i = \pm 1, i=1,\dots 3$. With $m_4 = m_1$ and $m_6 = m_2$ we obtain
$\lambda_1 = m^2_3\,(m^2_3 - 4\,m^2_1)$ and 
$\lambda_2 = m^2_3\,(m^2_3 - 4\,m^2_2)$,
so that $m^2_3$ can be factorized on both sides of \eqn{seqm}.
Next, by putting $m_3 = 0$ we obtain
%--
\bq
m^2_1\,( p^2_2 + m^2_2 + m^2_5 ) +
s_4\,m_1\,m_2\,( p^2_1 + m^2_1 + m^2_5) = 0,
\eq
%--
from which we recognize the familiar infrared configuration with
%--
%\bq
$p^2_1 = -m^2_1$, $p^2_2 = - m^2_2$, and $m_4 = m_1$,
$m_6 = m_2, m_3 = m_5 = 0$.
%\eq
%--
which corresponds to configuration b) in \fig{fig231}. If $p_2^2$ is 
unconstrained (the case with $p_1^2$ leads to the same configuration because 
of the symmetry of the diagram), we must have $m_4 = 0$.
Searching for configurations with two photons, there is only one possibility
compatible with the standard model, $m_2=0$. Inserting these conditions in the
second relation of Eq.~(321) of III we obtain $m_1 = m_3 = m_6$ and
$p_1^2 = -m_5^2$. Then the first relation of Eq.~(321) requires 
$P^2 = -m_6^2$, leading to configuration d) of \fig{fig231}.
In the following we classify the sub-leading singularities.
%--
For the sub-leading singularity corresponding to $\alpha_1 = \alpha_2 = 
\alpha_3 = 0$ the reduced diagram is a one-loop three-point function where the
classification of infrared singular configurations is already known:
we obtain three solutions which correspond to the two configurations shown 
in \fig{fig231}: The first is $m_4 = 0, \, P^2 = -m^2_6, \, p^2_1 = - m^2_5$,
the second is $m_5 = 0, \, p^2_1 = - m^2_4, \, p^2_2 = - m^2_6$, with a
third one given by $m_6 = 0, \, P^2 = - m^2_4, \, p^2_2 = - m^2_5$.
%--
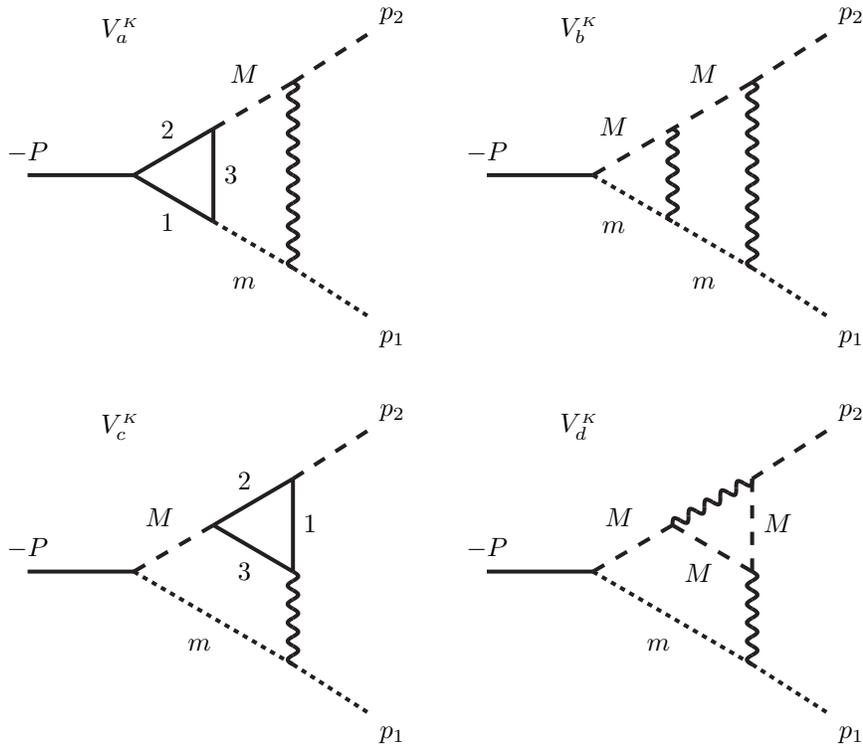
\begin{figure}[ht]
\begin{center}
\begin{picture}(150,75)(0,0)
 \SetWidth{1.5}
 \Line(0,0)(40,0)                     \Text(0,5)[cb]{$-P$}
 \DashLine(128,-53)(100,-35){2}       \Text(138,-65)[cb]{$p_1$}
 \DashLine(128,53)(100,35){5}         \Text(138,57)[cb]{$p_2$}
 \Line(70,-17.5)(40,0)                \Text(53,-21)[cb]{$1$}
 \Line(70,17.5)(40,0)                 \Text(53,14)[cb]{$2$}
 \Line(70,-17.5)(70,17.5)             \Text(77,-3)[cb]{$3$}
 \DashLine(100,-35)(70,-17.5){2}      \Text(82,-43)[cb]{$m$}
 \Photon(100,-35)(100,35){2}{10}
 \DashLine(100,35)(70,17.5){5}        \Text(82,35)[cb]{$M$}
 \Text(35,50)[cb]{$V^{\bca}_a$}
\end{picture}
\hspace{0.6cm}
\begin{picture}(150,75)(0,0)
 \SetWidth{1.5}
 \Line(0,0)(40,0)                    \Text(0,5)[cb]{$-P$}
 \DashLine(128,-53)(100,-35){2}      \Text(138,-65)[cb]{$p_1$}
 \DashLine(128,53)(100,35){5}        \Text(138,57)[cb]{$p_2$}
 \DashLine(100,35)(40,0){5}          \Text(82,35)[cb]{$M$}\Text(48,15)[cb]{$M$}
 \DashLine(100,-35)(40,0){2}         \Text(82,-43)[cb]{$m$}\Text(48,-22)[cb]{$m$}
 \Photon(70,-17.5)(70,17.5){2}{5}
 \Photon(100,-35)(100,35){2}{10}
 \Text(35,50)[cb]{$V^{\bca}_b$}
\end{picture}
\end{center}
%\end{figure}
\vspace{1.9cm}
%\begin{figure}[ht]
\begin{center}
\begin{picture}(150,75)(0,0)
 \SetWidth{1.5}
 \Line(0,0)(40,0)                     \Text(0,5)[cb]{$-P$}
 \DashLine(128,-53)(100,-35){2}       \Text(138,-65)[cb]{$p_1$}
 \DashLine(128,53)(100,35){5}         \Text(138,57)[cb]{$p_2$}
 \Line(100,35)(100,0)                 \Text(107,15)[cb]{$1$}
 \Line(100,35)(70,17.5)               \Text(82,31)[cb]{$2$}
 \Line(100,0)(70,17.5)                \Text(82,-3)[cb]{$3$}
 \DashLine(100,-35)(40,0){2}          \Text(65,-30)[cb]{$m$}
 \DashLine(70,17.5)(40,0){5}          \Text(50,17)[cb]{$M$}
 \Photon(100,0)(100,-35){2}{5}
 \Text(35,50)[cb]{$V^{\bca}_c$}
\end{picture}
\hspace{0.6cm}
\begin{picture}(150,75)(0,0)
 \SetWidth{1.5}
 \Line(0,0)(40,0)                     \Text(0,5)[cb]{$-P$}
 \DashLine(128,-53)(100,-35){2}       \Text(138,-65)[cb]{$p_1$}
 \DashLine(128,53)(100,35){5}         \Text(138,57)[cb]{$p_2$}
 \DashLine(100,-35)(40,0){2}          \Text(65,-30)[cb]{$m$}
 \DashLine(70,17.5)(40,0){5}          \Text(50,17)[cb]{$M$}
 \DashLine(100,35)(100,0){5}          \Text(110,15)[cb]{$M$}
 \DashLine(100,0)(70,17.5){5}         \Text(80,-4)[cb]{$M$}
 \Photon(100,0)(100,-35){2}{5}
 \Photon(100,35)(70,17.5){2}{5}
 \Text(35,50)[cb]{$V^{\bca}_d$}
\end{picture}
\end{center}
\vspace{2cm}
\caption[]{The $V^{\bca}$ infrared configurations. The photon line represents 
a general massless particle while dashed and solid lines refer to
different massive particles. The mass of the three particles in the 
internal triangle are $m_1$, $m_2$ and $m_3$.}
\label{fig231}
\end{figure}
%--
\subsection{Evaluation of the $V^{\bca}$ cases \label{Evalbca}}
%--
In order to evaluate infrared configurations for this family of diagrams we 
use the parametrization of \eqn{defbca}.
At first we combine propagators $[1]_{\bca}-[3]_{\bca}$ with Feynman 
parameters $x_1,x_2$,
%--
\bq
\pi^4\,V^{\bca} = 
\mu^{2\ep}\,\egam{3}\,\int\,d^nq_1 d^nq_2\,\dsimp{2}(\{x\})\,
\frac{1}{[4]_{\bca}[5]_{\bca}[6]_{\bca}}\,
\frac{1}{(q^2_1 + 2\,\spro{R_x}{q_1} + Q^2_x)^3},
\eq
%--
where $x$-dependent quantities are
%--
\bq
R_x = (1-x_1)\,P - x_2\,q_2,  \quad
Q^2_x = x_1\,( m^2_1 - m^2_2) + x_2\,( q^2_2 + m^2_3 - m^2_1) + m^2_2 +
(1-x_1)\,P^2.
\eq
%--
Integration over $q_1$ gives
%--
\bqa
\pi^2\,V^{\bca} &=& 
i\,\mu^{2\ep}\,\frac{\egam{1+\ep/2}}{\pi^{\ep/2}}\!
\dssimp{2}(\{x\})\,\Bigl[x_2(1-x_2)\Bigr]^{-1-\ep/2}\!\!
\int\!\frac{d^n q_2}{(q^2_2+2\,\spro{P_x}{q_2} +
M^2_x)^{1+\ep/2}\,[4]_{\bca}[5]_{\bca}[6]_{\bca}},
\eqa
%--
with new auxiliary quantities defined by
%--
\bq
P_x = \frac{1-x_1}{1-x_2}\,P = X\,P,
\qquad\quad
M_x^2 = 
\frac{\chi(x_1\,;\,P^2\,;\,m^2_2\,,\,m^2_1) + x_2\,( m^2_3 - m^2_1)}
     {x_2\,(1 - x_2)}.
%M^2_x = \frac{- P^2\,x^2_1 + x_1\,( P^2 + m^2_1 - m^2_2) +
%x_2\,( m^2_3 - m^2_1) + m^2_2}{x_2\,(1 - x_2)}.
\label{defXwi}
\eq
%--
Secondly, we combine the remaining propagators with Feynman parameters
$y_i,\, i=1,\dots,3$: it follows
%--
\bqa
\pi^2\,V^{\bca} &=&
 i\,\mu^{2\ep}\,\pi^{-\ep/2}\,\egam{4+\frac{\ep}{2}}\,
\dsimp{2}(\{x\})\,[x_2\,(1-x_2)]^{-1-\ep/2}
\dsimp{3}(\{y\})\,y^{\ep/2}_3
\nl
{}&\times& \intmomi{n}{q_2}\,\Bigl[
y_3\,[123]_{\bca} + ( y_2-y_3 )\,[4]_{\bca} + ( 1 - y_1 )\,[5]_{\bca} + 
( y_1 - y_2 )\,[6]_{\bca}\Bigr]^{-4-\ep/2},
\eqa
%--
where $[123]_{\bca} = q^2_2+2\,\spro{P_x}{q_2} + M^2_x$. After performing the
$q_2$-integration we obtain the following result:
%--
\bqa
V^{\bca} &=& 
-\,\left(\frac{\mu^2}{\pi}\right)^{\ep}\egam{2+\ep}\,
\dsimp{2}(\{x\})\,[x_2\,(1-x_2)]^{-1-\ep/2}\,
\dsimp{3}(\{y\})\,y^{\ep/2}_3\,U^{-2-\ep}_{\bca},
\label{startVbca}
\eqa
%--
where the quadratic form $U_{\bca}$ is given by
%--
\bq
U_{\bca} =
- [ p_2\,y_1 - P\,(y_2-X\,y_3) + p_1 ]^2
+ ( P^2 - p_1^2 + m_6^2 - m_5^2 )\,y_1
- ( P^2 + m_6^2 - m_4^2 )\,y_2 + ( M_x^2 - m_4^2 )\,y_3 + p_1^2 + m_5^2
\label{defU}
\eq
%--
To proceed in our derivation we introduce $\bX = 1 - X$ 
(with $X$ given in \eqn{defXwi}) and perform a change of variables, 
$y_2 = y'_2 + X\,y_3$, obtaining
%--
\bqa
V^{\bca} &=&
- \left(\frac{\mu^2}{\pi}\right)^{\ep}\egam{2+\ep}\,\dsimp{2}(\{x\})\,
  [x_2\,(1-x_2)]^{-1-\ep/2}\,\intfx{y_1}\,
\spliti{0}{\bX\,y_1}{y_1}{y_2}{{\cal Y}}{{\bca}}
%\Bigl(\, \int_0^{\bX\,y_1}\!\!\!\!dy_2\,{\cal Y}_{\bcan{1}}
%+ \int_{\bX\,y_1}^{y_1}\!\!\!\!dy_2\,{\cal Y}_{\bcan{2}}\,\Bigg)
\nl
{\cal Y}_{\bcan{i}} &=& \int_0^{a_i}\!\!\!\!dy_3
\,y_3^{\ep/2}\,(A_{\bca}\,y_3 + B_{\bca})^{-2-\ep}, \qquad\qquad
a_1= \frac{y_2}{\bX}, \qquad a_2= \frac{y_1-y_2}{X}
\label{hypcalY}
\eqa
%--
where $A_{\bca}$ and $B_{\bca}$ are defined by
%--
\bqa
A_{\bca} &=& M_x^2 - \bX\,m_4^2 - ( P^2 + m_6^2 )\,X
\nl
B_{\bca} = - (\, p_2\,y_1 - P\,y_2 + p_1 \,)^2
\,&+&\, ( P^2 - p_1^2 + m_6^2 - m_5^2 )\,y_1
\,-\, ( P^2 + m_6^2 - m_4^2 )\,y_2 \,+\, p_1^2 \,+\, m_5^2
\label{defAB}
\eqa
%--
The ${\cal Y}_{\bcan{i}}$ functions of \eqn{hypcalY} can be expressed in terms 
of an hypergeometric function,
%--
\bq
{\cal Y}_{\bcan{i}}=
\frac{\egams{1+\ep/2}}{\egam{2+\ep}}\,(A_{\bca}\,B_{\bca})^{-1-\ep/2}
-\frac{2}{2+\ep}\,
(a_i\,A^2_{\bca})^{-1-\ep/2}\,
\hyper{2+\ep}{1+\frac{\ep}{2}}{2+\frac{\ep}{2}}
{-\frac{B_{\bca}}{a_i\,A_{\bca}}}.
\label{calYexp}
\eq
%--
The first term in \eqn{calYexp} does not depend on $a_i$; therefore, we 
obtain,
%--
\bqa
V^{\bca} &=& 
- \left(\frac{\mu^2}{\pi}\right)^{\ep}\dsimp{2}(\{x\})\,
\Bigl[x_2\,(1 - x_2)\Bigr]^{-1-\ep/2}
\nl
{}&\times& 
\Big[\,
\egams{1+\frac{\ep}{2}}\dssimp{2}(\{y\})\,(A_{\bca}\,B_{\bca})^{-1-\ep/2}
- 2\,\frac{\egam{2+\ep}}{2+\ep}\intsx{y_1}\,
\spliti{0}{\bX\,y_1}{y_1}{y_2}{J}{{\bca}}
\Big],
\label{commonbca}
\eqa
%--
where $J_{\bcan{i}}$ is defined by
%--
\bq
J_{\bcan{i}} = \lpar a_i\,A^2_{\bca}\rpar^{-1-\ep/2}\,
\hyper{2+\ep}{1+\frac{\ep}{2}}{2+\frac{\ep}{2}}{-\frac{B_{\bca}}{a_iA_{\bca}}}.
\eq
%\bqa
%V^{\bca} &=& 
%- \left(\frac{\mu^2}{\pi}\right)^{\ep}\dsimp{2}(\{x\})\,
%\Bigl[x_2\,(1 - x_2)\Bigr]^{-1-\ep/2}
%\nl
%&& 
%\Big[\,
%\egams{1+\frac{\ep}{2}}\dssimp{2}(\{y\})\,(A_{\bca}\,B_{\bca})^{-1-\ep/2}
%- 2\,\frac{\egam{2+\ep}}{2+\ep}\intsx{y_1}\,\Big( I_{\bcan{1}} + I_{\bcan{2}}\Big)
%\Big],
%\label{commonbca}
%\eqa
%%--
%where $I_{\bcan{i}}$ is defined by
%%--
%\bq
%I_{\bcan{1}} + I_{\bcan{2}} = \spliti{0}{\bX\,y_1}{y_1}{y_2}{J}{{\bca}},
%\quad
%J_{\bcan{i}} = \lpar a_i\,A^2_{\bca}\rpar^{-1-\ep/2}\,
%\hyper{2+\ep}{1+\frac{\ep}{2}}{2+\frac{\ep}{2}}
%{-\frac{B_{\bca}}{a_i\,A_{\bca}}}.
%\eq
%--
Having derived the general result of \eqn{commonbca} we proceed
with separate evaluations of the four cases of \fig{fig231}.
%--
\subsubsection{Evaluation of $V^{\bca}_a$ \label{evalbcaa}}
%--
For the configuration a) of \fig{fig231} we have:
%--
\bqa
A_{\bcan{a}} = \frac{V_{\bcan{a}}(x_1,x_2)}{x_2\,(1-x_2)},
\qquad && \qquad
B_{\bcan{a}} = \beta\lpar y_2\,,\,y_1\,;\,P^2\,,\,M^2\,,\,m^2\rpar, 
\\
V_{\bcan{a}} =  - P^2\,x_1^2 + (P^2-m^2+M^2)\,x_1\,x_2 + m^2\,x_2^2
&+& (P^2+m_1^2-m_2^2)\,x_1 - (P^2+M^2-m_3^2+m_1^2)\,x_2 + m_2^2.
\nn
\label{defABa}
\eqa
%--
After a transformation of variable, $y_2\to y_1\,y_2$, in the first integral of
\eqn{commonbca} the $y_1$-integration can be carried out giving raise to the 
infrared pole. After that we put $\ep = 0$ in $I_{\bcan{1}}$ 
and $I_{\bcan{2}}$, thus obtaining:
%--
\bq
V_a^{\bca} =
\left(\frac{\mu^2}{\pi}\right)^{\ep}\!\!
\egams{1+\frac{\ep}{2}}\dssimp{2}(\{x\})\,
\bigg[
\frac{1}{\ep}\intsx{y_2}\,\Big[V_{\bcan{a}}\,\chi(y_2)\Big]^{-1-\ep/2}
\!\!+\, V_{\bcan{a}}^{-1}\intsx{y_1}
\sspliti{0}{\bX\,y_1}{y_1}{y_2}{C^{-1}}{{\bcan{a}}}
\bigg],
\eq
%\bqa
%V_a^{\bca} &=&
%\frac{1}{\ep}\,\left(\frac{\mu^2}{\pi}\right)^{\ep}
%\egams{1+\frac{\ep}{2}}\,\dsimp{2}(\{x\})\,\Bigl\{
%\intfx{y_2}\,V_{\bcan{a}}^{-1-\ep/2}(x_1,x_2)\,\chi^{-1-\ep/2}(y_2)
%\nl
%&+& V_{\bcan{a}}^{-1}(x_1,x_2)\,\intfx{y_1}\,
%\spliti{0}{\bX\,y_1}{y_1}{y_2}{C^{-1}}{{\bca}}\Bigl\},
%\qquad
%C_{i\,;\,\bca} = a_i\,A_{\bcan{a}} + B_{\bcan{a}},
%%\int_0^{\bar X\,y_1}\!\!\!\!dy_2\,(a_1\,A_{\bcan{a}} + B_{\bcan{a}})^{-1}
%%+\int_{\bar X\,y_1}^{y_1}\!\!\!\!dy_2\,(a_2\,A_{\bcan{a}} + 
%%B_{\bcan{a}})^{-1},\Bigr]\Bigl\},
%\eqa
%--
where $C_{i,\bcan{a}} = a_i\,A_{\bcan{a}} + B_{\bcan{a}}$ and $\chi$ is defined 
in \eqn{chidef}.
Henceforth we change variables in the last two integrals, $y_2= y_1\,y'_2$;  
the $y_1$-integration can be performed, giving
%--
\bqa
V_a^{\bca} &=&
\left(\frac{\mu^2}{\pi}\right)^{\ep}\!\!
\egams{1+\frac{\ep}{2}}\,\dsimp{2}(\{x\})\,
\bigg\{
\frac{1}{\ep}\intfx{y_2}\,\Bigl[V_{\bcan{a}}\,\chi(y_2)\Bigr]^{-1-\ep/2}
\nl
&+& \sum_{i=1}^{2}\,
\int_0^{b_i}\!\!dy_2\,\frac{1}{\alpha_i(y_2)\,V_{\bcan{a}}}\,
\ln \Bigl[ 1 + \frac{x_2\,(1-x_2)\,b_i\,\alpha_i(y_2)}{y_2\,
V_{\bcan{a}}} \Bigr]
\bigg\},
\label{eqwithtt}
\eqa
%--
where we have introduced
%--
\bq
\alpha_1(y_2) = \bchi(y_2), \qquad
\alpha_2(y_2) = \chi(y_2),   \qquad
b_1= \bX, \qquad
b_2= X.
\label{alpha}
\eq
%--
In the last two integrals of \eqn{eqwithtt} we change, once again, variables:
%--
\bqa
x_1 &=& x'_1 + x_2, \qquad x_2 = 1 - x'_2, \qquad x_1 \leftrightarrow x_2,
\qquad \mbox{for} \quad i = 1,
\nl
x_1 &=& 1 - x'_1, \qquad x_2 = 1 - x'_2, \qquad x_1 \leftrightarrow x_2,
\qquad \mbox{for} \quad i = 2,
\eqa
%--
thus obtaining
%--
\bq
V_a^{\bca}= \frac{1}{\ep}\,\left(\frac{\mu^2}{\pi}\right)^{\ep}
\egams{1+\frac{\ep}{2}}\,{\cal X}_{\bcan{a}}\,
{\cal Y}_{\bca} + J_{\bcan{a,1}} + J_{\bcan{a,2}},
\label{eqXYJJa}
\eq
%--
where the various ingredients are defined by
%--
\bqa
{\cal X}_{\bcan{a}} &=& \dsimp{2}(\{x\})\,V_{\bcan{a}}^{-1-\ep/2}(x_1,x_2),
\qquad
{\cal Y}_{\bca} = \intfx{y_2}\,\chi^{-1-\ep/2}(y_2)
\nl
J_{\bcan{a,i}} &=& \dssimp{2}(\{x\})\int_0^{x_2/x_1}\!\!\!\!dy_2\,
\frac{1}{\alpha_i(y_2)\,V_{\bcan{a,i}}}\,
\ln \Bigl[ 
1 + \frac{x_2\,(1-x_1)\,\alpha_i(y_2)}{y_2\,V_{\bcan{a,i}}} 
\Bigr].
\label{XYJJa}
\eqa
%--
The quadratic forms introduced in \eqn{XYJJa} are:
%--
\bq
V_{\bcan{a,i}}(x_1,x_2) = x^t\,H_{a,i}\,x + 2\,K_{a,i}^t\,x + L_{a,i}, 
\qquad\qquad i= 1,\,2;
\eq
%--
\bqa
V_{\bcan{a,1}} &=&
  M^2\,x_1^2 + (P^2-M^2+m^2)\,x_1\,x_2 - P^2\,x_2^2
+ (m_2^2-m_3^2-M^2)\,x_1 + (m_1^2-m_2^2+M^2-m^2)\,x_2 + m_3^2,
\nl
V_{\bcan{a,2}} &=&
  m^2\,x_1^2 + (P^2-m^2+M^2)\,x_1\,x_2 - P^2\,x_2^2
+ (m_1^2-m_3^2-m^2)\,x_1 + (m_2^2-m_1^2+m^2-M^2)\,x_2 + m_3^2.
\nn
\eqa
%--
To compute ${\cal X}_{\bcan{a}}$ and ${\cal Y}_{\bca}$, defined in 
\eqn{eqXYJJa}, we use a BST relation. 
Given $V_{\bcan{a}}= x^t\,H_{a}\,x + 2\,K_{a}^t\,x + L_{a}$ we introduce 
$X_{\bcan{a}}^0 = 1$, $X_{\bcan{a}}^3 = 0$ and also
%--
\bq
B_{\bcan{a}} = L_a - K_a^t\,H_a^{-1}\,K_a, \qquad\qquad
X_{\bcan{a}}^i = -\,\Big( K_a^t\,H_a^{-1} \Big)^i \qquad i= 1,\,2;
\eq
%--
we obtain for 
${\cal X}_{\bcan{a}} = {\cal X}_{\bcan{a}}^1 + {\cal X}_{\bcan{a}}^2\,\ep$
%--
\bq
{\cal X}_{\bcan{a}}^n=
\frac{(-1)^{n+1}}{2^n\,n!\,B_{\bcan{a}}}\,
\bigg[\,
2\!\dssimp{2}\,\ln^{n-1}V_{\bcan{a}}\,\Big( \ln V_{\bcan{a}} + n \Big)
- \sum_{i=0}^{2}\,(X_{\bcan{a}}^i - X_{\bcan{a}}^{i+1})
  \intsx{x}\,\ln^n V^i_{\bcan{a}}\,
\bigg],
\eq
%the following relation holds:
%--
%\bqa
%{\cal X}_{\bcan{a}} &=& \frac{1}{2\,B_{\bcan{a}}}\,
%\Bigg[\,2\dsimp{2}\,\ln V_{\bcan{a}}
%- \intsx{x}\,\sum_{i=0}^{2}\,
%  (X_{\bcan{a}}^i - X_{\bcan{a}}^{i+1})\,\ln V_{\bcan{a}}^i + 1\,\Bigg]
%\nl
%&-& \frac{\ep}{8\,B_{\bcan{a}}}\,
%\Bigg[\,2\dsimp{2}\,\ln V_{\bcan{a}}\,\Big( \ln V_{\bcan{a}} + 2 \Big)
%- \intsx{x}\,\sum_{i=0}^{2}\,(X_{\bcan{a}}^i - X_{\bcan{a}}^{i+1})\,
%\ln^2 V^i_{\bcan{a}}\,\Bigg],
%\nl
%\eqa
%--
where the $V^i_{\bcan{a}}$ quadratic forms are
%--
\bq
V^1_{\bcan{a}} = V_{\bcan{a}}(1,x), \quad
V^2_{\bcan{a}} = V_{\bcan{a}}(x,x), \quad
V^3_{\bcan{a}} = V_{\bcan{a}}(x,0).
\eq
%--
Similarly, for ${\cal Y}_{\bca}$, we get
%--
\bq
{\cal Y}_{\bca} = {\cal Y}_{\bca}^1 + {\cal Y}_{\bca}^2\,\ep,
\qquad\qquad
{\cal Y}_{\bca}^n= 
\,\frac{(-1)^{n+1}}{2^n\,n!\,\Bbt}\,
\bigg\{ 
\intsx{y}\,\ln^{n-1}\chi(y)\,\Big[ \ln\chi(y) + 2\,n \Big] - L^n_{\chi}
\bigg\},
\label{bcaY}
\eq
%\bqa
%{\cal Y}_{\bca} &=& \frac{1}{2\,\Bbt}\,
%\Bigl[ \intfx{y}\,\ln\chi(y) - L^1_{\chi} + 2 \Bigr]
%- \frac{\ep}{8\,\Bbt} \Bigl\{ \intfx{y}\,\ln\chi(y)\,
%\Bigl[ \ln\chi(y) + 4 \Bigr] - L^2_{\chi}\Bigr\},
%\label{bcaY}
%\eqa
%--
where BST factor and co-factor are defined in \eqn{btfactors} and $L^n_{\chi}$
in \eqn{Racan}. To compute $J_{\bcan{a,i}}$ we use \eqn{BTli1} which in the 
present case reads:
%--
\bqa
\frac{1}{V_{\bcan{a,i}}}\,
\ln\Big(1-\frac{A_{\bcan{a,i}}}{V_{\bcan{a,i}}}\Big) &=&
\frac{1}{B_{\bcan{a}}}\,\Bigg\{\frac{1}{A_{\bcan{a,i}}}\,
\ln\Big(1-\frac{A_{\bcan{a,i}}}{V_{\bcan{a,i}}}\Big)\,
\Big[ 1 - \frac{1}{2}\,\sum_{j=1}^2\,
(x_j-X_{\bcan{a,i}}^j)\,\partial_{x_j} \Big]\,A_{\bcan{a,i}}
\nl
{}&-& \frac{1}{2}\,\sum_{j=1}^2\,(x_j-X_{\bcan{a,i}}^j)\,\partial_{x_j}\,
\li{2}{\frac{A_{\bcan{a,i}}}{V_{\bcan{a,i}}}}\Bigg\}.
\eqa
%--
where $A_{\bcan{a,i}}= -x_2\,(1-x_1)\,\alpha_i/y_2$ and BST factor 
and co-factors are
%--
\bq
B_{\bcan{a}} = 
L_{a,1} - K_{a,1}^t\,H_{a,1}^{-1}\,K_{a,1} = 
L_{a,2} - K_{a,2}^t\,H_{a,2}^{-1}\,K_{a,2} =
L_a - K_a^t\,H_a^{-1}\,K_a,
\qquad\quad
X_{\bcan{a,i}} = - K_{a,i}^t\,H_{a,i}^{-1}.
\eq
%--
After integration by parts we obtain
%--
\bqa
J_{\bcan{a,i}} &=&
\frac{1}{B_{\bcan{a}}}\,\intfx{y_2}\,\Bigl\{
\intfx{x_1}\int_{x_1\,y_2}^{x_1}\!\!dx_2\,
\frac{1}{\alpha_i(y_2)}\,\Bigg[\,\frac{1}{2}\,
\left( \frac{1-X_{\bcan{a,i}}^1}{1-x_1} + 
\frac{X_{\bcan{a,i}}^2}{x_2} \right)\,
\ln( 1 + \eta_{\bcan{a,i}} ) 
+ \li{2}{-\eta_{\bcan{a,i}}} \Bigg]
\nl
{}&-& \frac{1}{2}\,\intfx{x}\,\Bigg[
  \frac{X_{\bcan{a,i}}^1-X_{\bcan{a,i}}^2}{\alpha_i(y_2)}\,
  \li{2}{-\eta_{\bcan{a,i}}^1}
- \frac{y_2\,X_{\bcan{a,i}}^1-X_{\bcan{a,i}}^2}{\alpha_i(y_2)}\,
  \li{2}{-\eta_{\bcan{a,i}}^2}\Bigg]\Bigr\},
\eqa
%--
where the $\eta$-functions are defined by
%--
\bq
\eta_{\bcan{a,i}}= 
\frac{x_2\,(1-x_1)\,\alpha_i(y_2)}{y_2\,V_{\bcan{a,i}}(x_1,x_2)},
\qquad\qquad
\eta_{\bcan{a,i}}^1= 
\frac{x\,(1-x)\,\alpha_i(y_2)}{y_2\,V_{\bcan{a,i}}(x,x)},
\qquad\qquad
\eta_{\bcan{a,i}}^2= 
\frac{x\,(1-x)\,\alpha_i(y_2)}{V_{\bcan{a,i}}(x,x\,y_2)}.
\eq
%--
Collecting all pieces together, we get:
%--
\bq
\ovalbox{\boldmath $V^{\bca}_a$} =
- \lpar \frac{\mu^2}{\pi} \rpar^{\ep}\,\egam{1+\ep}\,
\Big(\, \frac{1}{\ep}\,V^{\bcan{a}}_{-1} \,+\, V^{\bcan{a}}_0 \,\Big)
\eq
%--
\bq
V^{\bcan{a}}_{-1}= - {\cal X}_{\bcan{a}}^1\,{\cal Y}_{\bca}^1,
\qquad\qquad
V^{\bcan{a}}_0= 
- {\cal X}_{\bcan{a}}^1\,{\cal Y}_{\bca}^2
- {\cal X}_{\bcan{a}}^2\,{\cal Y}_{\bca}^1
- J_{\bcan{a,1}} - J_{\bcan{a,2}}.
\eq
%--
In the following section we will consider the b) configuration of \fig{fig231}.
%--
\subsubsection{Evaluation of $V^{\bca}_b$ \label{Evalbcab}}
%--
Configuration b) of \fig{fig231} is a special case of a). Here the polynomial 
$V$ takes the form ($\beta$ is defined in \eqn{betadef})
%--
\bq
V_{\bcan{b}}(x_1,x_2) = \beta\lpar 1-x_1\,,\,1-x_2\,;\,P^2\,;\,m^2\,,\,M^2\rpar
\eq
%--
which has a double zero at $x_1 = x_2 = 1$.
The BST factor for the polynomial $V_{\bcan{b}}$ is zero, revealing again the 
presence of a new infrared pole. For this reason we cannot 
put $\ep = 0$ from the beginning; instead we reexamine \eqn{hypcalY} which,
in the present configuration, reads:
%--
\bqa
V^{\bca}_b &=&
- \Big( \frac{\mu^2}{\pi}\Big)^{\ep}\,\egam{2+\ep}\,
  \dsimp{2}(\{x\})\,[x_2\,(1-x_2)]^{-1-\ep/2}\,\intfx{y_1}\,
\nl
{}&\times& 
\triagi{0}{\bX}{y_1}\,dy_2 dy_3 \,y_3^{\ep/2}\,\Big[\,
\frac{\beta(1-x_1,1-x_2)}{x_2\,(1-x_2)}\,y_3 + \bbeta(y_2,y_1) 
\,\Big]^{-2-\ep},
%\Bigg(\, 
%  \int_0^{\bX\,y_1}\!\!\!\!dy_2\,\int_0^{y_2/\bX}\!\!\!\!dy_3
%+ \int_{\bX\,y_1}^{y_1}\!\!\!\!dy_2\,\int_0^{(y_1-y_2)/X}\!\!\!\!dy_3
%\,\Bigg)\,y_3^{\ep/2}\,\Big[\,
%\frac{\beta(1-x_1,1-x_2)}{x_2\,(1-x_2)}\,y_3 + \bbeta(y_2,y_1) 
%\,\Big]^{-2-\ep},
\label{fands}
\eqa
%--
In the first(second) integral of \eqn{fands} we change variables according to:
%--
\bq
x_1 \rightarrow x_1 + x_2, \quad 
x_2 \rightarrow 1 - x_2, \quad 
x_1 \rightarrow x_1\,x_2\, \qquad \Bigl(
x_1 \rightarrow 1 - x_1, \quad 
x_2 \rightarrow 1 - x_2, \quad 
x_1 \rightarrow x_1\,x_2\Bigr),
\eq
%--
thus obtaining
%--
\bqa
V^{\bca}_b 
&=&
- \Big( \frac{\mu^2}{\pi}\Big)^{\ep}\egam{2+\ep}\!
  \dscub{3}(\{x\},y_1)\,
\frac{[x_2\,(1-x_2)]^{-\ep/2}}{1-x_2}\,
\sum_{i=1}^{2}\,\int_{a_i}^{b_i}\!\!\!\!\!dy_2\!\int_{0}^{d_i}\!\!\!\!\!dy_3\,
y_3^{\ep/2}\,
\Big[\,\frac{x_2\,\alpha_i(x_1)}{1-x_2}\,y_3 + \bbeta(y_2,y_1) \,\Big]^{-2-\ep}
\nl
&&
a_1= 0, \quad
b_1= x_1y_1, \quad
d_1= \frac{y_2}{x_1}, \quad\qquad
a_2= (1-x_1)y_1, \quad
b_2= y_1, \quad
d_2= \frac{y_1-y_2}{x_1}.
\eqa
%--
where the functions $\alpha_i$ are defined in \eqn{alpha}.
%\bqa
%V^{\bca}_b &=&
%- \Big( \frac{\mu^2}{\pi}\Big)^{\ep}\,\egam{2+\ep}\,
%  \dcub{3}(x_1,x_2,y_1)\,x_2^{-\ep/2}\,(1-x_2)^{-1-\ep/2}\,
%\nl
%{}&\times& \Bigg\{\,   \int_0^{x_1 y_1}\!\!\!\!dy_2\,
%  \int_0^{y_2/x_1}\!\!\!\!dy_3\,y_3^{\ep/2}\,\Big[\,
%\frac{x_2\,\bchi(x_1)}{1-x_2}\,y_3 + \bbeta(y_2,y_1) \,\Big]^{-2-\ep}
%\nl
%{}&+& \int_{(1-x_1)y_1}^{y_1}\!\!\!\!dy_2\,
%\int_0^{(y_1-y_2)/x_1}\!\!\!\!dy_3\,y_3^{\ep/2}\,\Big[\,
%\frac{x_2\,\chi(x_1)}{1-x_2}\,y_3 + \bbeta(y_2,y_1) \,\Big]^{-2-\ep}\,\Bigg\}
%\eqa
%--
We also transform the integration domain of $y_2$ and $y_3$ in the 
first(second) integral of \eqn{fands} according to:
%--
\bq
y_3 \rightarrow y_2\,y_3/x_1, \quad 
y_2 \rightarrow y_1\,y_2,
\qquad \Bigl(
y_3 \rightarrow (y_1-y_2)\,y_3/x_1, \quad 
y_2 \rightarrow y_1\,(1-y_2)\Bigr).
\eq
%--
After this transformation we obtain
%--
\bqa
V^{\bca}_b &=&
- \Big( \frac{\mu^2}{\pi}\Big)^{\ep}\,\egam{2+\ep}\,\sum_{i=1}^2
  \dcub{3}(x_1,x_2,y_3)\,\int_0^{x_1}\!dy_2\,
\Bigl[ x_1\,(1-x_2)\,y_2\Bigr]^{1+\ep/2}\,\lpar\frac{y_3}{x_2}\rpar^{\ep/2}\,
%x_1^{1+\ep/2}\,x_2^{-\ep/2}\,(1-x_2)^{1+\ep/2}\,
%y_2^{1+\ep/2}\,y_3^{\ep/2}\,
{\cal Y}_{\bcan{b,i}}
\nl
{\cal Y}_{\bcan{b,i}} &=& \intsx{y_1}\,y_1^{-\ep/2}\,\Big[\,
x_2\,y_2\,y_3\,\alpha_i(x_1) + x_1\,(1-x_2)\,y_1\,\alpha_i(y_2)\,\Big]^{-2-\ep}.
\eqa
%--
${\cal Y}_{\bcan{b,i}}$ is then expressed in terms of an hypergeometric 
function:
%--
\bqa
{\cal Y}_{\bcan{b,i}} &=&
\frac{2}{2-\ep}\,[\,x_2\,y_2\,y_3\,\alpha_i(x_1)\,]^{-2-\ep}\,
\hyper{2+\ep}{1-\frac{\ep}{2}}{2-\frac{\ep}{2}}
{-\frac{x_1\,(1-x_2)\,\alpha_i(y_2)}{x_2\,y_2\,y_3\,\alpha_i(x_1)}}
\nl
{}&=& B(1-\frac{\ep}{2},1+\frac{3}{2}\,\ep)\,\,
[\, x_2\,y_2\,y_3\,\alpha_i(x_1) \,]^{-1-3\,\ep/2}\,
[\, x_1\,(1-x_2)\,\alpha_i(y_2) \,]^{-1+\ep/2}
\nl
{}&-& \frac{2}{2+3\,\ep}\,[\, x_1\,(1-x_2)\,\alpha_i(y_2) \,]^{-2-\ep}\,
\hyper{2+\ep}{1+\frac{3}{2}\,\ep}{2+\frac{3}{2}\,\ep}
{-\frac{x_2\,y_2\,y_3\,\alpha_i(x_1)}{x_1\,(1-x_2)\,\alpha_i(y_2)}}.
\eqa
%--
Substituting this partial result in the expression for $V^{\bca}_b$ and 
setting $\ep = 0$ where possible, we get:
%--
\bq
V^{\bca}_b = - \Big( \frac{\mu^2}{\pi}\Big)^{\ep}\,
  \egam{1-\frac{\ep}{2}}\,\egam{1+\frac{3}{2}\,\ep}\,
  \Big(\, V^{\bca}_{b,\ep} + V^{\bca}_{b,0} \Big)
\eq
%--
where we have defined
%--
\bqa
V^{\bca}_{b,\ep} &=&
\sum_{i=1}^2\dcub{3}(x_1,x_2,y_3)\,\int_0^{x_1}\!dy_2\,
  x_1^{\ep}\,x_2^{-1-2\,\ep}\,(1-x_2)^{\ep}\,
  y_2^{-\ep}\,y_3^{-1-\ep}\,
  \alpha_i^{-1-3/2\,\ep}(x_1)\,\alpha_i^{-1+\ep/2}(y_2)
\nl
V^{\bca}_{b,0} &=&
- \sum_{i=1}^2\dcub{3}(x_1,x_2,y_3)\,\int_0^{x_1}\!dy_2\,
\frac{y_2}{\alpha_i(y_2)}\,
[\, x_2\,y_2\,y_3\,\alpha_i(x_1) + x_1\,(1-x_2)\,\alpha_i(y_2) \,]^{-1}.
\label{functof}
\eqa
%--
After some straightforward integration the function of \eqn{functof} become
%--
\bqa
V^{\bca}_{b,\ep} &=&
\frac{1}{2\,\ep^2}\,\frac{\egam{1-2\,\ep}\egam{1+\ep}}{\egam{1-\ep}}\,
\sum_{i=1}^2\,\intsxy{x}{y}\,x^{\ep}\,y^{-\ep}\,
  \alpha_i^{-1-3/2\,\ep}(x)\,\alpha_i^{-1+\ep/2}(y)
\nl
V^{\bca}_{b,0} &=& - \sum_{i=1}^2\dcub{2}(x_1,x_2)\,\int_0^{x_1}\!dy\,
\frac{\ln( 1 + \eta_{\bcan{\,b,i}} )}{\alpha_i(x_1)\,\alpha_i(y)\,x_2},
\qquad\quad
\eta_{\bcan{b,i}}= \frac{x_2\,y\,\alpha_i(x_1)}{x_1\,(1-x_2)\,\alpha_i(y)}
\eqa
%--
After a Laurent expansion in $\ep$, $V^{\bca}_{b,\ep}$ has the following form:
%--
\bq
V^{\bca}_{b,\ep}= \frac{1}{2}\,\sum_{n=-2}^{0}\,{\cal R}^{\bcan{b}}_n\,\ep^n,
\eq
%--
%with coefficients ${\cal R}$ given by
%--
\bqa
{\cal R}^{\bcan{b}}_{-2} &=&
\Big[ \intsx{x}\,\chi^{-1}(x) \Big]^2 = [J_0^0]^2
\qquad
{\cal R}^{\bcan{b}}_{-1} =
- J_0^0\,J_0^1 + \sum_{i=1}^{2}\,\Bigl[ J_{10}^{00}(i) - J_{01}^{00}(i) \Bigr]
\nl
{\cal R}^{\bcan{b}}_0 &=&
-\,\frac{3}{4}\,[J_0^1]^2 + \frac{5}{4}\,J_0^2\,J_0^0
+ 2\,\zeta(2)\,[J_0^0]^2+ \frac{1}{2}\,\sum_{i=1}^{2}\,\Bigl[ 
- 2\,J_{11}^{00}(i)
\nl
{}&+& J_{20}^{00}(i) + J_{02}^{00}(i) - 3\,J_{10}^{10}(i) + 
J_{10}^{01}(i) + 3\,J_{01}^{10}(i) - J_{01}^{01}(i)\Bigr].
\label{AGlast}
\eqa
%--
In \eqn{AGlast} we have introduced $J_n^k$ and $J_{n m}^{k h}(i)$ 
defined by:
%--
\bq
J_n^k= \intsx{z}\,\ln^n z\,\frac{\ln^k\chi(z)}{\chi(z)} \qquad
J_{n m}^{k h}(i)= \intsxy{x}{y}\,\ln^{n}x\,\ln^{m}y\,
\frac{\ln^{k}\alpha_i(x)}{\alpha_i(x)}\,
\frac{\ln^{h}\alpha_i(y)}{\alpha_i(y)}
\eq
%--
$J$ integrals of the first kind have already been considered in 
section~\ref{evaladac} (see \eqn{evalJnk}).
For $J$ integrals of the second kind it is easy to verify that the following
property holds:
%--
\bq
\sum_{i=1}^{2}\,J_{00}^{k_1k_2}(i)= J_0^{k_1}\,J_0^{k_2}
\eq
%--
The general analysis of $J$ is reported in appendix~\ref{app:Jnnkk};
here it is enough to say that for $J$ we are always able to find a smooth 
integral representation.
Finally, note that the residue of the double pole is exactly one half of 
the square of the residue for the single pole in the one-loop case.
For $V^{\bca}_{b,0}$ we apply \eqn{BTli1}:
%--
\bqa
\frac{1}{\alpha_i(y)}\,
\ln\Big( 1 - \frac{b\,y}{\alpha_i(y)} \Big) &=& \frac{1}{\Bbt}\,\Bigg\{
\Big( 1 - \frac{y-Z_i}{2\,y} \Big)
\ln\Big( 1 - \frac{b\,y}{\alpha_i(y)} \Big)
-\frac{y-Z_i}{2}\,\partial_{y}\,
\li{2}{\frac{b\,y}{\alpha_i(y)}}\Bigg\},
\eqa
%--
where $b= -x_2\,\alpha_i(x_1)/x_1/(1-x_2)$, $Z_1= \bXbt$ and $Z_2= \Xbt$
After integrating by parts, we obtain:
%--
\bq
V^{\bca}_{b,0} =
- \,\frac{1}{2\,\Bbt}\,\sum_{i=1}^2\dscub{2}(\{x\})\,\int_0^{x_1}\!\!\!dy\,
\frac{1}{x_2\,\alpha_i(x_1)}\,
\bigg[ \Big( 1 + \frac{Z_i}{y} \Big)\,
\ln( 1 + \eta_{\bcan{b,i}}) + \li{2}{-\,\eta_{\bcan{b,i}}} \bigg]
\eq
%--
Collecting all pieces together, we get:
%--
\bq
\ovalbox{\boldmath $V^{\bca}_b$} =
- \lpar \frac{\mu^2}{\pi} \rpar^{\ep}\egam{1+\ep}\,
\Big(\, 
      \frac{1}{\ep^2}\,V^{\bcan{b}}_{-2} 
\,+\, \frac{1}{\ep}\,V^{\bcan{b}}_{-1} 
\,+\, V^{\bcan{b}}_0 
\,\Big)
\eq
%--
\bq
V^{\bcan{b}}_{-2}= \frac{1}{2}\,{\cal R}^{\bcan{b}}_{-2}
\qquad\qquad
V^{\bcan{b}}_{-1}= \frac{1}{2}\,{\cal R}^{\bcan{b}}_{-1}
\qquad\qquad
V^{\bcan{b}}_0= 
  \frac{1}{2}\,{\cal R}^{\bcan{b}}_{0}
+ \frac{3}{8}\,\zeta(2)\,{\cal R}^{\bcan{b}}_{-2} + V^{\bca}_{b,0}
\eq
%--
\subsubsection{Evaluation of $V^{\bca}_c$ \label{Evalbcac}}
%--
For configuration c) of \fig{fig231} we have
%--
\bqa
&{}&
A_{\bcan{c}}= \frac{V_{\bcan{c}}(x_1,x_2)}{x_2\,(1-x_2)},
\qquad V_{\bcan{c}}(x_1,x_2)=
\chi(x_1\,;\,-M^2\,;\,m^2_2\,,\,m^2_1) + x_2\,( m^2_3 - m^2_1)
%M^2\,x_1^2 + (m_1^2-m_2^2-M^2)\,x_1 + (m_3^2-m_1^2)\,x_2 + m_2^2,
\nl
&{}& B_{\bcan{c}}(y_1,y_2) =
\beta\lpar y_1 - y_2\,,\,1 - y_2\,;\,P^2\,;\,m^2\,,\,M^2\rpar.
\label{defABc}
\eqa
%--
For this configuration we transform variables according to $y_1 = y'_1 + y_2$ 
and $y_2= 1 - y'_2$ in the first integral of \eqn{commonbca} which will have 
the same form as the similar integral in case a) and we obtain:
%--
\bq
V_c^{\bca}= 
\lpar \frac{\mu^2}{\pi} \rpar^{\ep}
\frac{1}{\ep}\,\egams{1+\frac{\ep}{2}}\,
{\cal X}_{\bcan{c}}\,{\cal Y}_{\bca} + J_{\bcan{c,1}} + J_{\bcan{c,2}},
\label{eqXYJJc}
\eq
%--
where the various components are defined by
%--
\bq
{\cal X}_{\bcan{c}} = \dsimp{2}(\{x\})\,V_{\bcan{c}}^{-1-\ep/2}(x_1,x_2),
\qquad
{\cal Y}_{\bca} = \intfx{y_2}\,\chi^{-1-\ep/2}(y_2),
\eq
%--
\bq
\sum_i\,J_{\bcan{c,i}} = \dsimp{2}(\{x\})\,\intfx{y_1}\,
\spliti{0}{\bX\,y_1}{y_1}{y_2}{C^{-1}}{{\bcan{c}}}, \qquad
C_{i,\bcan{c}} = a_i\,A_{\bcan{c}} + B_{\bcan{c}}.
\eq
%--
%\bq
%J_{\bcan{c,1}}= \dsimp{2}(\{x\})\,\intfx{y_1}\,\int_0^{\bX\,y_1}\!\!\!\!dy_2\,
%V_{\bcan{c}}^{-1}\,(a_1\,A_{\bcan{c}} + B_{\bcan{c}})^{-1},
%\eq
%%--
%\bq
%J_{\bcan{c,2}}=
%\dsimp{2}(\{x\})\,\intfx{y_1}\,\int_{\bX\, y_1}^{y_1}\!\!\!\!dy_2\,
%V_{\bcan{c}}^{-1}\,(a_2\,A_{\bcan{c}} + B_{\bcan{c}})^{-1}.
%\eq
%--
and $\chi(x)$ is defined in \eqn{chidef}.
${\cal Y}_{\bca}$ has already been computed in \eqn{bcaY}, while for computing 
${\cal X}_{\bcan{c}}$ we can use the following BST relations:
%--
\bq
V_{\bcan{c}}^{\mu}=
\frac{1}{m_3^2-m_1^2}\,\frac{1}{\mu+1}\,\partial_{x_2}\,V_{\bcan{c}}^{\mu+1},
\qquad\qquad
V_{\bcan{c}}^{\mu}= \frac{1}{b_{\bcan{c}}}\,\Bigg\{
1 - \frac{1}{\mu+1}\,    \Big[\,
  \frac{1}{2}\,(x_1-X_{\bcan{c}})\,\partial_{x_1} + x_2\,\partial_{x_2} \,\Big]
\Bigg\}\,V_{\bcan{c}}^{\mu+1},
\eq
%--
\bq
b_{\bcan{c}}= -\,\frac{\lambda(M^2,m_1^2,m_2^2)}{4\,M^2}, \qquad\qquad
X_{\bcan{c}}= \frac{M^2-m_1^2+m_2^2}{2\,M^2}, \qquad\quad
\bX_{\bcan{c}}= 1 - X_{\bcan{c}}.
\eq
%--
After intergration by parts, we obtain the following two results:
%--
\bq
{\cal X}_{\bcan{c}}= 
\frac{1}{m_3^2-m_1^2}\,
\Big( {\cal X}_{\bcan{c}}^{11} - \frac{\ep}{4}\,{\cal X}_{\bcan{c}}^{12} \Big)
\;\; \mbox{if} \quad m_1\neq m_3,
\qquad
{\cal X}_{\bcan{c}}= 
\frac{1}{2\,b_{\bcan{c}}}\,
\Big( {\cal X}_{\bcan{c}}^{21} - \frac{\ep}{4}\,{\cal X}_{\bcan{c}}^{22} \Big)
\;\; \mbox{if} \quad b_{\bcan{c}} \neq 0.
\eq
%--
%\bq
%{\cal X}_{\bcan{c}}= 
%\left\{
%\ba{ll}
%\frac{1}{m_3^2-m_1^2}\,
%\Big( {\cal X}_{\bcan{c}}^{10} - \frac{\ep}{4}\,{\cal X}_{\bcan{c}}^{11} \Big)
%\qquad &\mbox{if} \quad m_1\neq m_3
%\nl\nl
%\frac{1}{2\,b_{\bcan{c}}}\,
%\Big( {\cal X}_{\bcan{c}}^{20} - \frac{\ep}{4}\,{\cal X}_{\bcan{c}}^{21} \Big)
%\qquad &\mbox{if} \quad b_{\bcan{c}} \neq 0
%\ea
%\right.
%\eq
%--
\bq
{\cal X}_{\bcan{c}}^{1n} =
\intsx{x}\,\Big[ \ln^n V_{\bcan{c}}(x,x) - \ln^n V_{\bcan{c}}(x,0) \Big];
\eq
%--
\bq
{\cal X}_{\bcan{c}}^{2n} =
\dssimp{2}\,\ln^{n-1} V_{\bcan{c}}\,\Big( 3\,\ln V_{\bcan{c}} + 2\,n \Big)
- \intsx{x}\,\Big[   \bX_{\bcan{c}}\,\ln^n V_{\bcan{c}}(1,x)
  + (x+X_{\bcan{c}})\,\ln^n V_{\bcan{c}}(x,x)  \Big].
\eq
%--
%\bqa
%{\cal X}_{\bcan{c}}^{10} &=&
%\intsx{x}\,\Big[ \ln V_{\bcan{c}}(x,x) - \ln V_{\bcan{c}}(x,0) \Big],
%\qquad\quad {\cal X}_{\bcan{c}}^{11}=
%\intsx{x}\,\Big[ \ln^2 V_{\bcan{c}}(x,x) - \ln^2 V_{\bcan{c}}(x,0) \Big];
%\\
%{\cal X}_{\bcan{c}}^{20} &=& 3\,\dsimp{2}\,\ln V_{\bcan{c}}
%- \intsx{x}\,  \Big[ 
% (1-X_{\bcan{c}})\,\ln V_{\bcan{c}}(1,x) + 
%(x+X_{\bcan{c}})\,\ln V_{\bcan{c}}(x,x)  \Big] + 1,
%\nl
%{\cal X}_{\bcan{c}}^{21} &=&
%\dsimp{2}\,\ln V_{\bcan{c}}\,\Big( 3\,\ln V_{\bcan{c}} + 4 \Big)- \intsx{x}\,
%  \Big[   (1-X_{\bcan{c}})\,\ln^2 V_{\bcan{c}}(1,x)
%  + (x+X_{\bcan{c}})\,\ln^2 V_{\bcan{c}}(x,x)  \Big].
%\nn
%\eqa
%--
Therefore, our result is also valid when $m_1 = m_3$ or 
$\lambda(M^2,m_1^2,m_2^2) = 0$, but not when the two conditions occur 
simultaneously. In the standard model this can actually happen only when
$m_1 = m_3 = M$ and $m_2 = 0$ which corresponds to configuration d) to be 
analyzed in the next subsection.

In order to compute $J_{\bcan{c,1}}$ and $J_{\bcan{c,2}}$, we change variables 
according to:
%--
\bqa
x_1 &=& x'_1 + x_2, \qquad x_2 = 1 - x'_2, \qquad x_1 \leftrightarrow x_2,
\qquad y_2= y_1\,y_2', \qquad\qquad \mbox{for} \quad i = 1,
\nl
x_1 &=& 1 - x'_1, \qquad x_2 = 1 - x'_2, \qquad x_1 \leftrightarrow x_2,
\qquad y_2= y_1\,(1-y_2'), \qquad\qquad \mbox{for} \quad i = 2,
\label{transf}
\eqa
%--
thus obtaining
%--
\bq
J_{\bcan{c,i}}=
\dsimp{2}(\{x\})\,\intfx{y_1}\,\int_0^{x_2/x_1}\!\!\!\!dy_2\,\,
\frac{1}{V_{\bcan{c,i}}}\,\,\,
\frac{x_2\,(1-x_1)\,y_1}{y_1\,y_2\,V_{\bcan{c,i}} + 
x_2\,(1-x_1)\,B_{\bcan{c,i}}},
\eq
%--
where we have introduced new quadratic forms $V_{\bcan{c,i}}$:
%--
\bqa
V_{\bcan{c,1}} &=&
\chi(1-x_1+x_2\,;\,-M^2\,;\,m^2_2\,,\,m^2_1) + (1-x_1)\,( m^2_3 - m^2_1),
%-\,M^2\,(1-x_1+x_2)\,(x_1-x_2) + m_1^2\,(1-x_1+x_2) + m_2^2\,(x_1-x_2)
%+ (m_3^2-m_1^2)\,(1-x_1),
\qquad\,
B_{\bcan{c,1}}= B_{\bcan{c}}(y_1,\,y_1y_2)
\\
V_{\bcan{c,2}} &=&
\chi(1-x_2\,;\,-M^2\,;\,m^2_2\,,\,m^2_1) + (1-x_1)\,( m^2_3 - m^2_1),
%-\,M^2\,x_2\,(1-x_2) + m_1^2\,(1-x_2) + m_2^2\,x_2 + (m_3^2-m_1^2)\,(1-x_1),
\qquad\qquad\,\,
B_{\bcan{c,2}}= B_{\bcan{c}}(y_1,\,y_1(1-y_2)).
\nn
\eqa
%--
For $V_{\bcan{c,1}}$ and $V_{\bcan{c,2}}$ two BST relations are 
available. Looking at the expression for $J_{\bcan{c,i}}$, we immediately 
see that the required BST relations are of the form of \eqn{BTli0}.
We have
%--
\bqa
\frac{1}{V_{\bcan{c,i}}}\,\frac{A}{V_{\bcan{c,i}}-A} &=&
\frac{1}{m_3^2-m_1^2}\,
\Big[  \frac{1}{V_{\bcan{c,i}}-A}\,\,{\cal P}_{i,1}\,A
+ {\cal P}_{i,1}\,\ln\Big(1-\frac{A}{V_{\bcan{c,i}}}\Big)\Big],
\nl
\frac{1}{V_{\bcan{c,i}}}\,\frac{A}{V_{\bcan{c,i}}-A} &=&
\frac{1}{b_{\bcan{c}}}\,\Big[
\frac{1}{V_{\bcan{c,i}}-A}\,\,(1+{\cal P}_{i,2})\,A
+ {\cal P}_{i,2}\,\ln\Big(1-\frac{A}{V_{\bcan{c,i}}}\Big)\Big],
\eqa
%--
where the explicit form of ${\cal P}_{i,j}$ is
%--
\bq
{\cal P}_{1,1} = - \partial_{x_1} - \partial_{x_2}, \qquad
{\cal P}_{1,2} = 
(1-x_1)\,\partial_{x_1} + 
\frac{1}{2}\,(1-x_1-x_2+X_{\bcan{c}})\,\partial_{x_2},
\eq
%--
\bq
{\cal P}_{2,1}=  - \partial_{x_1}, \qquad {\cal P}_{2,2}=
(1-x_1)\,\partial_{x_1} + \frac{1}{2}\,(1-x_2-X_{\bcan{c}})\,\partial_{x_2}.
\eq
%--
In our case it is $A = -x_2\,(1-x_1)\,B_{\bcan{c,i}}/(y_1 y_2)$, and we can
define $a_{i,j}$ such that
%--
\bq
{\cal P}_{i,1}\,A= a_{i,1}\,\frac{B_{\bcan{c,i}}}{y_1\,y_2},
\qquad\qquad
(1+{\cal P}_{i,2})\,A= \frac{a_{i,2}}{2}\,\frac{B_{\bcan{c,i}}}{y_1\,y_2}.
\eq
%--
For $a_{ij}$ we obtain:
%--
\bqa
a_{1,1} &=& 1-x_1-x_2, \qquad
a_{1,2} = (x_1+x_2-X_{\bcan{c}}-1)(1-x_1),
\nl
a_{2,1} &=&  -x_2, \qquad\qquad\quad
a_{2,2} = -\,(1-x_2-X_{\bcan{c}})\,(1-x_1).
\label{aij}
\eqa
%--
After integration by parts the $J_{\bcan{c,i}}$ can be expressed as
%--
\bq
J_{\bcan{c,i}} = \frac{1}{m_3^2-m_1^2}\,\sum_{\ssA=a}^d\,J_{i,1}^{\ssA} 
\qquad \mbox{if} \quad m_1\neq m_3,
\qquad\qquad
J_{\bcan{c,i}} = \frac{1}{2\,b_{\bcan{c}}}\,\,\sum_{\ssA=a}^d\,J_{i,2}^{\ssA} 
\qquad \mbox{if} \quad b_{\bcan{c}} \neq 0.
\eq
%--
%\bq
%J_{\bcan{c,i}}=
%\left\{
%\ba{ll}
%\frac{1}{m_3^2-m_1^2}\,
%\Big( J_{i,1}^a + J_{i,1}^b + J_{i,1}^c + J_{i,1}^d \Big)
%\qquad &\mbox{if} \quad m_1\neq m_3
%\nl\nl
%\frac{1}{b_{\bcan{c}}}\,
%\Big( J_{i,2}^a + J_{i,2}^b + J_{i,2}^c + J_{i,2}^d \Big)
%\qquad &\mbox{if} \quad b_{\bcan{c}} \neq 0
%\ea
%\right.
%\eq
%--
\bqa
J_{i,j}^a &=&
-\dsimp{2}(\{x\})\,a_{i,j}\,\intfx{y_1}\,\int_0^{x_2/x_1}\!\!\!\!dy_2\,\,
\frac{y_1}{y_1\,y_2\,V_{\bcan{c,i}} + x_2\,(1-x_1)\,B_{\bcan{c,i}}},
\nl
J_{i,j}^b &=&
\dsimp{2}(\{x\})\,\intfx{y_1}\,\int_0^{x_2/x_1}\!\!\!\!dy_2\,\,b_{i,j}\,
\frac{y_1}{B_{\bcan{c,i}}}\,
\ln\Bigg( 1 + 
\frac{x_2\,(1-x_1)\,B_{\bcan{c,i}}}{y_1\,y_2\,V_{\bcan{c,i}}} \Bigg),
\nl
J_{i,j}^c &=&
\dcub{3}(x\,\,\{y\})\,c_{i,j}\,\frac{y_1}{B_{\bcan{c,i}}}\,
\ln\Bigg( 1 + 
\frac{x\,(1-x)\,B_{\bcan{c,i}}}{y_1\,y_2\,V_{\bcan{c,i}}(x,x)} \Bigg),
\nl
J_{i,j}^d &=&
\dcub{3}(x\,\,\{y\})\,d_{i,j}\,\frac{y_1}{B_{\bcan{c,i}}}\,
\ln\Bigg( 1 + 
\frac{x\,(1-x)\,B_{\bcan{c,i}}}{y_1\,V_{\bcan{c,i}}(x,x\,y_2)} \Bigg),
\label{Jabcd}
\eqa
%--
where $a_{i,j}$ is defined in \eqn{aij}, while the $b\,\cdots\,d$ 
coefficients are:
%--
\bq
\ba{lll}
b_{1,1}= 0, &\qquad\quad
c_{1,1}= 0, &\qquad\quad
d_{1,1}= -(1-y_2),
\nl
b_{2,1}= 0, &\qquad\quad
c_{2,1}= -1, &\qquad\quad
d_{2,1}= y_2,
\nl
b_{1,2}= -3, &\qquad\quad
c_{1,2}= 1-X_{\bcan{c}}, &\qquad\quad
d_{1,2}= 1-x-2\,y_2+x\,y_2+X_{\bcan{c}},
\nl
b_{2,2}= -3,  &\qquad\quad
c_{2,2}= 1-x+X_{\bcan{c}}, &\qquad\quad
d_{2,2}= 1-2\,y_2+x\,y_2-X_{\bcan{c}}.
\ea
\eq
%--
All the quantities in \eqn{Jabcd} but $J_{i,j}^a$ are stable under numerical 
integration. For this integral we perform back the transformations of 
\eqn{transf} and get:
%--
\bq
J_{i,j}^a= \dsimp{2}(\{x\})\,a_{i,j}'\,J_i^a
\eq
%--
where the new coefficients $a_{i,j}'$ are
%--
\bq
a_{1,1}'= x_1-2x_2, \qquad
a_{1,2}'= (2x_2-x_1+X_{\bcan{c}})\,x_2,
\qquad
%\nn
%\eq
%%--
%\bq
a_{2,1}'= x_1-x_2, \qquad
a_{2,2}'= (\bX_{\bcan{c}}-x_1+x_2)\,x_2.
\label{aij'}
\eq
%--
The new integrals $J^a_i$ are defined by
%--
\bq
J_1^a = \intsx{y_1}\,\int_0^{\bX\,y_1}\!\!\!\!dy_2\,
\frac{1}{y_2\,V_{\bcan{c}} + x_2\,(x_1-x_2)\,B_{\bcan{c}}},
\eq
%--
\bq
J_2^a =
\intsx{y_1}\,\int_{\bX\, y_1}^{y_1}\!\!\!\!dy_2\,
\frac{1}{(y_1-y_2)\,V_{\bcan{c}} + x_2\,(1-x_1)\,B_{\bcan{c}}}.
\eq
%--
We transform according to $y_1 = y'_1 + y_2\,,\,
y_2 = 1 - y'_2\,,\,y_1 \leftrightarrow y_2$, obtaining
%--
\bq
J_1^a = \int_X^1\!\!dy_1\,\int_{(1-y_1)\,X/\bX}^{y_1}dy_2\,
\frac{1}{(1-y_1)\,V_{\bcan{c}} + x_2\,(x_1-x_2)\,\beta(y_2,y_1)},
\label{eqJ1a}
\eq
%--
\bq
J_2^a = \Bigg[ \intsxy{y_1}{y_2}
- \int_X^1\!\!dy_1\,\int_{(1-y_1)\,X/\bX}^{y_1}dy_2\Bigg]\,
\frac{1}{y_2\,V_{\bcan{c}} + x_2\,(1-x_1)\,\beta(y_2,y_1)},
\eq
%--
where the quadratic form $\beta$ given in \eqn{betadef}.
For $J^a_2$ we further transform $y_2 = y_1\,y_2'$; as a
consequence the $y_1$ integration becomes trivial giving:
%--
\bq
J_2^a = \intsx{y}\,\frac{1}{x_2\,(1-x_1)\,\chi(y)}\,
\ln\left\{
1 + \frac{x_2\,(1-x_1)^2\,\chi(y)}{y\,[1-x_1+(x_1-x_2)\,y]\,V_{\bcan{c}}}
\right\}.
\eq
%--
The $\chi$ function is defined in \eqn{chidef} .
For $J_1^a$ we can use the fact that the $B$ factor of the polynomial 
$\beta$ is zero. In this way we can use \eqn{BT0li0}:
%--
\bq
\frac{1}{V-A}\,\Big( {\cal P}_0 + {\cal P}_1^t\,\partial_x \Big)A=
-\,{\cal P}_1^t\,\partial_x\,\ln\Big(1-\frac{A}{V}\Big).
\eq
%--
To derive $J^a_1$ we replace
%--
\bq
V \to \beta(y_2,y_1), \qquad
A \to -\,\frac{(1-y_1)\,V_{\bcan{c}}}{x_2\,(x_1-x_2)}, \qquad
{\cal P}_0 \to 2, \qquad
 {\cal P}_1^t\,\partial_x \to - \,(y_1\,\partial_{y_1} + y_2\,\partial_{y_2}),
\eq
%--
and obtain the equation
%--
\bq
\Big( {\cal P}_0 + {\cal P}_1^t\,\partial_x \Big)A=
-\,(2-y_1)\,\frac{V_{\bcan{c}}}{x_2\,(x_1-x_2)}.
\eq
%--
Substituting in the expression for $J_1^a$, \eqn{eqJ1a}, 
%we obtain
%%--
%\bq
%J_1^a = -\,\frac{1}{V_{\bcan{c}}}\,
%\int_X^1\!\!dy_1\,\int_{(1-y_1)\,X/\bX}^{y_1}\,
%\frac{dy_2}{2-y_1}\,(y_1\,\partial_{y_1} + y_2\,\partial_{y_2})\,
%\ln\left\{ 
%1 + \frac{(1-y_1)\,V_{\bcan{c}}}{x_2\,(x_1-x_2)\,\beta(y_2,y_1)} \right\}.
%\eq
%%--
after integration by parts, we get the final result for $J^a_1$,
%--
\bqa
J_1^a &=& \frac{1}{V_{\bcan{c}}}\,
\int_X^1\!\!dy_1\,\int_{(1-y_1)\,X/\bX}^{y_1}dy_2\,\frac{4-y_1}{(2-y_1)^2}\,
\ln\left\{ 
1 + \frac{(1-y_1)\,V_{\bcan{c}}}{x_2\,(x_1-x_2)\,\beta(y_2,y_1)} \right\}
\nl
{}&+& \frac{1}{V_{\bcan{c}}}\,\intsx{y}\,\frac{X}{1+\bX\,y}\,
\ln\left\{ 
1 + \frac{y\,\bX\,V_{\bcan{c}}}{x_2\,(x_1-x_2)\,\beta(X\,y,1-\bX\,y)} \right\}.
\eqa
%--
Collecting all pieces together, we get:
%--
\bq
\ovalbox{\boldmath $V^{\bca}_c$} =
- \lpar \frac{\mu^2}{\pi} \rpar^{\ep}\egam{1+\ep}\,
\Big(\, \frac{1}{\ep}\,V^{\bcan{c}}_{-1} \,+\, V^{\bcan{c}}_0 \,\Big)
\eq
%--
\bqa
V^{\bcan{c}}_{-1} &=& 
-\frac{1}{m_3^2-m_1^2}\,{\cal X}_{\bcan{c}}^{11}\,{\cal Y}_{\bca}^1
\qquad
V^{\bcan{c}}_0= 
-\frac{1}{m_3^2-m_1^2}\,
\Big(
  {\cal X}_{\bcan{c}}^{11}\,{\cal Y}_{\bca}^2
+ {\cal X}_{\bcan{c}}^{12}\,{\cal Y}_{\bca}^1
+ \sum_{i=1}^2\sum_{\ssA=a}^d\,J_{i,1}^{\ssA} 
\Big)
\qquad \mbox{if} \quad m_1\neq m_3,
\nl
V^{\bcan{c}}_{-1} &=&
-\frac{1}{2\,b_{\bcan{c}}}\,{\cal X}_{\bcan{c}}^{21}\,{\cal Y}_{\bca}^1
\qquad\quad
V^{\bcan{c}}_0 =
-\frac{1}{2\,b_{\bcan{c}}}\,
\Big(
  {\cal X}_{\bcan{c}}^{21}\,{\cal Y}_{\bca}^2
+ {\cal X}_{\bcan{c}}^{22}\,{\cal Y}_{\bca}^1
+ \sum_{i=1}^2\sum_{\ssA=a}^d\,J_{i,1}^{\ssA} 
\Big)
\qquad \mbox{if} \quad b_{\bcan{c}} \neq 0.
\eqa
%--
\subsubsection{Evaluation of $V^{\bca}_d$ \label{Evalbcad}}
%--
Configuration d) of \fig{fig231} is a special case of c).
In fact the polynomial $V_{\bca}$ takes the form $V_{\bcan{d}} = M^2\,x_1^2$,
which has a double zero for $x_1 = 0$.
The BST factor for the polynomial $V_{\bcan{d}}$ is also zero revealing the 
presence of a new infrared pole; we are not allowed to put $\ep = 0$
but reexamine instead \eqn{hypcalY} which now is:
%--
\bqa
V_d^{\bca} &=& 
- \lpar \frac{\mu^2}{\pi} \rpar^{\ep}\egam{2+\ep}\,\dsimp{2}(\{x\})\,
  [x_2\,(1-x_2)]^{1+\ep/2}\,\intfx{y_1}\,
\spliti{0}{\bX\,y_1}{y_1}{y_2}{{\cal Y}}{{\bcan{d}}},
%\Bigl(\, \int_0^{\bX\,y_1}\!\!\!\!dy_2\,{\cal Y}_{\bcan{d,1}}
%+ \int_{\bX\,y_1}^{y_1}\!\!\!\!dy_2\,{\cal Y}_{\bcan{d,2}}\,\Bigg)
\nl
{\cal Y}_{\bcan{d,i}} &=& 
\int_0^{a_i}\!dy_3\,\,y_3^{\ep/2}\,
\Big[ M^2\,x_1^2\,y_3 + x_2\,(1-x_2)\,B_{\bcan{d}} \Big]^{-2-\ep}
\nl
{}&=& 
\frac{2}{2+\ep}\,[x_2\,(1-x_2)]^{-2-\ep}\,
\lpar \frac{a_i}{B^2_{\bcan{d}}}\rpar^{\!\!1+\ep/2}\,
%B_{\bcan{d}}^{-2-\ep}\,a_i^{1+\ep/2}\,
\!\!\!\!\!\!\!\hyperf{-\frac{a_iM^2x_1^2}{x_2(1-x_2)B_{\bcan{d}}}}.
%\!\!\!\!\!\!\hyper{2+\ep}{1+\frac{\ep}{2}}{2+\frac{\ep}{2}}
%{-\frac{a_i\,M^2\,x_1^2}{x_2\,(1-x_2)\,B_{\bcan{d}}}}.
\eqa
%--
where 
$a_1= y_2/\bX$, $a_2= (y_1-y_2)/X$ and $B_{\bcan{d}}(y_1,y_2) =
\beta( y_1 - y_2\,,\,1 - y_2\,;\,P^2\,;\,m^2\,,\,M^2)$,
with $\beta$ defined in \eqn{betadef}. 
We have also defined the shorthand notation
$\hyperf{x}\equiv \hyper{2+\ep}{1+\ep/2}{2+\ep/2}{x}$.
Now we split the integration region,
$[\bX y_1\,,\,y_1] = [0\,,\,y_1]\,\ominus\,[0\,,\,\bX y_1]$, 
and use the properties of the hypergeometric functions to obtain
%--
\bq
V_d^{\bca}= 
\lpar \frac{\mu^2}{\pi} \rpar^{\ep}
\Big(J_{\bcan{d,0}} + J_{\bcan{d,1}} + J_{\bcan{d,2}} \Big)
\eq
%--
\bqa
J_{\bcan{d,0}} &=&
- \,\frac{\egam{2+\ep}}{1+\ep/2}\!
\dssimp{2}(\{x\})\!\dssimp{2}(\{y\})
\left[ \frac{y_1-y_2}{x_2(1-x_1)B_{\bcan{d}}^2} \right]^{1+\ep/2}
%[x_2\,(1-x_1)]^{-1-\ep/2}\,(y_1-y_2)^{1+\ep/2}\,B_{\bcan{d}}^{-2-\ep}\,
%\nl
%&{}&\times\, 
\!\!\!\!\!\!\!\!\!\!
\hyperf{-\frac{M^2\,x_1^2\,(y_1-y_2)}{x_2(1-x_1)B_{\bcan{d}}}},
%\hyper{2+\ep}{1+\frac{\ep}{2}}{2+\frac{\ep}{2}}
%{-\frac{M^2\,x_1^2\,(y_1-y_2)}{x_2(1-x_1)B_{\bcan{d}}}},
\nl
J_{\bcan{d,1}} &=& \frac{1}{M^2}\,
\dsimp{2}(\{x\})\,\intfx{y_1}\,\int_0^{\bX\,y_1}\!\!\!\!dy_2\,
\frac{x_2\,(x_1-x_2)}{x_1^2}\,
\frac{1}{M^2\,x_1^2\,y_2+x_2\,(x_1-x_2)\,B_{\bcan{d}}},
\nl
J_{\bcan{d,2}} &=& -\,\frac{1}{M^2}\,
\dsimp{2}(\{x\})\,\intfx{y_1}\,\int_0^{\bX\,y_1}\!\!\!\!dy_2\,
\frac{x_2\,(1-x_1)}{x_1^2}\,
\frac{1}{M^2\,x_1^2\,(y_1-y_2)+x_2\,(1-x_1)\,B_{\bcan{d}}}.
\eqa
%--
We make the transformations
$x_2 = x_1\,x'_2 \,,\, y_1 = y'_1 + y_2 \,,\, y_2 = 1 - y'_2 \,,\,,
y_1 \leftrightarrow y_2$, obtaining
%--
\bqa
J_{\bcan{d,0}} &=&
- \frac{\egam{2+\ep}}{1+\ep/2}\!\dscub{2}(\{x\})\!\dssimp{2}(\{y\})\,
x_1^{-\ep/2}\,
\left[ \frac{y_2}{x_2(1-x_1)\beta^2(y_2,y_1)} \right]^{1+\ep/2}
\!\!\!\!\!\hyperf{-\frac{M^2\,x_1\,y_2}{x_2(1-x_1)\beta(y_2,y_1)}},
%[x_2\,(1-x_1)]^{-1-\ep/2}\,y_2^{1+\ep/2}\,
%\nl
%&{}&\times\, \beta(y_2,y_1)^{-2-\ep}\,
%\hyper{2+\ep}{1+\frac{\ep}{2}}{2+\frac{\ep}{2}}
%{-\frac{M^2\,x_1\,y_2}{x_2\,(1-x_1)\,\beta(y_2,y_1)}},
\nl
J_{\bcan{d,1}} &=& \frac{1}{M^2}\,
\dcub{2}(\{x\})\,\int_{X'}^1\!\!dy_1\,\int_{(1-y_1)\,X'/\bX'}^{y_1}dy_2\,
\frac{x_2\,(1-x_2)}{x_1}\,
\frac{1}{M^2\,(1-y_1)+x_2\,(1-x_2)\,\beta(y_2,y_1)},
\nl
J_{\bcan{d,2}} &=& -\,\frac{1}{M^2}\,
\dcub{2}(\{x\})\,\int_{X'}^1\!\!dy_1\,\int_{(1-y_1)\,X'/\bX'}^{y_1}dy_2\,
\frac{x_2\,(1-x_1)}{x_1}\,
\frac{1}{M^2\,x_1\,y_2+x_2\,(1-x_1)\,\beta(y_2,y_1)},
\label{eqJd0}
\eqa
%--
where $\beta$ is defined in \eqn{betadef} and
%--
\bq
X' = \frac{1-x_1}{1-x_1\,x_2}, \qquad\qquad
\bX' = 1 - X' = \frac{x_1\,(1-x_2)}{1-x_1\,x_2}.
\eq
%--
For $J_{\bcan{d,1}}$ ($J_{\bcan{d,2}}$) we apply the same technique 
already used in the previous subsection to treat $J_1^a$ ($J_2^a$). We obtain
%--
\bqa
J_{\bcan{d,1}} &=& \frac{1}{M^4}\,\intsx{x_2}\dsimp{2}{\{y\}}\,
\frac{4-y_1}{(2-y_1)^2}\,x_2\,(1-x_2)\,
\ln\left( 1 + \frac{(1-x_2)\,y_2}{1-y_1} \right)\,\ln\left\{ 
1 + \frac{(1-y_1)\,M^2}{x_2\,(1-x_2)\,\beta(y_2,y_1)} \right\}
\nl
{}&+& \,\frac{1}{M^4}\,
\dcub{3}(\{x\},y)\,\frac{x_2\,(1-x_2)}{x_1}\,\frac{X'}{1+\bX'\,y}\,
\ln\left\{ 
1 + \frac{x_1\,y\,M^2}{x_2\,(1-x_1\,x_2)\,\beta(X'\,y,1-\bX'\,y)} \right\},
\\
J_{\bcan{d,2}} &=& \frac{1}{M^2}\,\dcub{3}(\{x\},y)\,\frac{1}{x_1\,\chi(y)}\,
\ln\left\{
1 - \frac{x_1\,x_2\,(1-x_1)\,(1-x_2)\,y\,\chi(y)}
 {[1-x_1+x_1\,(1-x_2)\,y]\,[M^2\,x_1\,y+x_2\,(1-x_1)\,\chi(y)]}\right\}.
\nn
\eqa
%--
To extract the infrared poles from $J_{\bcan{d,0}}$ of \eqn{eqJd0} we first 
transform according to $y_2 = y_1\,y_2'$ and then perform a sector 
decomposition of the unit square in $x_1$ and $y_1$:
$[0,1]^2 = [0,1]\,\otimes\,[0,x_1]\,\oplus\,[0,1]\,\otimes\,[0,y_1]$.
Each sector is then mapped back into the unit square giving
$J_{\bcan{d,0}}= J_{\ep,1} + J_{\ep,2}$, with
%--
\bqa
J_{\ep,1} &=&
-\,(M^2)^{-1-\ep/2}\,\egams{1+\ep/2}\,
\dscub{3}(x,\{y\})\,\,x_1^{-1-2\,\ep}\,y_1^{-1-\ep}\,
\chi(y_2)^{-1-\ep/2}
\\
{}&+& \,
%(M^2)^{-2-\ep}\,
\frac{\egam{2+\ep}}{1+\ep/2}\dscub{4}(\{x\},\{y\})\,\,
x_1^{-1-2\,\ep}\,y_1^{-\ep/2}\,
\left[ \frac{x_2(1-x_1)}{y_2\,M^4} \right]^{1+\ep/2}
\!\!\!\!\!\hyperf{-\frac{x_2\,(1-x_1)\,y_1\,\chi(y_2)}{M^2\,y_2}},
%[x_2\,(1-x_1)]^{1+\ep/2}\,
%\nl
%&{}&\times\, y_2^{-1-\ep/2}\,
%\hyper{2+\ep}{1+\frac{\ep}{2}}{2+\frac{\ep}{2}}
%{-\frac{x_2\,(1-x_1)\,y_1\,\chi(y_2)}{M^2\,y_2}},
\nl
J_{\ep,2} &=&
- \frac{\egam{2+\ep}}{1+\ep/2}
  \dscub{4}(\{x\},\{y\})\,x_1^{-\ep/2}\,y_1^{-1-2\,\ep}\,
  \left[ \frac{y_2}{x_2(1-x_1y_1)\chi^2(y_2)} \right]^{1+\ep/2}
  \!\!\!\!\!\hyperf{-\frac{M^2x_1y_2}{x_2(1-x_1y_1)\chi(y_2)}}.
\nn
%  [x_2\,(1-x_1\,y_1)]^{-1-\ep/2}\,
%  y_2^{1+\ep/2}\,
%\nl
%&{}&\times\, \chi(y_2)^{-2-\ep}\,
%\hyper{2+\ep}{1+\frac{\ep}{2}}{2+\frac{\ep}{2}}
%{-\frac{M^2\,x_1\,y_2}{x_2\,(1-x_1\,y_1)\,\chi(y_2)}}.
\label{eqJepi}
\eqa
%--
For $J_{\ep,1}$ we have used the properties of the hypergeometric 
functions of related arguments (see app.~\ref{app:hyp}).
In the first term of $J_{\ep,1}$, the integration over $x_1$ and $y_1$ 
is trivially performed giving the double infrared pole.
The two terms containing the hypergeometric function are also divergent 
for $\ep \to 0$. The corresponding pole can be extracted as in \eqn{poleext}.
Setting $\ep = 0$ when possible, we get:
%--
\bqa
J_{\bcan{d,0}} &=& 
-\,\frac{1}{2}\,\frac{1}{\ep^2}\,
   (M^2)^{-1-\ep/2}\,\egams{1+\frac{\ep}{2}}\,{\cal Y}_{\bca}\,
-\,\frac{\egam{2+\ep}}{1+\ep/2}
\dscub{4}(\{x\},\{y\})\,{\tilde J}_{\bcan{d,0}},
\nl
{\tilde J}_{\bcan{d,0}} &=& 
\frac{1}{2\,\ep}\,
\Bigg[
y_1^{-\ep/2}\,
\lpar\frac{x_2}{y_2 M^4}\rpar^{1+\ep/2}\,
\!\!\!\!\!\hyperf{-\frac{x_2\,y_1\,\chi(y_2)}{M^2\,y_2}}
- x_1^{-\ep/2}\,
  \lpar\frac{y_2}{x_2\chi^2(y_2)}\rpar^{1+\ep/2}\,
  \!\!\!\!\!\hyperf{-\frac{M^2\,x_1\,y_2}{x_2\,\chi(y_2)}}
\Bigg]
\nl
{}&-& 
  \frac{x_2}{M^2}\,\frac{1}{x_1}\,
  \frac{1-x_1}{x_2\,(1-x_1)\,y_1\,\chi(y_2)+M^2\,y_2} \Bigg|_+
+ \frac{y_2}{\chi(y_2)}\,\frac{1}{y_1}\,
  \frac{1}{x_2\,(1-x_1\,y_1)\,\chi(y_2)+M^2\,x_1\,y_2} \Bigg|_+.
\eqa
%--
${\cal Y}_{\bca}$ has already been considered in section~\ref{evalbcaa}.
The coefficients of its expansion
${\cal Y}_{\bca} = 
{\cal Y}_{\bca}^1 + {\cal Y}_{\bca}^2\,\ep + {\cal Y}_{\bca}^3\,\ep^2$ 
are given in \eqn{bcaY}.
Now we expand the hypergeometric 
functions around $\ep = 0$ according to \eqn{hypep} and perform everywhere 
the $x_1$ and $y_1$ integrations.
Using properties of the dilogarithm we obtain:
%--
\bqa
{\tilde J}_{\bcan{d,0}} &=& 
\frac{\egam{1+\ep}}{M^2}\,\int_0^1\,\frac{dx\,dy}{\chi(y)}\Bigl\{
-\,\frac{1}{2\,\ep}\,\ln \eta_{\bcan{d}}\,
+ \frac{1}{2}\,\Big[ \frac{1}{2}\,\ln \frac{x}{y} + \ln \chi(y) \Big]\,
\ln\eta_{\bcan{d}}\nl
{}&-& \li{2}{\eta_{\bcan{d}}}
- \ln\eta_{\bcan{d}}\,\ln(1-\eta_{\bcan{d}}) + \frac{\zeta(2)}{2}
\Bigr\}.
\qquad\qquad
\eta_{\bcan{d}}= \frac{x\,\chi(y)}{y\,M^2}
\label{notyet}
\eqa
%--
%\bqa
%{\tilde J}_{\bcan{d,0}} &=&
%-\,\frac{1}{2\,\ep}\,\pi^{-\ep}\,\egam{1+\ep}\,\frac{1}{M^2}\,
%\int_0^1\!\!dx\,dy\,\frac{\ln \eta_{\bcan{d,0}}}{\chi(y)}\,
%+ \frac{1}{M^2}\,\int_0^1\!\!dx\,dy\,\frac{1}{\chi(y)}\,
%\nl
%{}&\times& \Big\{
%\frac{1}{2}\,\Big[ \frac{1}{2}\,\ln \frac{x}{y} + \ln \chi(y) \Big]\,
%\ln\eta_{\bcan{d,0}} - \li{2}{\eta_{\bcan{d,0}}}
%- \ln\eta_{\bcan{d,0}}\,\ln(1-\eta_{\bcan{d,0}}) + \frac{\zeta(2)}{2}\Big\}.
%\label{notyet}
%\eqa
%--
The result of \eqn{notyet} can be cast in a form suited for 
numerical evaluation, by introducing the integrals
%--
%\bq
%J_n^k= \intsx{y}\,\ln^n y\,\frac{\ln^k\chi(y)}{\chi(y)}
%\eq
%%--
%which have already been treated in section~\ref{evaladac} (see \eqn{evalJnk}).
%
$J_n^k$ already treated in section~\ref{evaladac} (see \eqn{evalJnk}).
Combining all terms together we finally write $J_{\bcan{d,0}}$ as
%--
\bq
\ovalbox{\boldmath $V^{\bca}_d$} = 
- \lpar \frac{\mu^2}{\pi} \rpar^{\ep}\egam{1+\ep}\,
\Big(\, 
      \frac{1}{\ep^2}\,V^{\bcan{d}}_{-2} 
\,+\, \frac{1}{\ep}\,V^{\bcan{d}}_{-1} 
\,+\, V^{\bcan{d}}_0 
\,\Big)
\eq
%--
\bqa
V^{\bcan{b}}_{-2} &=& \frac{J_0^0}{2\,M^2},
\qquad
V^{\bcan{b}}_{-1}= 
- \,\frac{1}{2\,M^2}\,\bigg[
\Big( \frac{3}{2}\,\ln M^2 + 1\Big)\,J_0^0  - \frac{1}{2}\,J_0^1 + J_1^0
\bigg],
\nl
V^{\bcan{b}}_0 &=&
\frac{1}{2\,M^2}\,\bigg\{
\frac{1}{8}\,( \ln^2 M^2 - 4\,\ln M^2 - 10\,\zeta(2) - 8 )\,J_0^0
+ \frac{1}{4}\,( 5\,\ln M^2 + 6 )\,J_0^1
- \frac{1}{2}\,( \ln M^2 + 2 )\,J_1^0
\nl
{}&+& \frac{3}{2}\,J_1^1 - \frac{7}{8}\,J_0^2 - \frac{1}{2}\,J_2^0
+ 2\,\int_0^1\!\!dx\,dy\,\frac{1}{\chi(y)}\,\Big[
\li{2}{\eta_{\bcan{d}}} + \ln\eta_{\bcan{d}}\,\ln(1-\eta_{\bcan{d}})
\Big]
\bigg\}
- J_{\bcan{d,1}} - J_{\bcan{d,2}}.\qquad
\label{resV231d}
\eqa
%--
\subsection{The $V^{\bbb}$ diagram \label{vbbb}}
%--
Finally, we consider the $V^{\bbb}$-family given in \fig{TLvertbbb} which 
is representables as 
%--
\bq
\pi^4\,V^{\bbb} =
\mu^{2\ep}\,\int\,d^nq_1 d^nq_2\,
\frac{1}{[1]_{\bbb}[2]_{\bbb}[3]_{\bbb}[4]_{\bbb}[5]_{\bbb}[6]_{\bbb}},
\label{v222prop}
\eq
%--
with propagators
%--
\bq
[1]_{\bbb} = q^2_1 + m^2_1,  \quad
[2]_{\bbb} = (q_1-p_2)^2+ m^2_2,  \quad
[3]_{\bbb} = (q_1-q_2 +p_1)^2 + m^2_3,
\eq
%--
\bq
[4]_{\bbb} = (q_1-q_2-p_2)^2 + m^2_4, \quad
[5]_{\bbb} = q^2_2 + m^2_5,  \quad
[6]_{\bbb} = (q_2-p_1)^2 + m^2_6,
\label{def222}
\eq
%--
\begin{figure}[ht]
\begin{center}
\begin{picture}(150,75)(0,0)
 \SetWidth{1.5}
 \Line(0,0)(40,0)         \LongArrow(0,8)(20,8)          \Text(-11,7)[cb]{$-P$}
 \Line(128,-53)(100,-35)  \LongArrow(128,-63)(114,-54)   \Text(138,-70)[cb]{$p_1$}
 \Line(128,53)(100,35)    \LongArrow(128,63)(114,54)     \Text(138,62)[cb]{$p_2$}
 \Line(70,-17.5)(100,35)                            \Text(97,10)[cb]{$1$}
 \Line(100,35)(70,17.5)                             \Text(82,31)[cb]{$2$}
 \Line(70,-17.5)(40,0)                              \Text(53,-21)[cb]{$3$}
 \Line(70,17.5)(40,0)                               \Text(53,14)[cb]{$4$}
 \Line(100,-35)(82,-3.5)\Line(78,3.5)(70,17.5)      \Text(97,-16)[cb]{$5$}
 \Line(100,-35)(70,-17.5)                           \Text(82,-39)[cb]{$6$}
\end{picture}
\end{center}
\vspace{2cm}
\caption[]{The irreducible two-loop vertex diagrams $V^{\bbb}$. External 
momenta are flowing inwards.} 
\label{TLvertbbb}
\end{figure}
%--

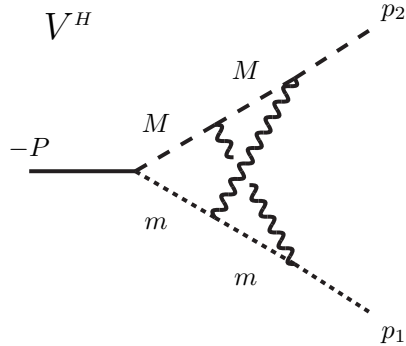
\begin{figure}[ht]
\begin{center}
\begin{picture}(150,75)(0,0)
 \SetWidth{1.5}
 \Line(0,0)(40,0)                    \Text(0,5)[cb]{$-P$}
 \DashLine(128,-53)(100,-35){2}      \Text(138,-65)[cb]{$p_1$}
 \DashLine(128,53)(100,35){5}        \Text(138,57)[cb]{$p_2$}
 \DashLine(70,17.5)(40,0){5}         \Text(48,15)[cb]{$M$}
 \DashLine(70,-17.5)(40,0){2}        \Text(48,-22)[cb]{$m$}
 \DashLine(100,35)(70,17.5){5}       \Text(82,35)[cb]{$M$}
 \DashLine(100,-35)(70,-17.5){2}     \Text(82,-43)[cb]{$m$}
 \Photon(70,-17.5)(100,35){2}{9}
 \Photon(100,-35)(83,-5.25){2}{5}\Photon(77,5.25)(70,17.5){2}{2}
 \Text(15,50)[cb]{\Large $V^{\bbb}$}
\end{picture}
\end{center}
\vspace{2cm}
\caption[]{The $V^{\bbb}$ infrared configurations. The photon line represents 
a general massless particle while dashed and solid lines refer to 
different massive particles.}
\label{fig222}
\end{figure}

This diagram has only one infrared configuration (shown in \fig{fig222}) 
which corresponds to
%--
%\bq
$m_1= m_5= 0$, and $p_1^2= - m_3^2= -m_6^2= -m^2$
$p_2^2= - m_2^2= -m_4^2= -M^2$
%\eq
%--
%--
At first we combine propagators $[1]_{\bbb}-[4]_{\bbb}$ with Feynman 
parameters $x_1,x_2,x_3$,
%--
\bq
\pi^4\,V^{\bbb} = 
\mu^{2\ep}\,\egam{4}\,\int\,d^nq_1 d^nq_2\,\dsimp{3}(\{x\})\,
\frac{1}{[5]_{\bbb}[6]_{\bbb}}\,
\frac{1}{(q^2_1 + 2\,\spro{R_x}{q_1} + Q_x^2)^4},
\eq
%--
where $x$-dependent quantities are
%--
\bq
R_x = - x_2\,q_2 - x_1\,p_2 + x_3\,P,
\qquad  
Q^2_x = x_2\,q_2^2 + 2\,x_2\,\spro{p_2}{q_2} - 2\,x_3\,\spro{P}{q_2}
\eq
%--
Integration over $q_1$ gives
%--
\bq
\pi^2\,V^{\bbb} = 
i\,\frac{\mu^{2\ep}}{\pi^{\ep/2}}\,\egam{2+\frac{\ep}{2}}\dsimp{3}(\{x\})\,
%i\,\pi^{-\ep/2}\,\egam{2+\frac{\ep}{2}}\,\dsimp{3}(\{x\})\,
\Bigl[x_2(1-x_2)\Bigr]^{-2-\ep/2}\,
\int\,
\frac{d^nq_2}{(q^2_2+2\,
\spro{P_x}{q_2} +M^2_x)^{2+\ep/2}\,[5]_{\bbb}[6]_{\bbb}},
%\intmomi{n}{q_2}\,
%\frac{1}{(q^2_2+2\,\spro{P_x}{q_2} +M^2_x)^{2+\ep/2}\,[5]_{\bbb}[6]_{\bbb}},
\eq
%--
with new auxiliary quantities defined by
%--
\bq
P_x = \frac{1}{x_2\,(1-x_2)}\,\Big[ x_2\,(1-x_1)\,p_2 - x_3\,(1-x_2)\,P \Big]
\qquad
M^2_x = -\frac{1}{x_2\,(1 - x_2)}\,\Big[ x_1\,p_2 - x_3\,P \Big]^2.
\eq
%--
Secondly, we combine the remaining propagators with Feynman parameters
$y_1,y_2$: it follows
%--
\bq
\pi^2V^{\bbb} =
 i\frac{\mu^{2\ep}}{\pi^{\ep/2}}\,\egam{4+\frac{\ep}{2}}\!
\dssimp{3}(\{x\})\,[x_2\,(1-x_2)]^{-2-\ep/2}\!\!
\dssimp{2}(\{y\})\,y_2^{1+\ep/2}\!\!
\int\!\!\frac{d^nq_2}{(q_2^2 + 2\,\spro{R_y}{q_1} + Q_y^2)^{4+\ep/2}},
\eq
%--
where $y$-dependent quantities are $R_y = y_2\,P_x - (y_1-y_2)\,p_1$ and 
$Q_y^2 = y_2\,M_x^2$.
Integration over $q_2$ followed by a transformation $y_2 = y_1\,y'_2$, gives
%--
\bq
V^{\bbb} = 
-\,\lpar \frac{\mu^2}{\pi} \rpar^{\ep}\!\egam{2+\ep}\!
\dssimp{3}(\{x\})\!\dscub{2}(\{y\})\,[x_2\,(1-x_2)]^{-2-\ep/2}\,
y_1^{-\ep/2}\,y_2^{1+\ep/2}\,(A_{\bbb}+B_{\bbb}\,y_1)^{-2-\ep}_{\bca},
\eq
%--
where $A_{\bbb}$ and $B_{\bbb}$ are
%--
\bq
A_{\bbb}\! =
M_x^2\,y_2\! = \frac{y_2}{x_2(1\!-\!x_2)}\,\bbeta(x_3,x_1),
\qquad
B_{\bbb}\! =
- \Big[ P_xy_2 - p_1(1\!-\!y_2) \Big]^2\!\!\! =\,
\bbeta\Big( 1 - \frac{x_2\!-\!x_3}{x_2}\,y_2 , 
            1 - \frac{x_1\!-\!x_2}{1\!-\!x_2}\,y_2 \Big)
\label{defABbbb}
\eq
%--
and the quadratic $\bbeta$ is defined in \eqn{betabdef}.
To proceed in the evaluation of the diagram we perform the following changes 
of variables
%--
\bq
x_3 \rightarrow x_2\,x_3, \qquad
y_2 \rightarrow \frac{1-x_2}{x_1-x_2}\,y_2, \qquad
x_2 \rightarrow x_1\,x_2, \qquad
x_1 \rightarrow \frac{x_1}{1-(1-x_1)\,x_2}, \qquad
y_2 \rightarrow x_1\,y_2,
\eq
%--
and derive
%--
\bqa
V^{\bbb} &=& 
-\,\lpar \frac{\mu^2}{\pi} \rpar^{\ep}\egam{2+\ep}\,
\dcub{4}(\{x\}\,,\,y_2)\,
x_1^{-\ep/2}\,[1-(1-x_1)\,x_2]^{\ep}\,[x_2\,(1-x_2)]^{-1-\ep/2}\,
y_2^{1+\ep/2}\,{\cal Y}_{\bbb}
\nl
{\cal Y}_{\bbb} &=&
\int_0^1\!\!dy_1\,y_1^{-\ep/2}\,
\Big( {\cal A}_{\bbb} + {\cal B}_{\bbb}\,y_1  \Big)^{-2-\ep}=
\frac{2}{2-\ep}\,{\cal A}_{\bbb}^{-2-\ep}\,
\hyper{2+\ep}{1-\frac{\ep}{2}}{2-\frac{\ep}{2}}
{-\frac{{\cal B}_{\bbb}}{{\cal A}_{\bbb}}}
\eqa
%--
where ${\cal A}_{\bbb}, {\cal B}_{\bbb}$ factors are
%--
\bq
{\cal A}_{\bbb} = \frac{x_1\,y_2}{x_2\,(1-x_2)}\,\bchi(x_2\,x_3),
\qquad\qquad
{\cal B}_{\bbb} = \bbeta( 1 - (1-x_3)\,y_2 , 1 - x_1\,y_2 ).
\eq
%--
The quadratic form $\bchi$ is defined in \eqn{chibdef}.
Using properties of the hypergeometric function and setting $\ep = 0$ whenever 
possible, we obtain:
%--
\bq
V^{\bbb} = 
-\,\lpar \frac{\mu^2}{\pi} \rpar^{\ep}\egam{1-\frac{\ep}{2}}\,
\egam{1+\frac{3}{2}\,\ep}\,\dcub{4}(\{x\},y_2)\,
{\cal B}_{\bbb}^{-1+\ep/2}\,{\cal V}^{\bbb}
\eq
%--
\bq
{\cal V}^{\bbb} = 
  x_1^{-1-2\,\ep}\,\left[\frac{1-(1-x_1)\,x_2}{y_2}\right]^{\ep}\,
  [x_2\,(1-x_2)]^{\ep}\,\bchi^{-1-3\,\ep/2}(x_2\,x_3)
+ \frac{y_2}{x_1\,y_2\,\bchi(x_2\,x_3)+x_2\,(1-x_2)\,{\cal B}_{\bbb}}.
\label{seebel}
\eq
%--
The first term in \eqn{seebel} is an integral of the type shown in 
\eqn{poleext} and the $V^{\bbb}$ diagram can be rewritten as
%--
\bq
V^{\bbb}= 
\lpar \frac{\mu^2}{\pi} \rpar^{\ep}\egam{1-\frac{\ep}{2}}\,
\egam{1+\frac{3}{2}\,\ep}\,
\Big[ \frac{1}{2\,\ep}\,\int_0^1\!\!dx_3\,J_x\,J_y - J \Big]
\label{Jfirst}
\eq
%--
\bq
J_x = \intfx{x_2}\,x_2^{\ep}\,(1-x_2)^{2\,\ep}\,
\bchi^{\,-1-3\,\ep/2}(x_2\,x_3), \qquad\qquad
J_y = \intfx{y_2}\,y_2^{-\ep}\,\bchi^{\,-1+\ep/2}(1-(1-x_3)\,y_2),
\eq
%--
\bq
J = \dcub{4}(\{x\}\,,\,y_2)\,\frac{1}{x_1\,\bchi(x_2\,x_3)}\,
\Bigg[  \,
\frac{x_2\,(1-x_2)}{x_1\,y_2\,\bchi(x_2\,x_3)+x_2\,(1-x_2)\,{\cal B}_{\bbb}}\,
- \,\frac{1}{{\cal B}_{\bbb}^0}\,\Bigg].
\label{Jsecond}
\eq
%--
Here we define 
${\cal B}_{\bbb}^0={\cal B}_{\bbb}(x_1=0)=\bchi(1-(1-x_3)\,y_2)$.
The quadratic forms corresponding to $J_x$ and $J_y$ can be treated according 
to:
%--
\bqa
\bchi^{\,\mu}(x_2\,x_3) &=& \frac{1}{\Bbt}\, \Bigg[ 
1 - \frac{1}{2\,(\mu+1)}\,\Big( x_2 - \frac{\bXbt}{x_3} \Big)\,\partial_{x_2}
\Bigg]\, \bchi^{\,\mu+1}(x_2\,x_3)
\nl
\bchi^{\,\mu}(1-(1-x_3)\,y_2) &=& \frac{1}{\Bbt}\,\Bigg[ 
1 - \frac{1}{2\,(\mu+1)}\,\Big( y_2 - \frac{\Xbt}{1-x_3} \Big)\,\partial_{y_2}
\Bigg]\, \bchi^{\,\mu+1}(1-(1-x_3)\,y_2).
\label{btJxJy}
\eqa
%--
After integration by parts we write $J_{x,y}$ as
%--
\bq
J_x= \frac{1}{2\,\Bbt}\,\int_0^1\!\!dx_2\,[I_x^0 + I_x^1\,\ep]
\qquad\qquad
J_y= \frac{1}{2\,\Bbt}\,\int_0^1\!\!dy_2\,[I_y^0 + I_y^1\,\ep]
\eq
%--
\bqa
I_x^0 &=&    \ln\bchi(x_2\,x_3)
+ \frac{\bXbt}{x_3}\,\ln\frac{\bchi(x_3)}{\bchi(0)} - \ln\bchi(x_3) + 2
\nl
I_x^1 &=& - \frac{3}{4}\,\ln^2\bchi(x_2\,x_3)
+ (\ln x_2 + 2\,\ln(1-x_2))\,\ln\bchi(x_2\,x_3)
- \frac{\bXbt}{x_2\,x_3}\,\ln\frac{\bchi(x_2\,x_3)}{\bchi(0)}
\nl
{}&-& \frac{2}{1-x_2}\,\Big(1-\frac{\bXbt}{x_3}\Big)\,
  \ln\frac{\bchi(x_2\,x_3)}{\bchi(x_3)} 
- \frac{3}{4}\,\frac{\bXbt}{x_3}\,\ln\frac{\bchi(x_3)}{\bchi(0)}\,
  \Big[ \ln\bchi(x_3) + \ln\bchi(0) \Big]
+ \frac{3}{4}\,\ln^2\bchi(x_3) - 6
\nl
I_y^0 &=&   \ln\bchi(1-(1-x_3)\,y)
+ \frac{\Xbt}{1-x_3}\,\ln\frac{\bchi(x_3)}{\bchi(1)} - \ln\bchi(x_3) + 2
\nl
I_y^1 &=& \frac{1}{4}\,\ln^2\bchi(1-(1-x_3)\,y)
- \ln y\,\ln\bchi(1-(1-x_3)\,y)
+ \frac{\Xbt}{y\,(1-x_3)}\,\ln\frac{\bchi(1-(1-x_3)\,y)}{\bchi(1)}
\nl
{}&+& \frac{1}{4}\,\frac{\Xbt}{1-x_3}\,\ln\frac{\bchi(x_3)}{\bchi(1)}\,
  \Big[ \ln\bchi(x_3) + \ln\bchi(1) \Big] - \frac{1}{4}\,\ln^2\bchi(x_3) + 2
\label{IxIy}
\eqa
%--
The computation of $J$, \eqns{Jfirst}{Jsecond}, is actually more involved. 
First of all we transform variables according to $x'_1= y_2\,x_1$.
%--
%\bq
%x_1 \rightarrow \frac{x_1}{y_2},
%\qquad \Rightarrow \qquad
%\int_0^1\!\!dy_2\,\int_0^1\!\!\frac{dx_1}{x_1} \rightarrow 
%\int_0^1\!\!dy_2\,\int_0^{y_2}\!\!\frac{dx_1}{x_1}=
%\int_0^1\!\!dx_1\,\int_{x_1}^1\!\!dy_2\,\frac{1}{x_1}.
%\eq
%%--
For the term proportional to ${\cal B}_{\bbb}$ we write
$1/x_1 = (1-x_1)^2/x_1 + (2 - x_1)$. Now $J$ is given by:
%--
\bq
J = 
\dcub{3}(\{x\})\,\int_{x_1}^1\!\!dy_2\,\,
\frac{1}{\bchi}\,\Bigg\{\,
\frac{1}{x_1}\,\bigg[\,
\frac{x_2\,(1-x_2)\,(1-x_1)^2}{x_1\,\bchi+x_2\,(1-x_2)\,\bbeta}\,
- \,\frac{1}{\bbeta_{x_1}}\,\bigg]
\,+\,
\frac{x_2\,(1-x_2)\,(2-x_1)}{x_1\,\bchi+x_2\,(1-x_2)\,\bbeta}
\,\Bigg\},
\label{J12}
\eq
%--
We have introduced shorthand notations:
%--
\bq
\bchi= \bchi(\,x_2\,x_3\,), \qquad\qquad
\bbeta= \bbeta(\,1-(1-x_3)\,y_2\,,\,1-x_1\,), \qquad\qquad
\bbeta_{x_1}= \bchi(\,1-(1-x_3)\,y_2\,).
\eq
%--
Both terms of \eqn{J12} are split, i.e. 
%--
\bqa
J = J_{11} + J_{12} + J_{21} + J_{22}, &\qquad&
J_{ij} = \dcub{3}(\{x\})\,J'_{ij},
\nl
J'_{11} = \int_{x_1}^1\!\!dy_2\,\,
\frac{1}{x_1\,\bchi}\,\Bigg[\,
\frac{(1-x_1)^2}{\bbeta}\, - \,\frac{1}{\bbeta_{x_1}}\,\Bigg],
&\qquad&
J'_{12} = -\,\int_{x_1}^1\!\!dy_2\,\,
\frac{1}{\bbeta}\,\frac{(1-x_1)^2}{x_1\,\bchi+x_2\,(1-x_2)\,\bbeta},
\nl
J'_{21} = \int_{x_1}^1\!\!dy_2\,\,\,
\frac{1-x_3}{\bchi}\,\,\,
\frac{x_2\,(1-x_2)\,(2-x_1)}{x_1\,\bchi+x_2\,(1-x_2)\,\bbeta},
&\qquad&
J'_{22} = \int_{x_1}^1\!\!dy_2\,\,\,
\frac{x_3}{\bchi}\,\,\,
\frac{x_2\,(1-x_2)\,(2-x_1)}{x_1\,\bchi+x_2\,(1-x_2)\,\bbeta},
\eqa
%--
In order to compute $J_{11}$, we use \eqn{btJxJy} to increase the
power of $\bchi$ by one unit, while for $\bbeta$ and $\bbeta_{x_1}$ we use:
%--
\bq
\frac{(1-x_1)^2}{\bbeta} - \frac{1}{\bbeta_{x_1}}=
-\,\frac{1}{2\,\Bbt}\,\Bigg[ 
\Big( y_2 - \frac{1-(1-x_1)\,\bXbt}{1-x_3} \Big)\,\partial_{y_2}\,\ln\bbeta
- \Big( y_2 - \frac{1-\bXbt}{1-x_3} \Big)\,\partial_{y_2}\,\ln\bbeta_{x_1}
\Bigg]
\label{btJ11}
\eq
%--
The presence of $(1-x_1)^2$ is crucial in preventing spurious singularities at
$x_1 = 1$. This is the reason of our splitting
$1/x_1 = (1-x_1)^2/x_1 + (2 -x_1)$.
After integration by parts we get:
%--
\bq
J_{11}=
\frac{1}{4\,\Bbt^2}\,\dcub{4}(\{x\}\,,\,y_2)\,\,I_x^0\,\,I_{11}
\eq
%--
\bq
I_{11}=
  \frac{1-x_1}{x_1}\,\Bigl[ \ln\frac{\bbeta_0}{\bbeta_{00}}
- \ln\frac{\bbeta_2}{\bbeta_{20}}\Bigr]
+ \frac{x_3-\bXbt}{x_1\,(1-x_3)}\,
  \ln\frac{\bbeta_1\,\bbeta_{20}}{\bbeta_2\,\bbeta_{10}}
+ \frac{\bXbt}{1-x_3}\,\ln\frac{\bbeta_1}{\bbeta_2}
\eq
%--
where $I_x^0$ is given in \eqn{IxIy} and the new quadratic forms are given by
%--
\bq
\ba{lll}
\bbeta_0 = \bbeta(\, x_3 + (1-x_1)\,(1-x_3)\,y \,,\, 1-x_1 \,),
&\qquad&
\bbeta_{00} = \bchi(\, x_3 + (1-x_1)\,(1-x_3)\,y \,),
\nl
\bbeta_1 = \bbeta(\, x_3 \,,\, 1-x_1 \,),
&\qquad&
\bbeta_{10} = \bchi(\, x_3 \,),
\nl
\bbeta_2 = \bbeta(\, 1 - (1-x_3)\,x_1 \,,\, 1-x_1 \,),
&\qquad&
\bbeta_{20} = \bchi(\, 1 - (1-x_3)\,x_1 \,),
\ea
\eq
%--
It can be easily seen that the logarithms vanish when the denominator of the 
corresponding factor is zero. The result is then smooth enough to be 
integrated numerically.

For $J_{12}$, we use \eqn{BTli0} which in the present case reads as follows:
%--
\bq
\frac{1}{\bbeta}\,\frac{(1-x_1)^2}{\bxi} = \frac{1}{\Bbt}\,\Bigg[
\,\frac{1}{\bxi}\, + \,\frac{1}{2\,x_1\,\bchi}\,
  \Big( y_2 - \frac{1-(1-x_1)\,\bXbt}{1-x_3} \Big)\,\partial_{y_2}\,
  \ln\Big(1+\frac{x_1\,\bchi}{x2\,(1-x_2)\,\bbeta}\Big)\Bigg],
\eq
%--
\bq
\bxi = x_1\,\bchi(x_2\,x_3) + x_2\,(1-x_2)\,\bbeta(1-(1-x_3)\,y_2,1-x_1).
\eq
%--
Now we integrate by parts, obtaining:
%--
\bq
J_{12} =  \frac{1}{2\,\Bbt}\,\dcub{4}(\{x\}\,,\,y_2)\,\,I_{12}
- \frac{1}{\Bbt}\,J_{s1},
\qquad
J_{s1} = \dcub{3}(\{x\})\,\int_{x_1}^1\!\!dy_2\,\,\bxi^{-1},
\eq
%--
\bqa
I_{12} &=& 
\frac{1-x_1}{x_1\,\bchi}\,\Bigl[
\ln\Big( 1 + \frac{x_1\,\bchi}{x_2\,(1-x_2)\,\bbeta_0} \Big)
- \ln\Big( 1 + \frac{x_1\,\bchi}{x_2\,(1-x_2)\,\bbeta_2} \Big)\Bigr]
\nl
{}&+& 
 \Big( \bXbt + \frac{x_3-\bXbt}{x_1} \Big)\,\frac{1}{(1-x_3)\,\bchi}\,  \Big[
  \ln\Big( 1 + \frac{x_1\,\bchi}{x_2\,(1-x_2)\,\bbeta_1} \Big)
  -\ln\Big( 1 + \frac{x_1\,\bchi}{x_2\,(1-x_2)\,\bbeta_2} \Big)  \Big]
\eqa
%--
%\bq
%J_{s1} = \dcub{3}(\{x\})\,\int_{x_1}^1\!\!dy_2\,\,\bxi^{-1},
%\eq
%--
To compute $J_{s1}$ we perform the change of variable
$y_2 \to (1-y_2)/(1-x_3)$, i.e. 
%--
\bq
\int_0^1\!\!dx_3\,\int_{x_1}^1\!\!dy_2 \rightarrow
\Big[   \int_0^1\!\!dy_2\,\int_0^{y_2}\!\!dx_3
- \int_{1-x_1}^1\!\!dy_2\,\int_0^{1-(1-y_2)/x_1}\!\!dx_3 \Big]\,\frac{1}{1-x_3}
\nn
\label{chJs}
\eq
%--
and obtain:
%--
\bq
J_{s1}=
\dcub{3}(\{x\})\,\int_{x_3}^{1-(1-x_3)\,x_1}\,
\frac{dy_2}{(1-x_3)\,\bxi_s},
\qquad\quad \bxi_s= x_1\,\bchi \,+\, x_2\,(1-x_2)\,\bbeta_s,
\qquad \bbeta_s= \bbeta(\,y_2\,,\,1-x_1\,).
\eq
%--
For $\bxi_s$ we have the following BST relation:
%--
\bq
\bxi_s^{-1} = \frac{1}{\Bbt\,\rho}\,\Bigg\{ 1 - \frac{1}{2}\,    \Bigg[
    \Big( x_3 - \frac{\bXbt}{x_2} \Big)\,\partial_{x_3}
  + \Big( y_2 - (1-x_1)\,\bXbt \Big)\,\partial_{y_2} \Bigg]\,\ln\bxi_s\Bigg\},
\qquad
\rho = x_1 + x_2\,(1-x_2)\,(1-x_1)^2.
\label{btJs}
\eq
%--
Integration by parts gives:
%--
\bq
J_{s1}=
\frac{1}{2\,\Bbt}\,\dcub{4}(\{x\}\,,\,y_2)\,\,\frac{I_{s1}}{\rho}
\eq
%--
\bqa
I_{s1} &=&   
 \frac{1-x_1}{1-x_3}\,\Bigl[
\Big( 1 - \frac{\bXbt}{x_2} \Big)\,\ln\frac{\bxi_0}{\bxi_{01}}
-\,\bXbt\,\ln\frac{\bxi_1}{\bxi_2}\Bigr]
+ \frac{1}{1-x_3}\,\Bigl[ \ln\frac{\bxi_{11}}{\bxi_{21}}
+ \frac{\bXbt}{x_2}\,\ln\frac{\bxi_1}{\bxi_{11}}
\nl
{}&-& \Big( 1 - x_1 + x_1\,\frac{\bXbt}{x_2} \Big)\,
  \ln\frac{\bxi_2}{\bxi_{21}}\Bigr]
+ 
(1-x_1)\,\Bigl[ \ln\bxi_0 - \frac{\bXbt}{x_2}\,\ln\frac{\bxi_3}{\bxi_{31}}
- \ln\bxi_{31} + 2\Bigr],
\eqa
%--
%\bqa
%I_{s1} &=&   (1-x_1)\,\ln\bxi_0
%+ \frac{1-x_1}{1-x_3}\,\Big( 1 - \frac{\bXbt}{x_2} \Big)\,
%  \ln\frac{\bxi_0}{\bxi_{01}}
%- \frac{1-x_1}{1-x_3}\,\bXbt\,\ln\frac{\bxi_1}{\bxi_2}
%+ \frac{1}{1-x_3}\,\ln\frac{\bxi_{11}}{\bxi_{21}}
%+ \frac{1}{1-x_3}\,\frac{\bXbt}{x_2}\,\ln\frac{\bxi_1}{\bxi_{11}}
%\nl
%{}&-& \frac{1}{1-x_3}\,\Big( 1 - x_1 + x_1\,\frac{\bXbt}{x_2} \Big)\,
%  \ln\frac{\bxi_2}{\bxi_{21}}
%- (1-x_1)\,\frac{\bXbt}{x_2}\,\ln\frac{\bxi_3}{\bxi_{31}}
%- (1-x_1)\,\ln\bxi_{31} + 2\,(1-x_1)
%\eqa
%--
where the new quadratic forms, $\bxi_0$ etc, are given by
%--
\bq
\ba{lll}
\bxi_0 = x_1\,\bchi + x_2\,(1-x_2)\,\bbeta_0,
&\qquad&
\bxi_{01} = x_1\,\bchi(x_2) + x_2\,(1-x_2)\,\bbeta_0,
\nl
\bxi_1 = x_1\,\bchi + x_2\,(1-x_2)\,\bbeta_1,
&\qquad&
\bxi_{11} = x_1\,\bchi(x_2) + x_2\,(1-x_2)\,\bbeta_1,
\nl
\bxi_2 = x_1\,\bchi + x_2\,(1-x_2)\,\bbeta_2,
&\qquad&
\bxi_{21} = x_1\,\bchi(x_2) + x_2\,(1-x_2)\,\bbeta_2,
\nl
\bxi_3 = x_1\,\bchi(0) + x_2\,(1-x_2)\,\bbeta_3,
&\qquad&
\bxi_{31} = x_1\,\bchi(x_2) + x_2\,(1-x_2)\,\bbeta_3,
\qquad\quad
\bbeta_3 = (1-x_1)^2\,\bchi(\, x_3 \,).
\ea
\eq
%--
For $J_{21}$ we perform the same change of variable made for $J_{s1}$ 
(\eqn{chJs}), obtaining
%--
\bq
J_{21} = \int_0^1\!\!dx_1\,dx_2\, \Big[
  \int_0^1\!\!dy_2\,\int_0^{y_2}\!\!dx_3
- \int_{1-x_1}^1\!\!dy_2\,\int_0^{1-(1-y_2)/x_1}\!\!dx_3 \Big]\,
\frac{x_2\,(1-x_2)\,(2-x_1)}{\bchi\,\bxi_s}.
\eq
%--
Secondly, we use \eqn{BTli0}
%--
\bq
\frac{x_2\,(1-x_2)}{\bchi\,\bxi_s}=
\frac{1}{\Bbt}\,\Bigg[\,\frac{x_2\,(1-x_2)}{\bxi_s}\,
+ \,\frac{1}{2\,\bbeta_s}\,
  \Big( x_3 - \frac{\bXbt}{x_2} \Big)\,\partial_{x_3}\,
  \ln\Big(1+\frac{x_2\,(1-x_2)\,\bbeta_s}{x_1\,\bchi}\Big)\Bigg],
\eq
%--
Similarly for $J_{22}$ we use:
%--
\bq
\frac{x_2\,(1-x_2)}{\bchi\,\bxi}=
\frac{1}{\Bbt}\,\Bigg[\,\frac{x_2\,(1-x_2)}{\bxi}\,
+ \,\frac{1}{2\,\bbeta_s}\,
  \Big( x_2 - \frac{\bXbt}{x_3} \Big)\,\partial_{x_2}\,
  \ln\Big(1+\frac{x_2\,(1-x_2)\,\bbeta}{x_1\,\bchi}\Big)\Bigg],
\eq
%--
After integration by parts, we collect the results obtaining:
%--
\bq
J_2\equiv J_{21} + J_{22}= 
\frac{1}{2\,\Bbt}\,\dcub{4}(\{x\}\,,\,y_2)\,\,(2-x_1)\,I_2
\,\,+\,\, \frac{1}{2\,\Bbt}\,J_{s2},
\eq
%--
\bqa
I_2 &=& - \frac{1-x_1}{\bbeta_0}\,
  \ln\Big( 1 + \frac{x_2\,(1-x_2)\,\bbeta_0}{x_1\,\bchi} \Big)
+ \Big( x_3 - \frac{\bXbt}{x_2} \Big)\,\frac{1}{\bbeta_1}\,
  \ln\Big( 1 + \frac{x_2\,(1-x_2)\,\bbeta_1}{x_1\,\bchi} \Big)
\nl
{}&-& \Big( x_3 - \frac{\bXbt}{x_2} \Big)\,\frac{x_1}{\bbeta_2}\,
  \ln\Big( 1 + \frac{x_2\,(1-x_2)\,\bbeta_2}{x_1\,\bchi} \Big)
+ \frac{\bXbt}{x_2}\,\frac{1-x_1}{\bbeta_3}\,
  \ln\Big( 1 + \frac{x_2\,(1-x_2)\,\bbeta_3}{x_1\,\bchi(0)} \Big),
\eqa
%--
\bq
J_{s2} = \dcub{3}(\{x\})\,\int_{x_1}^1\!\!dy_2\,\,
(2-x_1)\,\frac{x_2\,\Xbt + (1-x_2)\,\bXbt + x_2\,(1-2\,x_2)\,(1-x_3)}
     {x_1\,\bchi+x_2\,(1-x_2)\,\bbeta}.
\eq
%--
To compute $J_{s2}$ we use the same techniques already used for $J_{s1}$ and 
obtain
%--
\bqa
J_{s2} &=& \frac{1}{2\,\Bbt}\,\dcub{4}(\{x\}\,,\,y_2)\,\,(2-x_1)\,
\Bigg[\,  \frac{x_2\,\Xbt + (1-x_2)\,\bXbt}{\rho}\,I_{s1}
+ \frac{1-2\,x_2}{\rho}\,I_{s2}\,\Bigg],
\nl
I_{s2} &=&   2\,x_2\,(1-x_1)\,(1-x_3)\,\ln\bxi_0
+ \Big[ 1 - (1-x_1)\,x_2 \Big]\,\bXbt\,\ln\bxi_1
\nl
{}&-& \Big[ x_2\,(1-x_1)\,\Xbt + x_1\,\bXbt \Big]\,\ln\bxi_2
- (1-x_1)\,\bXbt\,\ln\bxi_3 + 2\,x_2\,(1-x_1)\,(1-x_3)
\eqa
%--
Summarysing, 
%we introduce $\Delta_{\ssU\ssV} = \gamma + \ln\pi - 
%\ln \tHss/\mid P^2\mid$ and 
we write the result for $V^{\bbb}$ as
%--
\bq
\ovalbox{\boldmath $V^{\bbb}$} =
-\,\left( \frac{\mu^2}{\pi} \right)^{\ep}\,
\egam{1+\ep}\,
\Big(\, \frac{1}{\ep}\,V^{\bbb}_{-1} + V^{\bbb}_0 \,\Big)
\eq
%--
\bqa
V^{\bbb}_{-1} 
&=& 
- \frac{1}{8\,\Bbt^2}\,\dcub{4}(\{x\}\,,\,y_2)\,I_x^0\,I_y^0,
\nl
V^{\bbb}_0 
&=& 
\dcub{4}(\{x\}\,,\,y_2)\,\Bigg\{
  \frac{1}{2\,\Bbt}\,\Big[ I_{12} + (2-x_1)\,I_2 \Big]
+ \frac{1}{4\,\Bbt^2}\,\bigg[
- \frac{1}{2}\,(\,I_x^0\,I_y^1 + I_x^1\,I_y^0\,)
\nl
{}&+& I_x^0\,I_{11} + \frac{(2-x_1)\,[x_2\,\Xbt + (1-x_2)\,\bXbt]-2}{\rho}\,I_{s1}
+ (2-x_1)\,\frac{1-2\,x_2}{\rho}\,I_{s2} \bigg]
\Bigg\},
\eqa
%--
\subsection{Collinear limits of $V^{\bbb}$} \label{v222coll}
%--
In the previous section we have derived a suitable integral representation
for the $V^{\bbb}$ diagram in a generic infrared configuration and for 
arbitrary value of the masses.
The resulting representation is stable and can be computed numerically
even if it contains several polynomial denominators depending on the 
integration variables. This is made possible because each denominator is 
multiplied by a logarithm or a polylogarithm which vanishes exactly at the 
zeros of the denominator.
However it can happen that, for some values of the masses, this 
compensation is delayed and there we encounter numerical instabilities, 
revealing a region where the integrand has strong peaks. This is always the 
case when one of the two masses in the diagram  (or both) is vanishing, 
i.e. in the collinear region.

For example, if we consider the expression of $I_x^0$ of \eqn{IxIy} the 
second term is proportional to:
%--
\bq
\frac{1}{x_3}\,\ln\frac{\bchi(x_3)}{\bchi(0)} =
\frac{1}{x_3}\,\ln\left( 1 + \frac{-P^2\,x_3 + P^2-M^2+m^2}{M^2}\,x_3 \right)
\eq
%--
In this term the stability around $x_3 = 0$ is at stake when $M^2$ is small 
compared to $|P^2|$.

In all these cases one has to compute explicitly the leading behaviour of
the diagram, leaving a stable remainder. We have considered the following 
collinear limits of $V^{\bbb}$:
$ a) \; m^2 = M^2 \ll |P^2|$ and $ b) \; m^2 \ll  M^2 \ll |P^2|$.
In both cases, we have found that it is better to change Feynman 
parametrization and proceed in the following way (see 
also~\cite{Ferroglia:2003yj} section 10.2 ).
Starting from \eqn{v222prop}, we first combine propagators 
$[1]_{\bbb}-[2]_{\bbb}$ with parameter $z_1$, propagators 
$[3]_{\bbb}-[4]_{\bbb}$ with $z_2$ and $[5]_{\bbb}-[6]_{\bbb}$ with $z_3$:
%--
\bq
\pi^4\,V^{\bbb}_0 = \mu^{2\ep}\,\dcub{3}(\{z\})\,
\int\,d^nq_1 d^nq_2\,\frac{1}{[12]_{\bbb}^2 [34]_{\bbb}^2 [56]_{\bbb}^2},
\eq
%--
\bq
[12]_{\bbb}= [1]_{\bbb}\,(1-z_1) + [2]_{\bbb}\,z_1, \qquad
[34]_{\bbb}= [4]_{\bbb}\,(1-z_2) + [3]_{\bbb}\,z_2, \qquad
[56]_{\bbb}= [5]_{\bbb}\,(1-z_3) + [6]_{\bbb}\,z_3.
\eq
%--
Next we combine $[12]_{\bbb}-[34]_{\bbb}$ with parameter $x$ and integrate 
in $q_1$, obtaining:
%--
\bq
\pi^2\,V^{\bbb}_0= 
\mu^{2\ep}\,i\,\pi^{-\ep/2}\,\egam{2+\frac{\ep}{2}}\,\intsx{x}
\dcub{3}(\{z\})\,[x\,(1-x)]^{-1-\ep/2}\,\intmomi{n}{q_2}\,
\frac{1}{[1234]_{\bbb}^{2+\ep/2}[56]_{\bbb}^2},
\eq
%--
\bq
[1234]_{\bbb}= 
  \big[ q_2 + (1-z_1)\,p_2 - z_2\,P \big]^2 
+ \frac{\chi(z_1;p_2^2;m_1^2,m_2^2)}{x}
+ \frac{\chi(z_2;P^2;m_4^2,m_3^2)}{1-x},
\eq
%--
%\bq
%\chi_{\ssB}(p^2,m_i^2,m_j^2,z)=\, p^2 z\,(1-z) + m_i^2\,(1-z) + m_j^2\,z
%\eq
%--
with $\chi$ defined in \eqn{chidef}. Finally we introduce the parameter $y$ 
for the remaining two denominators and integrate in $q_2$:
%--
\bq
V^{\bbb}_0= 
- \,\left(\frac{\mu^2}{\pi}\right)^{\ep}\!\egam{2+\ep}\,
\intsx{x}\,\intsx{y}\dcub{3}(\{z\})\,
[x\,y\,(1-x)]^{1+\ep/2}\,(1-y)\,\,U_{\bbb}^{-2-\ep}
\eq
%--
\bqa
U_{\bbb} &=&  x\,y\,(1-x)\,(1-y)\,  \big[
  - P^2\,(z_2\!-\!z_3)\,(1\!-\!z_1\!-\!z_2)
  + p_1^2\,(z_2\!-\!z_3)\,(1\!-\!z_1\!-\!z_3)
  + p_2^2\,(1\!-\!z_1\!-\!z_3)\,(1\!-\!z_1\!-\!z_2)  \big]
\nl
{}&+&\,  y\,(1-x)\,\,\chi(z_1;p_2^2;m_1^2,m_2^2)
\,+\, x\,y\,\,\chi(z_2;P^2;m_4^2,m_3^2)
\,+\, x\,(1-x)\,\,(1-y)\,\chi(z_3;p_1^2;m_5^2,m_6^2)
\eqa
%--
Inserting the values corresponding to the infrared configuration of 
\fig{fig222} we get (after the change of variable $z_3 \to 1-z_3$):
%--
\bq
V^{\bbb}_0= 
- \,\left(\frac{\mu^2}{P^2\,\pi}\right)^{\!\ep}\,
\frac{\egam{2+\ep}}{P^4}\dscub{5}(x,y,\{z\})\,
[x\,y\,(1-x)]^{1+\ep/2}\,(1-y)\,\,
( a_0 + a_m\,\ep_m + a_{\ssM}\,\ep_{\ssM} )^{-2-\ep}
\eq
%--
\bqa
a_0 &=& x\,y\,\Big\{
(1\!-\!x)\,(1\!-\!y)\,[z_1z_2z_3 + (1\!-\!z_1)(1\!-\!z_2)(1\!-\!z_3)]
+ (x+y-xy)\,z_2\,(1\!-\!z_2)\Big\}
\\
a_m &=& x\,\Big\{
y\,(1\!-\!x)\,(1\!-\!y)\,[z_1z_2 + z_2(1\!-\!z_3) + (1\!-\!z_1)(1\!-\!z_3)]
+ (1\!-\!x)(1\!-\!y)^2(1\!-\!z_3)^2 + y\,(x+y-xy)\,z_2 \Big\}
\nl
a_{\ssM} &=& y\, \Big\{
x\,(1\!-\!x)\,(1\!-\!y)\,[z_1z_3 + z_1(1\!-\!z_2) + (1\!-\!z_2)(1\!-\!z_3)]
+ (1\!-\!x)(1\!-\!x+xy)\,z_1^2 + x\,(x+y-xy)\,(1\!-\!z_2) \Big\}
\nn
\eqa
%--
We have also introduced the (small) ratios $\ep_m = m^2/P^2$ and 
$\ep_{\ssM}=  M^2/P^2$.
The polynomials $a_0$, $a_1$ and $a_2$ are positive definite in the 
integration region, but vanish at the hedge.
These zeros are responsible for the infrared pole and the collinear 
divergencies.

At first we perform a Mellin-Barnes splitting, once for case a) and twice
for case b):
%--
\bqa
\mbox{a) } \qquad
V^{\bbb}_0 &=&
- \,\left(\frac{\mu^2}{P^2\,\pi}\right)^{\ep}\,
\frac{\egam{2+\ep}}{P^4}\,\frac{1}{2\pi i}
\int_{-i\infty}^{+i\infty}\!\!\!\!\!\!\!\!\!\!ds\,\,\,\Beta{s}{2+\ep-s}\,
\nl
{}&\times& \dcub{5}(x,y,\{z\})\,
[x\,y\,(1-x)]^{1+\ep/2}\,(1-y)\,\,\,
a_0^{-s}\,(a_m + a_{\ssM})^{s-2-\ep}\,\ep_m^{s-2-\ep}
\nl
\mbox{b) } \qquad
V^{\bbb}_0 &=&
- \,\left(\frac{\mu^2}{P^2\,\pi}\right)^{\ep}\,
\frac{\egam{2+\ep}}{P^4}\,
\left(\frac{1}{2\pi i}\right)^2
\int_{-i\infty}^{+i\infty}\,ds\,dt\;
\Beta{s}{2+\ep-s}\,\Beta{t}{s-t}\,
\nl
{}&\times& \dcub{5}(x,y,\{z\})\,
[x\,y\,(1-x)]^{1+\ep/2}\,(1-y)\,\,\,
a_0^{-t}\,\lpar a_m\,\ep_m\rpar^{s-2-\ep}\,
\lpar a_{\ssM}\,\ep_{\ssM}\rpar^{t-s}
%a_0^{-t}\,a_m^{s-2-\ep}\,\ep_m^{s-2-\ep}\,a_{\ssM}^{t-s}\,\ep_{\ssM}^{t-s}
\eqa
%--
where $0 < \Reb\,s < 2+\ep$ and $0 < \Reb\,t < \Reb\,s$.
For the $s$ and $t$ integrations we close the integration contour 
in the positive real half-plane and compute the residues at the poles of 
$s$ and $t$ (which are $t \sim s \sim n + \alpha\ep$, $n\ge2$).
Since we are interested in the leading behaviour in $\ep_m \to 0$ and 
$\ep_{\ssM} \to 0$, we restrict our attention to the poles at $n = 2$ (higher 
values of $n$ would lead to terms proportional to $\ep_m^{n-2}$ or 
$\ep_{\ssM}^{n-2}$).

A problem connected to case b) is the appearance of poles at $t = 2+\ep$ 
together with poles at $t = s$ and $s = 2+\ep$. 
The residue of these poles in $t$ generates terms containing all powers
of $\ep_m/\ep_{\ssM}$; 
%--
%\bq
%\left(\frac{\ep_m}{\ep_{\ssM}}\right)^{s-2-\ep}\!\!\!\ln^m\ep_{\ssM}
%\eq
%--
without an assumption on the value of the ratio $\ep_m/\ep_{\ssM}$, we would
be forced to consider all poles at $s = n+\alpha\ep$ ($n\ge2$) and to resum 
the series.
For this reason we limit our analysis to $m^2 \ll M^2$ and neglect in this 
way all contributions proportional to $m^2/M^2$.

The poles in $s$ and $t$ come from the integration over Feynman 
parameters and are, obviously, related to the zeros of $a_0$, $a_1$ and $a_2$. 
These three polynomials can vanish only at the end point of the integration 
region; in order to bring all the zeros to the origin, we first split the 
integration domain of each variable (for $x$ and $y$ this is actually not 
necessary) according to the following rule:
%--
\bq
\intsx{z}f(z) = 
\bigg[ \int_0^{1/2}\!\!\!dz + \int_{1/2}^1\!\!\!dz \bigg]\,f(z) =
\frac{1}{2}\,\intsx{z}\,
\bigg[ f\Big(\frac{z}{2}\Big) + f\Big(\frac{2-z}{2}\Big) \bigg].
\eq
%--
In this way we disentangle  $z = 0$ from $z = 1$; then we remap each sector 
into $[0,1]$, moving all end-point singularities in $z = 0$.
Applying this decomposition to the $z_i$ integrals we obtain eight 
new integrals.
Consider one of the integrals generated in case a) where the poles are 
at $s = 2+\alpha\,\ep$,
%--
\bq
I(s)= \dcub{5}(x,y,\{z\})\,
(x\,y)^{1-s+\ep/2}\,(1-x)^{1+\ep/2}\,(1-y)\,\,\,
A_{a1}^{-s}\,\lpar B_{a1}\,\ep_m\rpar^{s-2-\ep}
\eq
%--
\bqa
A_{a1} &=&
(1\!-\!x)\,(1\!-\!y)\,[z_1 z_2z_3 + (2\!-\!z_1)(2\!-\!z_2)(2\!-\!z_3)]
+ 2\,(x+y-xy)\,z_2\,(2\!-\!z_2)
\\
B_{a1} &=& (1 - x)\,\Bigl[
2x y\,(1\!-\!y)(4\!-\!2 z_3 + z_1 z_3)
+ y\,(1\!-\!x+xy)\,z_1^2
+ x\,(1\!-\!y)^2\,(2\!-\!z_3)^2\Bigr]
+ 4xy\,(x+y\!-\!xy)
\nn
\eqa
%--
To extract the proper behaviour around $x = y = 0$, we apply a sector 
decomposition (see section~\ref{SD}):
%--
\bqa
I(s) &=& \dcub{5}(x,y,\{z\})\,(x\,y)^{1-s}\,(1 - x\,y)\,
\Big[ y^{\ep/2}\,(1-x)^{1+\ep/2}\,
A_{a1,1}^{-s}\,B_{a1,1}^{s-2-\ep}
\nl
{}&+& x^{\ep/2}\,(1-xy)^{\ep/2}\,(1-y)\,
A_{a1,2}^{-s}\,B_{a1,2}^{s-2-\ep}\Big]\,\ep_m^{s-2-\ep}
\eqa
%--
%I(s) &=& \dcub{5}(x,y,\{z\})\,
%\Big[ x^{1-s}\,y^{1-s+\ep/2}\,(1-x)^{1+\ep/2}\,(1-xy)\,\,\,
%A_{a1,1}^{-s}\,B_{a1,1}^{s-2-\ep}
%\nl
%&+& x^{1-s+\ep/2}\,y^{1-s}\,(1-xy)^{1+\ep/2}\,(1-y)\,\,\,
%A_{a1,2}^{-s}\,B_{a1,2}^{s-2-\ep}\Big]\,\ep_m^{s-2-\ep}
%\eqa
%--
\bqa
A_{a1,1} &=&
(1\!-\!x)\,(1\!-\!xy)\,[z_1z_2z_3 + (2\!-\!z_1)(2\!-\!z_2)(2\!-\!z_3)]
+ 2\,x(1+y-xy)\,z_2\,(2\!-\!z_2)
\nl
B_{a1,1} &=&
2xy\,(1\!-\!x)(1\!-\!xy)(4\!-\!2z_3+z_1z_3)
+ y\,(1\!-\!x)(1\!-\!x+x^2y)\,z_1^2
\nl
{}&+& \,(1\!-\!x)(1\!-\!xy)^2(2\!-\!z_3)^2 + 4 x^2 y\,(1+y\!-\!xy)
\nl
A_{a1,2} &=&
(1\!-\!xy)\,(1\!-\!y)\,[z_1z_2z_3 + (2\!-\!z_1)(2\!-\!z_2)(2\!-\!z_3)]
+ 2\,y(1+x-xy)\,z_2\,(2\!-\!z_2)
\nl
B_{a1,2} &=&
2xy\,(1\!-\!xy)(1\!-\!y)(4\!-\!2z_3+z_1z_3)
+ \,(1\!-\!xy)(1\!-\!xy+xy^2)\,z_1^2
\nl
&{}& + x\,(1\!-\!xy)(1\!-\!y)^2(2\!-\!z_3)^2 + 4 x y^2\,(1+x\!-\!xy)
\eqa
%--
The poles are at $s = 2 + \ep/2$ and $s = 2$ 
(both $B_{a1,1}$ and $B_{a1,2}$ are not vanishing at $x = y = 0$).
and we use \eqn{poleext} to extract the poles.
All integrals that appear in the evaluation of the diagram have been analyzed
according to this strategy. The coefficients of the collinear logarithms
are computed with a numerical integration with a result that agrees with 
all analytical expansions which have been presented in the literature.
Our numerical findings can be summarized by means of the following formulas:
%--
\bq
V^{\bbb}= 
-\,\left(\frac{\mu^2}{P^2\,\pi}\right)^{\ep}\,\egam{1+\ep}\,
\Big[ V^{\bbb}_{-1}\,\ep^{-1} + V^{\bbb}_0 \Big],
\eq
%--
For case a) we obtain
%--
\bq
V^{\bbb}_{i}= 
-\frac{1}{P^4}\,\sum_{n=0}^4\,a^{(i)}_{n}\ln^n\frac{m^2}{P^2}
+ {\cal O}\left( \frac{m^2}{P^2} \right) 
+ {\cal O}\left( \frac{m^2-M^2}{m^2+M^2} \right),
\qquad i= -1,0
\eq
%--
\bq
\ba{lllll}
a^{(-1)}_{0} = - 2.40409(3),
\quad&
a^{(-1)}_{1} = - 3.289868(2),
\quad&
a^{(-1)}_{2} = 0,
\quad&
a^{(-1)}_{3} = - \frac{2}{3},
\quad&
a^{(-1)}_{4} = 0,
\nl
a^{(0)}_{0} = -10.007(2),
\quad&
a^{(0)}_{1} = 1.2020(2),
\quad&
a^{(0)}_{2} = 4.934802(8),
\quad&
a^{(0)}_{3} = 0,
\quad&
a^{(0)}_{4} = \frac{1}{6}.
\ea
\eq
%--
\\
For case b)
%--
\bq
V^{\bbb}_{i}= 
-\frac{1}{P^4}\,\sum_{n,k=0}^4b^{(i)}_{n\,k}
\ln^n\frac{m^2}{P^2}\ln^k\frac{M^2}{P^2}
+ {\cal O}\left( \frac{m^2}{P^2} \right) 
+ {\cal O}\left( \frac{m^2}{M^2} \right),
\qquad i= -1,0
\eq
%--
\bq
\ba{lllll}
b^{(-1)}_{0\,0} = - 2.40411(3),
\quad&
b^{(-1)}_{1\,0} = b^{(-1)}_{1,0} = - 1.644834(1),
\quad&
\quad&
\quad&
\nl
b^{(-1)}_{2\,0} = b^{(-1)}_{1\,1} = b^{(-1)}_{0\,2} = 0,
\quad&
b^{(-1)}_{3\,0} = b^{(-1)}_{0\,3} = - \frac{1}{12},
\quad&
b^{(-1)}_{2\,1} = b^{(-1)}_{1\,2} = - \frac{1}{4},
\quad&
\quad&
\nl
b^{(-1)}_{4\,0} = b^{(-1)}_{3\,1} = b^{(-1)}_{2\,2} = 
b^{(-1)}_{1\,3} = b^{(-1)}_{0\,4} = 0,
\quad&
\quad&
\quad&
\quad&
\ea
\eq
%--
\bq
\ba{lllll}
b^{(0)}_{0\,0} = 0.8118(3),
\quad&
b^{(0)}_{1\,0} = 3.00514(3),
\quad&
b^{(0)}_{1,0} = -4.20720(5),
\quad&
\quad&
\nl
b^{(0)}_{2\,0} = 1.23370(1),
\quad&
b^{(0)}_{1\,1} = 0.82246(3),
\quad&
b^{(0)}_{0\,2} = 2.87863(1),
\quad&
b^{(0)}_{3\,0} = b^{(0)}_{2\,1} = b^{(0)}_{1\,2} b^{(0)}_{0\,3} = 0,
\quad&
\nl
b^{(0)}_{4\,0} = \frac{1}{24},
\quad&
b^{(0)}_{3\,1} = \frac{1}{12},
\quad&
b^{(0)}_{2\,2} = 0,
\quad&
b^{(0)}_{1\,3} = - \frac{1}{12},
\;\;&
b^{(0)}_{0\,4} = \frac{1}{8},
\ea
\eq
%--

\subsection{A detailed study of the $V^{\bbb}$ configurations} \label{IRLclass}
\begin{figure}[th]
\vspace{1.9cm}
\bqas  
{}&{}&
  \vcenter{\hbox{
  \begin{picture}(150,0)(0,0)
  \SetScale{0.6}
  \SetWidth{1.5}
  \Line(0,0)(50,0)
  \Line(50,0)(150,100)
  \Line(50,0)(150,-100)
  \Line(75,-25)(100,50)
  \Line(75,25)(100,-50)
  \Text(0,-25)[cb]{\Large $V^{\bbb}$}
  \Text(30,10)[cb]{$3$}
  \Text(30,-15)[cb]{$4$}
  \Text(48,25)[cb]{$1$}
  \Text(48,-30)[cb]{$5$}
  \Text(65,10)[cb]{$2$}
  \Text(65,-15)[cb]{$6$}
  \LongArrow(0,10)(30,10)   \Text(0,15)[cb]{$-P$}
  \LongArrow(150,115)(130,95)   \Text(110,70)[cb]{$p_2$}
  \LongArrow(150,-115)(130,-95)     \Text(110,-70)[cb]{$p_1$}  
  \end{picture}}}
%--
\qquad\qquad
%--
  \vcenter{\hbox{
  \begin{picture}(150,0)(0,0)
  \SetScale{0.6}
  \SetWidth{1.5}
  \Line(0,0)(50,0)
  \Line(50,0)(150,100)
  \Line(50,0)(150,-100)
  \Line(75,-25)(100,50)
  \Line(75,25)(100,-50)
  \Text(28,5)[cb]{$3$}
  \Text(38,15)[cb]{$4$}
  \Text(52,28)[cb]{$5$}
  \Text(38,-20)[cb]{$2$}
  \Text(52,-33)[cb]{$1$}
  \end{picture}}}
\eqas
\vspace{1.8cm}
\caption[]{Enumeration of internal masses and of internal vertices for 
$V^{\bbb}$.} 
\label{enumbbb}
\end{figure}
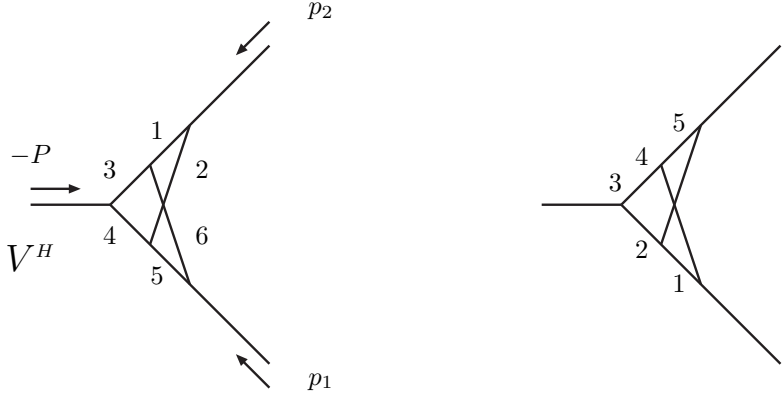

%--
We use $V^{\bbb}$ as a prototype for discussing the classification of
configurations that, for each family of diagrams, are infrared 
divergent (at least the QED-like). The tools are the corresponding Landau 
equations and a necessary condition for some configuration to be infrared 
divergent is that the Landau equations are satisfied. At first we enumerate 
internal masses and internal vertices of $V^{\bbb}$, as done in \fig{enumbbb}. 
%--
%--
Then we look for a non-trivial solution of the Landau equations where $P^2$ 
or $p^2_1(p^2_2)$ are unconstrained and one or two internal masses are zero, 
according to the rules of the standard model: there is no tri-linear vertex 
with two photon lines. In this operation particular care must be devoted in 
recognizing topologically equivalent diagrams.
%--
\begin{figure}[th]
\vspace{1.9cm}
\bqas  
{}&{}&
  \vcenter{\hbox{
  \begin{picture}(150,0)(0,0)
  \SetScale{0.6}
  \SetWidth{1.5}
  \Line(0,0)(50,0)
  \ArrowLine(50,0)(75,25)
  \ArrowLine(75,25)(100,-50)
  \ArrowLine(100,0)(50,0)
  \ArrowLine(100,50)(100,0)
  \ArrowLine(150,100)(100,50)
  \ArrowLine(100,-50)(150,-100)
  \Photon(75,25)(100,50){2}{5}
  \Photon(100,0)(100,-50){2}{5}
  \LongArrow(0,10)(30,10)   \Text(0,15)[cb]{$p_2$}
  \LongArrow(150,115)(130,95)   \Text(110,70)[cb]{$-P$}
  \LongArrow(150,-115)(130,-95)     \Text(110,-70)[cb]{$p_1$}  
  \end{picture}}}
%--
\qquad \equiv \qquad
%--
  \vcenter{\hbox{
  \begin{picture}(150,0)(0,0)
  \SetScale{0.6}
  \SetWidth{1.5}
  \Line(0,0)(50,0)
  \ArrowLine(150,100)(100,50)
  \ArrowLine(100,50)(75,25)
  \ArrowLine(75,25)(50,0)
  \ArrowLine(50,0)(75,-25)
  \ArrowLine(75,-25)(100,-50)
  \ArrowLine(100,-50)(150,-100)
  \Photon(75,25)(100,-50){2}{7}
  \Photon(100,50)(75,-25){2}{7}
  \LongArrow(0,10)(30,10)   \Text(0,15)[cb]{$p_2$}
  \LongArrow(150,115)(130,95)   \Text(110,70)[cb]{$-P$}
  \LongArrow(150,-115)(130,-95)     \Text(110,-70)[cb]{$p_1$}  
  \end{picture}}}
\eqas
\vspace{1.8cm}
\caption[]{Equivalence of two configurations of the $V^{\bbb}$-type containing
two internal photonic lines.} 
\label{topequivone}
\end{figure}
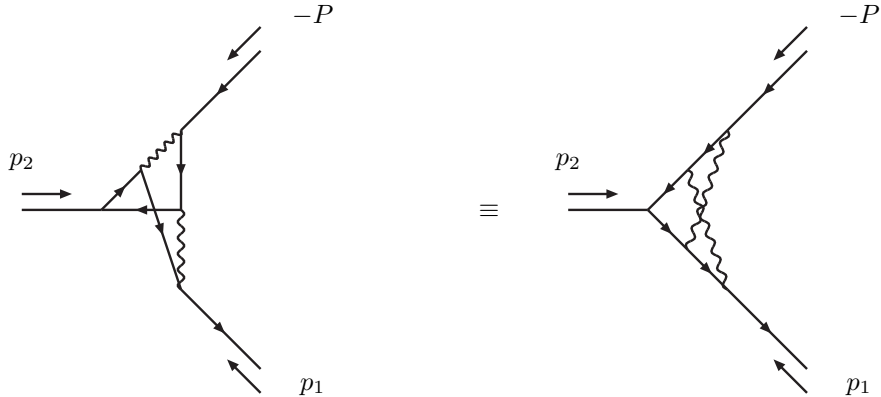 
%--
An example is illustrated in \fig{topequivone} where it would be enough to
compute our standard definition for $V^{\bbb}$ and to perform a permutation of
its arguments. The equivalence can be shown by simply enumerating the 
internal vertices and corresponds to $(12345) \leftrightarrow (34521)$.
A second example is given in \fig{topequivtwo} where the equivalence
is $(12345) \leftrightarrow (14325)$.
%--
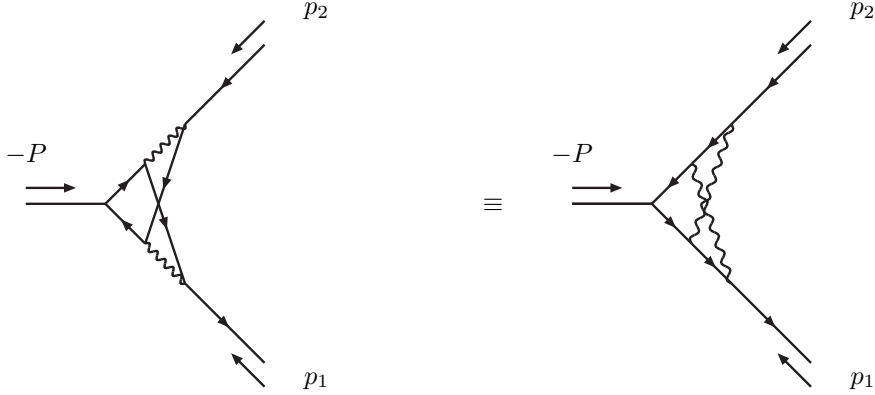
\begin{figure}[th]
\vspace{1.9cm}
\bqas  
{}&{}&
  \vcenter{\hbox{
  \begin{picture}(150,0)(0,0)
  \SetScale{0.6}
  \SetWidth{1.5}
  \Line(0,0)(50,0)
  \ArrowLine(150,100)(100,50)
  \ArrowLine(100,50)(75,-25)
  \ArrowLine(75,-25)(50,0)
  \ArrowLine(50,0)(75,25)
  \ArrowLine(75,25)(100,-50)
  \ArrowLine(100,-50)(150,-100)
  \Photon(75,-25)(100,-50){2}{5}
  \Photon(75,25)(100,50){2}{5}  
  \LongArrow(0,10)(30,10)   \Text(0,15)[cb]{$-P$}
  \LongArrow(150,115)(130,95)   \Text(110,70)[cb]{$p_2$}
  \LongArrow(150,-115)(130,-95)     \Text(110,-70)[cb]{$p_1$}  
  \end{picture}}}
%--
\qquad \equiv \qquad
%--
  \vcenter{\hbox{
  \begin{picture}(150,0)(0,0)
  \SetScale{0.6}
  \SetWidth{1.5}
  \Line(0,0)(50,0)
  \Line(0,0)(50,0)
  \ArrowLine(150,100)(100,50)
  \ArrowLine(100,50)(75,25)
  \ArrowLine(75,25)(50,0)
  \ArrowLine(50,0)(75,-25)
  \ArrowLine(75,-25)(100,-50)
  \ArrowLine(100,-50)(150,-100)
  \Photon(75,25)(100,-50){2}{7}
  \Photon(100,50)(75,-25){2}{7}
  \LongArrow(0,10)(30,10)   \Text(0,15)[cb]{$-P$}
  \LongArrow(150,115)(130,95)   \Text(110,70)[cb]{$p_2$}
  \LongArrow(150,-115)(130,-95)     \Text(110,-70)[cb]{$p_1$}  
  \end{picture}}}
\eqas
\vspace{1.8cm}
\caption[]{Equivalence of two configurations of the $V^{\bbb}$-type containing
two internal photonic lines.} 
\label{topequivtwo}
\end{figure} 
%--
For all diagrams but $V^{\bbb}$ we can solve the set of Landau equations,
looking for the leading or the sub-leading singularities, in the general
case and we have an explicit condition on internal masses and external momenta
which express the singular configurations. For $V^{\bbb}$ there is no 
known general expression and the set of eight equations that we can write 
must be examined configuration by configuration.
For instance, given propagators $[i]_{\ssH},\;i=1\cdots 6$, the 
{\em standard} configuration with
$p^2_1= -M^2, \; p^2_2 = -m^2$, $m_1 = m_3 = m$, and
$m_4 = m_5 = M, \; m_2 = m_6 = 0$,
has a solution with $P^2$ unconstrained, $\alpha_2, \alpha_6 \not = 0$ and 
$\alpha_1 = \alpha_3 = \alpha_4 = \alpha_5 = 0$. Clearly, this has to be put 
in correspondence with a non-leading singularity; we stress that to 
have a non-trivial solution is the necessary condition for a singularity, 
but the final presence of the singularity follows from infrared 
power-counting. 
Indeed, this configuration of $V^{\bbb}$ has a single pole at $\ep = 0$,
as well-known.

The configuration where $m_1 = m_5 = 0$ needs not to be considered, given the 
equivalence of \fig{topequivtwo}. This exhausts the configurations where we 
have two internal photons, no tri-linear couplings with two photonic legs and 
both photons attached to a current with one external, on-shell, line.
If the last condition is relaxed we can consider a configuration
$m_1 = m_4 = 0, \; m_3 = m_6 = m, \; m_2 = m_5 = M$
where we only require that both lines attached to a photon have the same mass.
From the system of Landau equations we observe a possible solution with
only $\alpha_1 \not= 0$ which requires $p^2_2 = -M^2$ (on-shell condition) and
some peculiar relation between $P^2$ and $p^2_1$, namely
$p^2_1= -2\,( - M\,m + 1/2\,M^2 - 1/2\,P^2)$. 
Finally, we consider the case of only one internal photon. The typical 
configuration has 
$p^2_1= -M^2$, $p^2_2 = -m^2$, and $m_1 = m_3 = m$, 
$m_4 = m_5 = M_1$, $m_2 = 0,\, m_6 = M_2$.
%\label{alsoNTS}
%\eq
%--
%Also in \eqn{alsoNTS} we have 
This is a non-trivial solution but infrared power 
counting shows that it does not correspond to a singularity. One starts from 
some parametrization of the diagram, performs sector decomposition and obtains
expressions of the form given in \eqn{IRpc} with $A \ge 0$.
%--
%\bq
%\intfx{x_i}\,x^{A_i+B_i\,\ep}_i\,F(x_i,\ep), \qquad
%\hbox{with} \quad A_i \ge 0.
%\eq
%--
\section{Behaviour of two - loop vertices in the collinear limit 
\label{sect:coll}}
%--
All results derived in this paper have a common property: diagrams are 
given in terms of integral representations where integrands are smooth 
functions thus allowing for a stable numerical integration.
These integral representations have been obtained in several ways:
we have used BST functional relations which force the appearance of $\Bbt$
factors in the denominator; their zeros correspond to anomalous (normal)
thresholds of the diagram. Far from threshold we register very good stability
in the numerical integration, but even near threshold we have been able to
find quite stable results.

For those cases where we have not been able to find the proper BST algorithm
we, nevertheless, succeeded in writing an integral representation of the 
following form:
%--
\bq
\dcub{k}(\{x\})\frac{1}{A}\,
\ln\left( 1 + \frac{A}{B} \right)
\qquad
\mbox{or}
\qquad
\dcub{k}(\{x\})\frac{1}{A}\,
\li{n}{\frac{A}{B}}
\label{general}
\eq
%--
where $A, B$ are multivariate polynomials in the Feynman parameters. These 
representations generalize the class of Nielsen polylogarithms where we only 
deal with monomials in one variable. In a word, two - loop diagrams are 
always reducible to combinations of integrals of the type given in 
\eqn{general} where the usual monomials that appear in the integral 
representation of Nielsen - Goncharov generalized polylogarithms are replaced 
by multivariate polynomials of arbitrary degree. We have made no attempt 
towards an analytical classification of these new transcendental functions; 
rather, we compute them numerically, possibly after the elimination of 
apparent singularities by means of sector decomposition techniques. 

Problems with numerical integration arise whenever $A$ and $B$ 
become small in the same region of Feynman parameters and this usually occurs 
when some mass is much smaller then the largest scale of the diagram, i.e. in 
the collinear regions.
In order to have a better understanding of the collinear limit, we consider 
the case of \eqn{resV231d}; here one of the terms in our representation is
%--
\bq
\int_0^1\!\!dx\,dy\,\frac{1}{\chi(y)}\,
\li{2}{\frac{x\,\chi(y)}{y\,M^2}}
\qquad\qquad
A \equiv x\,\chi(y) = x\,\Big[-P^2\,y^2 + (P^2\!-\!m^2\!+\!M^2)\,y + m^2 \Big], 
\qquad
B \equiv M^2\,y.
\eq
%--
Here numerical instabilities may arise from two specific regions:
$M^2$ is much smaller than $A$ which means that $B$ is always small,
$m^2$ is much smaller than $|P^2|$ so that one zero of $A$ is near $y = 0$, where 
also $B$ vanishes. Both regions are indeed of collinear nature.
%--
To conclude this section we list some useful expansions of one-dimensional 
integrals which often occur in our results.
The following formulae are used to extract collinear logarithms in order to 
obtain numerical stability in the collinear regions. 

The expansions have been implemented in the procedure of numerical integration
which decides if we need to compute the diagram in the collinear region or not 
and uses the collinear expansion or the initial integral representation.

All collinear divergencies occur at the border of the integration domain which,
for one-dimensional integrals means $x = 0$ or $x = 1$. 
The latter case can always be eliminated with a change of variable,
$x \to 1-x$. Therefore, all integrals in collinear regions which occur in 
our results can be cast in one of the following forms, where 
$Q(x)= a x^2 + b x + c - i\,\delta$, with $\delta \to 0_+$ 
and $R(x) = Q(x)/(d\,x)$:
%--
\bqa
&&I_0^{k,n} = \intsx{x}\,\frac{\ln^k x}{x}\,
\ln^n\frac{Q(x)}{d}\bmid_+, \qquad
I_1^{1} =
\int_0^1\!\frac{dx}{x}\,\ln\left(1+\frac{1}{R(x)}\right),
\qquad\
I_1^{n} = \int_0^1\!\frac{dx}{x}\,\li{n}{-\frac{1}{R(x)}},
\qquad
\\ 
&& I_2^{1} = \int_0^1\!\!\frac{dx}{Q(x)}\,
\ln\left(1\!+\!R(x)\right),
\qquad
I_2^{n} = \int_0^1\!\!\frac{dx}{Q(x)}\,
\li{n}{-R(x)}, 
\qquad
I_3 = \int_0^1\!\!\frac{dx}{Q(x)}\ln R(x)
\ln\!\left(1\!-\!R(x)\right).
\nn
\label{defbaspl}
\eqa
%--
In our example we find a $I_2^2$ and the two unstable cases correspond
respectively to $d \to 0$ and $c \to 0$.

When $d \to 0$, the behaviour of $I_0^{kn}$ is trivial and $I_1^i$ is 
regular. Furthermore, to extract the divergency for $I_2^1$ and $I_2^2$ it 
is enough to use well-known relations among (poly)logarithms of argument $z$
and $1/z$
%--
%\bq
%\ln\Big( 1+R(x) \,\Big) =
%\ln\Big( 1+\frac{1}{R(x)} \,\Big) + \ln R(x),
%\quad
%\li{2}{-R(x)} =
%- \,\li{2}{-\frac{1}{R(x)}}
%- \frac{1}{2}\,\ln^2 R(x) - \zeta(2)
%\eq
%--
and then apply the BST method (\eqn{BTli1} and \eqn{BTlin}).
After integration by parts, the logarithmic behaviour in $d$ arises 
naturally without any expansion.
The results, collected in appendix \ref{app:coll}, are therefore valid 
everywhere.
In order to extract the divergent behaviour for $c \to 0$, we proceed in 
the following way:
%--
\bq
\intsx{x}\,f(x,Q(x)) =
\intsx{x}\,[ f(x,Q(x)) - f(x,b x) ] + \intsx{x}\,f(x,b x + c) +
\intsx{x}\,[ f(x,b x) - f(x,b x + c) ],
\label{locdiv}
\eq
%--
%\bqa
%\intsx{x}\,f(x,Q(x)) &=&
%\intsx{x}\,[ f(x,Q(x)) - f(x,bx+c) ] + \intsx{x}\,f(x,bx+c) 
%\nl
%{}&=& \intsx{x}\,[ f(x,Q(x)) - f(x,bx) ] + \intsx{x}\,f(x,bx+c) + {\cal O}(c),
%\label{locdiv}
%\eqa
%--
where $f$ is one of the (poly)logarithms of \eqn{defbaspl}.
Only the second integral in \eqn{locdiv} is divergent when $c \to 0$, we can 
set $c = 0$ in the first (subtracted) term and the third one is $\ord{c}$.
The results for the expansion are again listed in appendix \ref{app:coll}.
%--
\section{Numerical Results \label{numres}}
%--
In this section we present numerical results for infrared configurations
corresponding to two-loop three-point scalar functions. 
All results are computed numerically using the analytical expressions
given in this paper. 
A particular attention has been devoted to collinear regions where the mass 
of the outgoing particles is small compared to the incoming momentum.
%In these regions the numerical integration is sometimes unstable and we were 
%forced to derive specific algorithms for extracting the collinear logarithms 
%from the total answer, leaving a sub-leading remainder which can be safely
%integrated.
%The technique used for this purpose has been described in \sect{sect:coll} 
%and the formulas implemented in the fortran code are listed in appendix 
%\ref{app:coll}.

In our general program, aimed to a numerical evaluation of multi-loop,
multi-leg Feynman diagrams we have developed a set of FORTRAN/77 codes
which go from standard $A_0,\,\dots\,,D_0$ functions to diagrams presented
in this paper. This new ensemble of programs which includes the treatment
of complex poles~\cite{Argyres:1995ym} will succeed to the corresponding 
Library of {\tt TOPAZ0}~\cite{Montagna:1993ai}.

The whole collection of codes is heavily based on the 
NAG-library~\cite{naglib}; while completing the analysis it became natural
to consider a migration of the whole set of programs to a stand-alone,
FORTRAN/95 version. The outcome of our decision is a brand-new version of
our numerical code, $\LB$~\cite{LB}, which is fully based on
quasi - Montecarlo methods (with a possible extension to a parallelized
version) and is presently under construction.
Our FORTRAN/95 version has been adapted from the automatic multi-dimensional 
integration subroutine DKBVRC, written by Alan~Genz; DKBVRC uses randomized 
Korobov rules~\cite{Genz} 

In the migration to FORTRAN/95 we have experienced huge gains in CPU-time, 
a better numerical precision and we foresee a future release of the code, 
although the numbers produced for this paper still rely on the (by now) old 
version. For present numerical results all the vertices are evaluated using 
the routine D01GDF or D01EAF~\cite{naglib}.
The first routine is based on the Korobov-Conroy~\cite{KC} number theoretic 
method with a Monte-Carlo error estimate, while the second one uses an 
adaptive subdivision strategy.

For those configurations where we could compare with analytical results
we have adapted our setup to match the known examples. Alternatively we have 
considered few physically relevant cases, selected among those not presented in 
the literature, and extracted from processes like 
$Z^* \to \barf f$, $H^* \to W^+ W^-$, $t \to W^+ b$, etc. 
In this paper we use the following input parameter set:
%--
\bqa
\mw &=& 80.380\,\GeV, \quad 
\mz = 91.1875\,\GeV, \quad 
\mh = 150\,\GeV \quad
\nl
\mt &=& 174.3\,\GeV, \quad 
m_b = 4\,\GeV, \quad
m_e= 0.510999\,\MeV
\label{NCsetup}
\eqa
%--
Comparisons have been performed with~\cite{Bonciani:2003te} and the results 
are shown in \tabn{tabV131aBMR}, \tabn{tabV131bBMR}, \tabn{tabV141aBMR}, 
\tabn{tabV141cBMR}, \tabn{tabV231bBMR}, \tabn{tabV231dBMR} and 
\tabn{tabV222BMR}; 
with~\cite{Davydychev:2003mv} and the results are shown in \tabn{tabDK}; 
with~\cite{Davydychev:2002hy} and the results are shown in \tabn{tabDS}.
In the comparison with the work of Bonciani - Mastrolia - 
Remiddi~\cite{Bonciani:2003te}, the evaluated finite part is defined 
through the following $\ep$ expansion~\footnote{Note that these authors 
define $n= 4 - 2\,\ep$. Therefore the $\ep$ of the following formula 
corresponds to $2\,\ep$ of~\cite{Bonciani:2003te}}:
%--
\bq
V_{\acan{a}} =  V_{\acan{a}}^{-2}\,\frac{1}{\ep^2} \,
+ V_{\acan{a}}^{-1}\,\frac{1}{\ep} + V_{\acan{a}}^0,
\qquad
{\rm etc.}
\eq
%--
For the comparison with the work of Davydychev - 
Kalmykov~\cite{Davydychev:2003mv} we use $V^{\bca}_a$ of \fig{fig231} with 
$m_1 = m_2 = m_3 \,(= 180\,\GeV)$, $m_i = 0, i \ge 4$ and $p^2_i = 0$ 
(which is related to $H^* \to gg$). We compute
%--
\bqa
V^{\bca}_a &=& 
\left(\frac{\mu^2}{\pi}\right)^{\ep}\,\egams{1+\frac{\ep}{2}}\,
( V_{\bcan{-2}}\,\frac{1}{\ep^2} + V_{\bcan{-1}}\,\frac{1}{\ep} + 
V_{\bcan{0}} ).
\label{DK}
\eqa
%--
Note that the presence of a double pole in \eqn{DK} reflects a collinear
singularity besides the infrared one.
For the comparison with the work of Davydychev-Smirnov~\cite{Davydychev:2002hy}
we consider the diagram $V^{\bca}_c$ of \fig{fig231} with $m_1 = m_4 = m$
and $m_2 = m_6 = M$. The analytical calculation is valid in the region 
$m^2 \ll M^2, |P|^2$, and we define
%--
\bqa
V^{\bca}_c &=& 
\left(\frac{\mu^2}{\pi}\right)^{\ep}\,\egam{2+\ep}\,
( V_{\bcan{-2}}\,\frac{1}{\ep^2} + V_{\bcan{-1}}\,\frac{1}{\ep} + V_{\bcan{0}}).
\label{DS}
\eqa
%--
For all comparisons performed in this paper we have found an excellent 
agreement with analytical calculations,\footnote{In several cases we had to 
code the analytical results since the authors did not present explicit 
numerical results} therefore signaling a satisfactory status of the overall 
goodness of our numerical algorithms. Needless to say, we have been able to 
produce results having no counter-examples in the literature.

In~\tabn{tabTensors} we give a sample of results for the 
numerical integration of tensor integrals. The relevant message is that
all analytical expressions which have been implemented in our code for tensor 
integrals has been derived using the same techniques already used for 
scalar configurations.

The topology chosen for tensor integrals is $V_{\bcan{c}}$, which enters in 
the computation of the fermionic corrections to $\sin^2\theta_{\rm eff}$ and 
these results have been already used in~\cite{Hollik:2005va}; this 
emphasizes the overall relevance of our results.

The tensor structure considered in this brief example is:
%--
\bqa
V_{\bcan{c}} &=& 
\frac{\mu^{2\ep}}{\pi^4}\,
\int\,\frac{d^nq_1 d^nq_2}{\prod_{l=1}^6\,[l]_{\bca}},
%\frac{1}{[1]_{\bca}[2]_{\bca}[3]_{\bca}[4]_{\bca}[5]_{bca}[6]_{\bca}},
\qquad\qquad
V_{\bcan{c}|i}^{\mu} =
\frac{\mu^{2\ep}}{\pi^4}
\int\!d^nq_1 d^nq_2\,
\frac{q_i^{\mu}}{\prod_{l=1}^6\,[l]_{\bca}} =
%\frac{q_i^{\mu}}
%{[1]_{\bca}[2]_{\bca}[3]_{\bca}[4]_{\bca}[5]_{\bca}[6]_{\bca}}=
\sum_{j=1}^2\,V_{\bcan{c}|ij}\,p_j^{\mu},
\nl
V_{\bcan{c}|i_1|i_2}^{\mu\nu} &=&
\frac{\mu^{2\ep}}{\pi^4}\,
\int\,d^nq_1 d^nq_2\,
\frac{q_{i_1}^{\mu}\,q_{i_2}^{\nu}}{\prod_{l=1}^6\,[l]_{\bca}} =
%\frac{q_{i_1}^{\mu}\,q_{i_2}^{\nu}}
%{[1]_{\bca}[2]_{\bca}[3]_{\bca}[4]_{\bca}[5]_{\bca}[6]_{\bca}}=
\sum_{j_1,j_2=1}^2 V_{\bcan{c}|i_1j_1|i_2j_2}\,p_{j_1}^{\mu}\,p_{j_2}^{\nu}
+ V_{\bcan{c}|i_10|i_20}\,\delta^{\mu\nu},
\label{tensS}
\eqa
%--
where $[l]_{\bca}$ has been defined in \eqn{defBCA}. The $\ep$ expansion for 
the coefficients is
$V_{\bcan{c}} =  V_{\bcan{c}}^{-1}\,/\ep + V_{\bcan{c}}^0$.
%--
\section{Conclusions \label{concu}}
%--
In this paper we have analyzed a special component of our project aimed to
a numerical evaluation of physical observables at the two-loop level: 
since QED/QCD are an integral part of any realistic theory and they are 
characterized by the exchange/emission of massless gauge-bosons leading to 
infrared divergent parts we had to prove that infrared and also collinear 
configurations can be treated within the same class of algorithms which we 
have used for massive configurations or, at least, within some simple 
extension of them.

The main result, therefore, has been to assemble relatively simple 
expressions for scalar two-loop vertices in a systematic and coherent manner 
so that they can be used for practical calculations. Our results introduce
integral representations which are well suited for numerical integration
and represent a generalization of the familiar Nielsen - Goncharov
multivariate polylogarithms.

Confining most of the paper to scalar configurations should not be
confused with a limitation of the method. Tensor integrals that arise in any
non-trivial theory, due to the spin structure, simply add extra polynomials of 
Feynman parameters in the integral representations of the diagrams. One can 
easily see, for instance from \eqn{innermost}, that these additional 
polynomials can only change the numerator structure inside \eqn{hereishyper} 
which is at the basis of our results: we simply get an hypergeometric function
with a different list of arguments and the whole derivation can be carried 
through along the same lines used for a scalar integrand.

Another important issue that has been addressed in this paper concerns the
systematization of any procedure for implementing infrared divergent graphs
in a realistic calculation, at least when using a modern language and when
considering QED (and also QCD) as embedded into a larger theory. Indeed,
QED alone with a massive regulator has been treated long ago in a seminal paper
by Cvitanovic and Kinoshita~\cite{Cvitanovic:1974sv}.

In the final stage of our project we will generate diagrams with the help of
$\GS$~\cite{GraphShot}: each diagram where an internal photon (or gluon) 
line appears will be subjected to a special investigation, namely the 
corresponding Landau equations will be examined. As soon as they are fulfilled
by the kinematical configuration that we are scanning and as soon as the
filter of infrared power-counting is passed we know that the configuration is
infrared singular and the appropriate subroutine will be initialized returning 
numerical answers for the residues and the finite part.
Understanding these motivations will hopefully explain our preference for
extending the numerical treatment to infrared divergent configuration
despite the recent, spectacular, progress in analytical evaluation.

For instance, one of our configurations, $V^{\bca}$, has been already
computed by Davydychev and Smirnov~\cite{Davydychev:2002hy} in the limit
$m^2 \ll M^2, |P^2|$; we have found excellent agreement and have able to extend
the numerical results to all values of $m$, therefore allowing for QCD
corrections to the top quark decay without approximations for internal masses.
For another setup of the same $V^{\bca}$ configuration we have another 
result by Davydychev and Kalmykov~\cite{Davydychev:2003mv} which is relevant
for Higgs decay into two photons or two gluons; again we found excellent 
agreement and are able to produce numerical results for the same diagram
embedded into the standard model, e.g. also with $\wb$-lines and not only 
gluons outside the inner massive quark loop.
Furthermore, we have been able to perform a numerical test of several 
analytical results by Bonciani, Mastrolia and Remiddi~\cite{Bonciani:2003te} 
corresponding to the whole set of two-loop topologies.

To summarize, we have been able to present all formulas that form the basis
for numerical evaluation of infrared residues and infrared finite parts of
arbitrary infrared configurations of two-loop vertices. The language may
sound unfamiliar but our results have far reaching consequences;
for instance, some of the results presented here have already been used for
computing two - loop electroweak 
pseudo - observables~\cite{Hollik:2005va}.
%--
\Acknowledgments
%--
We gratefully acknowledge discussions and comparisons with Andrei Davydychev, 
Misha Kalmykov and Volodya Smirnov. We would like to express our gratitude 
to Roberto Bonciani, Pierpaolo Mastrolia and Ettore Remiddi for discussions 
and comparisons. The contribution of Andrea Ferroglia to an early stage of 
this paper is also gratefully acknowledged. We recognize the role played by 
Stefano Actis in several steps of our global project. S.U. is indebted to
W.~Hollik for hospitality at MPI where part of the manuscript was written
and would like to thank Ulrich Meier for cross-checking some of the results.
%--
\clearpage
%--
\appendix
%--
\section{Taylor and Laurent expansion of Euler's functions}
%--
Here we collect results needed in expanding Euler's functions;
$\gamma$ is the Euler constant and $\zeta(x)$ is the Riemann zeta function.
%--
\bqa
\egam{1+z} &=& 1 + \sum_{n=1}^{\infty}\,G_n\,z^n, \qquad 
\egam{z} = \frac{\egam{1+z}}{z} = \frac{1}{z} + \sum_{n=0}^{\infty}\,
G_{n+1}\,z^n,
\nl
G_1 &=&  - \gamma,
\quad 
G_2 = \frac{1}{2}\,\Bigl[\gamma^2 + \zeta(2)\Bigr],
\quad
G_3 = - \frac{1}{6}\,\Bigl[ \gamma^3 + 3\,\zeta(2)\,\gamma + 
          2\,\zeta(3)\Bigr],
\nl
G_4 &=&  \frac{1}{24}\,\gamma^4 + \frac{1}{4}\,\zeta(2)\,\gamma^2 + 
  \frac{1}{8}\,\zeta^2(2) + \frac{1}{3}\,\zeta(3)\,\gamma + 
  \frac{1}{4}\,\zeta(4).
\eqa
%--
\section{Nielsen polylogarithms \label{app:Npoly}}
%--
Throughout the paper we have used ($n\,,\,p$ are positive integers)
%--
\bq
S_{n,p}(z) = \frac{(-1)^{n+p-1}}{(n-1)\,!\;p\,!}\,
\intfx{x}\,\frac{dx}{x}\,\ln^{n-1} x\;\ln^p (1 - z\,x), \qquad
S_{n-1,1}(z) = \li{n}{z}
\label{Npoly}
\eq
%--
\section{Properties of the hypergeometric function \label{app:hyp}}
%--
The Gauss hypergeometric function~\cite{ellip} is defined by:
%--
\bq
\hyper{a}{b}{c}{x}= \frac{\egam{c}}{\egam{b}\,\egam{c-b}}\,
\intsx{z}\,z^{b-1}\,(1-z)^{c-b-1}\,(1-x\,z)^{-a}, \qquad
\Reb\,b > \Reb\,c > 0.
\eq
%--
The special case $c = b+1$ frequently occurs in this paper.
Sometimes the Gauss hypergeometric series is used 
(circle of convergence $| x | = 1$):
%--
\bq
\hyper{a}{b}{b+1}{x}= 
\sum_{n=0}^{\infty}\,\frac{\egam{a+n}}{\egam{a}}\,
\frac{b}{b+n}\,\frac{x^n}{\egam{n+1}}.
\eq
%--
An important property of ${}_2F_1$ used in our paper is ($1-c,\, b-a,\, c-b-a$
are not integers):
%--
\bq
\hyper{a}{b}{b+1}{x}=
\frac{b}{b-a}\,(-x)^{-a}\,\hyper{a}{a-b}{a-b+1}{\frac{1}{x}}
+ \frac{\egam{b+1}\,\egam{a-b}}{\egam{a}}\,(-x)^{-b},
\eq
%--
where $|\arg(- x)| < \pi$. 
In other cases we need an expansion around a 
vanishing $\ep$; few examples are listed below 
(see also~\cite{Kalmykov:2006pu}):
%--
\bqa
\hyper{1+\alpha\,\ep}{1+\beta\,\ep}{2+\beta\,\ep}{x} &=&
(1+\beta\,\ep)\,\Bigg\{- \frac{\ln(1-x)}{x} + \ep\,\Big[\,
\frac{\alpha}{2}\,\frac{\ln^2(1-x)}{x} -\beta\,\frac{\li{2}{x}}{x}\,\Big]
\Bigg\},
\nl
\hyper{2+\alpha\,\ep}{1+\beta\,\ep}{2+\beta\,\ep}{x} &=&
\frac{1+\beta\,\ep}{1+\alpha\,\ep}\,\Bigg\{\frac{1}{1-x} + \ep\,\Big[\,
(\beta-\alpha)\,\frac{\ln(1-x)}{x} - \alpha\,\frac{\ln(1-x)}{1-x}\,\Big]
\Bigg\},
\nl
\hyper{2+\alpha\,\ep}{2+\beta\,\ep}{3+\beta\,\ep}{x} &=&
\frac{2+\beta\,\ep}{1+\alpha\,\ep}\,\Bigg\{
\frac{1}{x}\,\Big[\, \frac{1}{x}\,\ln(1-x) + \frac{1}{1-x} \,\Big]
\nl
{}&+& \ep\,\Big[\,\beta\,\frac{\li{2}{x}+\ln(1-x)}{x^2}
- \frac{\alpha}{2}\,\frac{\ln^2(1-x)}{x^2}
- \alpha\,\frac{\ln(1-x)}{x(1-x)}\,\Big]\Bigg\},
\nl
\hyper{2+\alpha\,\ep}{\beta\,\ep}{1+\beta\,\ep}{x} &=&
\frac{1}{1+\alpha\,\ep}\,\Bigg\{1 + \ep\,\Big[\, 
(\alpha-\beta) - \beta\,\ln(1-x) + \frac{\beta}{1-x} \,\Big]
+ \ep^2\,\beta\,\Big[\,(\alpha-\beta)\,\li{2}{x} 
\nl
{}&+& \frac{\alpha}{2}\,\ln^2(1-x)
- (\alpha-\beta)\,\ln(1-x) - \alpha\,\frac{\ln(1-x)}{1-x}\,\Big]\Bigg\}.
\label{hypep}
\eqa
%--
In several cases the hypergeometric function (with integer arguments) 
leads to elementary transcendental functions:
%--
\bq
\hyper{1}{2}{3}{z} = -\,\frac{2}{z}\,\Big[\, \frac{1}{z}\,\ln(1-z) + 1 \,\Big],
\qquad\qquad \hyper{1}{1}{2}{z} = -\,\frac{1}{z}\,\ln(1-z),
\eq
%--
\bq
\hyper{2}{2}{3}{z} = 
\frac{2}{z}\,\Big[\, \frac{1}{z}\,\ln(1-z) + 1 \,\Big] + \frac{2}{1-z} ,
\qquad\qquad \hyper{2}{1}{2}{z} = \frac{1}{1-z}.
\eq
%--
\section{Computation of the integral $J_{n m}^{k h}(i)$ 
         \label{app:Jnnkk}}
%--
The function $J_{n m}^{k h}(i)$ is defined by:
%--
\bq
J_{n m}^{k h}(i)= 
\intsxy{x}{y}\,\ln^{n}x\,\ln^{m}y\,
\frac{\ln^k\alpha_i(x)}{\alpha_i(x)}\,
\frac{\ln^h\alpha_i(y)}{\alpha_i(y)}, \qquad\qquad
\alpha_1(z)= \bchi(z), \qquad \alpha_2(z)= \chi(z),
\label{intfun}
\eq
%--
where $\chi$ and $\bchi$ are defined in \eqn{chidef} and \eqn{chibdef},
respectively. The integral in \eqn{intfun} can be computed by using
BST functional relations. We first use the following relation: 
%--
\bq
\alpha_i^{-1}(y)\,\ln^k \alpha_i(y)= \frac{1}{\Bbt}\,\Big[
\ln^k \alpha_i(y) - \frac{y-Z_i}{2\,(k+1)}\,\partial_y\,
  \big( \ln^{k+1}\alpha_i(y) - ln^{k+1}\alpha_i(0) \big)\Big],
\eq
%--
where $Z_1 = \bXbt$ and $Z_2 = \Xbt$ (see, once again, \eqn{chidef}).
With respect to the usual BST relation (see \eqn{BTlogn}) with have added 
a term, $\partial_y\,\ln^{k+1}\alpha_i(0)$, which is actually zero.
After the integration by parts, we treat the $x$-integral through 
another BST relation:
%--
\bqa
\alpha_i^{-1}(x)\,\ln^k \alpha_i(x) &=& \frac{1}{\Bbt}\,\Big[
\ln^k \alpha_i(x) - \frac{x-Z_i}{2\,(k+1)}\,\partial_x\,\ln^{k+1}\alpha_i(x)
\Big].
\nl
(x-Z_i)\,\alpha_i^{-1}(x)\,\ln^k\alpha_i(x) &=&
- \frac{1}{2\,P^2}\,\frac{\partial_x}{k+1}\,\ln^{k+1}\alpha_i(x).
\eqa
%--
The second integration by parts gives:
%--
\bq
J_{n m}^{k-1\,,\,h-1}(i)= \frac{1}{4\,\Bbt\,h}\,
\Big[ \frac{B_{n m}^{k-1\,,\,h-1}(i)}{\Bbt\,k}
+ \frac{H_{n m}^{k-1\,,\,h-1}(i)}{P^2\,(k+h)}\Big],
\eq
%--
where, new, auxiliary quantities have been introduced:
%--
\bqa
B_{n m}^{k-1\,,\,h-1}(i) &=& \intsxy{x}{y}\,L_{m}^{h-1}(y)\,
\ln^{n-1} x\,\bigg\{
\ln x\,l^{k-1}_i(x)\,\big[ l_i(x) + 2\,k \big]
+ n\,(x-Z_i)\,\frac{l^k_i(x)}{x}\bigg\}
\nl
{}&-& \,\sum_{p=0}^{m}\,\frac{m!}{p!}\,(-1)^{m-p}\,
l^h_i(0)\,\intsx{x}\,\bigg\{
x\,\ln^{n+p}x\,l^{k-1}_i(x)\,\big[ l_i(x) + 2\,k \big]
\nl
{}&+& (\ln x+n+p)\,(x-Z_i)\,\ln^{n+p-1}x\,l^k_i(x)\bigg\}
- \,\intsx{y}\,L_{m}^{h-1}(y)\nl
{}&\times&
\Big[\delta_{n,0}\,(1-Z_i)\,l^k_i(1)
- (y-Z_i)\,\ln^n y\,l^k_i(y)\Big]
+ \,\delta_{n,0}\,(-1)^{m}\,m!\,(1-Z_i)\,l^k_i(1)\,l^h_i(0)
\nl
H_{n m}^{k-1\,,\,h-1}(i) &=& (n+m)\,\intsx{x}\,\ln^{n+m-1}x\,
\bigg[ \frac{l^{k+h}_i(x)}{x}\bmid_+
- \Big( 1 + \frac{h}{k} \Big)\,l^h_i(0)\,
\frac{l^k_i(x)}{x}\bmid_+ \bigg]
\nl
{}&+& \delta_{n,0}\,\delta_{m,0}\,\bigg[ l^{k+h}_i(1)
- \Big( 1 + \frac{h}{k} \Big)\,l^k_i(1)\,
  l^h_i(0)
+ \frac{h}{k}\,l^{k+h}_i(0) \bigg].
\eqa
%--
Furthermore, we have defined:
%--
\bq
l_i(x) = \ln\alpha_i(x), \qquad
L_{n}^{k-1}(y)=
\ln^n y\,l^{k-1}_i(y)\,\big[ l_i(y) + 2\,k \big]
+ n\,(y-Z_i)\,\ln^{n-1} y\,\frac{l^k_i(y)}{y}\bmid_+
\eq
%--
\section{Useful expansions \label{app:coll}}
%--
Here we discuss various expansions of $I_0^{k,n}$ (defined in 
\sect{sect:coll}) for $c \to 0$ in terms of polylogarithms (\eqn{Npoly}). 
Let $Q(x) = a\,x^2 + b\,x +c$ and $e = a + b + c$; we have
%--
\bqa
I_0^{k,n}(a,b,c,d)
\equiv \intsx{x}\,\frac{\ln^k x}{x}\,
\ln^n\frac{Q(x)}{d}\bmid_+ 
&=& 
\sum_{h=1}^n\,(-1)^{k+h}\,n\,!\Bigl[ 
\sum_{l=0}^{n-h}\,\frac{(-1)^l(k+l)!}{l!\,(n-h-l)!}
\ln^{n-h-l}\Big(\frac{b}{d}\,\Big)\,S_{k+l+1,h}\Big( \!-\frac{a}{b} \,\Big)
\nl
{}&+& \,\frac{k!}{(n-h)!}
\ln^{n-h}\Big(\frac{c}{d}\,\Big)\,S_{k+1,h}\Big( \!-\frac{b}{c} \,\Big)\Bigr]
+ {\cal O}(c).
\eqa
%--
Then we consider a class of integrals which generalizes \eqn{defJnk}:
%--
\bq
J_n^k(a,b,c,d) \equiv \intsx{x}\,\frac{\ln^n x}{Q(x)}\ln^k\frac{Q(x)}{d},
\eq
%--
They can be computed by using the BST relation \eqn{BTlogn} which in the 
present case reads as follows:
%--
\bq
\frac{1}{Q(x)}\,\ln^k\frac{Q(x)}{d}= \frac{1}{B}\,\Big[ \ln^k\frac{Q(x)}{d} 
- \frac{x-X}{2\,(k+1)}\,\partial_x\,\ln^{k+1}\frac{Q(x)}{d}\Big],
\qquad
X= -\,\frac{b}{2\,a}, \qquad B= -\,\frac{b^2-4\,a\,c}{4\,a}.
\eq
%--
After integration by parts, we get:
%--
\bqa
J_n^0 &=& \frac{1}{2\,B}\,\bigg\{\intsx{x}\,
( n\,\ln^{n-1}x + \ln^n x )\,\ln\frac{Q(x)}{c} - n\,X\,I_0^{n-1,1}
+ 2\,(-1)^n\,\egam{n+1}\bigg\},
\nl 
J_n^{k-1} &=& \frac{1}{2B\,k}\bigg\{\intsx{x}\,
\bigg[  2\,k\,\ln^n \!x\,\ln^{k-1}\frac{Q(x)}{d}
+ ( n\,\ln^{n-1}\!x + \ln^n \!x )\,\ln^k\frac{Q(x)}{d}\bigg]
- n X I_0^{n-1,k}\bigg\},
\nl 
J_0^{k-1} &=& \frac{1}{2\,B\,k}\,
\bigg\{
\intsx{x}\,\ln^{k-1}\frac{Q(x)}{d}\,\bigg[ \ln\frac{Q(x)}{d} + 2\,k \bigg]
- (1-X)\,\ln^k\frac{e}{d} - X\,\ln^k\frac{c}{d}\bigg\},
\eqa
%--
where for the first two results $n \ge 1$.
Now we give the expansion of $I_1^n$ for $c \to 0$:
%--
\bqa
I_1^{1} &\equiv&
\int_0^1\frac{dx}{x}\,\ln\Big(1+\frac{d\,x}{Q(x)}\,\Big) 
= - \li{2}{\!-\frac{a}{b+d}} + \li{2}{\!-\frac{a}{b}}
+ \ln\Big(\frac{b}{c}\,\Big)\,\ln\Big(1+\frac{d}{b}\,\Big)
+ \frac{1}{2}\,\ln^2\Big(1+\frac{d}{b}\,\Big) + {\cal O}(c),
\nl
I_1^n &\equiv&
\int_0^1\frac{dx}{x}\,\li{n}{\!-\frac{d\,x}{Q(x)}}
= \int_0^1\frac{dx}{x}\,
\li{n}{\!-\frac{d}{ax+b}}\bmid_+
+ \ln\Big(\frac{b}{c}\,\Big)\,\li{n}{\!-\frac{d}{b}}
- S_{n-1,2}\Big( \!-\frac{d}{b} \,\Big) + {\cal O}(c), 
\eqa
%--
with $n \ge 2$.
Next we give the results for $I_2^1$ and $I_2^2$ for $d \to 0$.
The final expression has been derived following the strategy sketched 
in \sect{sect:coll} and is valid also for $d$ far from zero.
%--
\bqa
I_2^{1} &\equiv&
\int_0^1\,\frac{dx}{Q(x)}\,\ln\Big( 1+\frac{Q(x)}{d\,x} \,\Big)
= -\,\frac{1}{2B}\bigg\{ \intsx{x}\,\bigg[
  \ln\Big( 1+\frac{d\,x}{Q(x)} \,\Big) + \li{2}{-\frac{d\,x}{Q(x)}}\bigg]
\nl
{}&-& (1-X)\,\li{2}{-\frac{d}{e}} + X\,I_1^1 \bigg\}
- J_0^1 + J_1^0,
\nl
I_2^{2} &\equiv&
\int_0^1\,\frac{dx}{Q(x)}\,\li{2}{-\frac{Q(x)}{d\,x}}
= -\,\frac{1}{2B}\bigg\{\intsx{x}\,
\bigg[  \li{2}{-\frac{d\,x}{Q(x)}} - \li{3}{-\frac{d\,x}{Q(x)}}\bigg]
\nl
{}&+& (1-X)\,\li{3}{-\frac{d}{e}}
+ X\,I_1^2\bigg\} - \zeta(2)\,J_0^0
- \frac{1}{2}\,J_0^2 - \frac{1}{2}\,J_2^0 + J_1^1.
\eqa
%--
Finally, expanding $I_2^n$ for $c \to 0$ we obtain:
%--
\bqa
I_2^{1} &=& \frac{1}{b}\, \bigg\{
- \li{2}{-\frac{a}{b+d}} + \li{2}{-\frac{a+b}{d}}
+ \ln\frac{d}{c}\,\,\ln\Big(1+\frac{b}{d}\,\Big)
+ \frac{1}{2}\,\ln^2\Big( 1+\frac{b}{d} \,\Big) - 2\,\li{2}{-\frac{b}{d}}
+ {\cal O}(c) \bigg\},
\nl
I_2^n &=& \frac{1}{b}\, \bigg\{
\int_0^1\,\frac{dx}{x}\,
\li{n}{-\frac{ax+b}{d}}\bmid_+ - \li{n+1}{-\frac{a+b}{d}}
\qquad\qquad\qquad\qquad\qquad\qquad n\ge 2
\nl
&+& 
\ln\frac{d}{c}\,\li{n}{-\frac{b}{d}}
- S_{n-1,2}\Big( -\frac{b}{d} \,\Big) + (n+1)\,\li{n+1}{-\frac{b}{d}}
+ {\cal O}(c) \bigg\}.
\eqa
%--
\clearpage
%--
\section{Tables of numerical results \label{Tabnr}}
%--
In this section we collect a sample of our numerical results, based on the
setup of \eqn{NCsetup}.
%--
\vspace{1.cm}
\begin{table}[ht]\centering
\setlength{\arraycolsep}{\tabcolsep}
\renewcommand\arraystretch{1.2}
\begin{tabular}{|l|c|l|l|}
\hline 
    & $s$             & $\Reb\,V_{\acan{a}}^0$   & $\Imb\,V_{\acan{a}}^0$ \\
\hline 
Our & $[ 500\,\GeV ]^2$   & $-2.4142(1)\times 10^{-2}$  & $4.12(1)\times 10^{-3}$ \\
BMR &                 & $-2.414166\times 10^{-2}$   & $4.121472\times 10^{-3}$ \\
\hline 
Our & $M_Z^2$         & $-0.61103(4)$              & $9.447(6)\times 10^{-2}$ \\
BMR &                 & $-0.6110145 $              & $9.446769\times 10^{-2}$ \\
\hline 
Our & $100\,m_e^2$    & $-6.04493(6)\times 10^7$    & $3.0986(2)\times 10^7$ \\
BMR &                 & $-6.044903\times 10^7$      & $3.098533\times 10^7$ \\
\hline 
Our & $4.0001\,m_e^2$ & $-6.878781\times 10^{10}$   & $3.906172\times 10^{11}$ \\
BMR &                 & $-6.878781\times 10^{10}$   & $3.906172\times 10^{11}$ \\
\hline 
Our & $m_e^2$         & $8.533838\times 10^8$       & $0$ \\
BMR &                 & not available              & $0$ \\
\hline 
Our & $-100\,m_e^2$   & $5.463023\times 10^7$       & $0$ \\
BMR &                 & $5.463023\times 10^7$       & $0$ \\
\hline 
\end{tabular}
\vspace{0.3cm}
\caption[]{
{\bf Diagram ${\bf V_{\acan{a}}}$}.
Comparison with the results of~\cite{Bonciani:2003te} (BMR). 
The setup for \fig{fig131} is: $m_1 = m_2 = m= M= m_e$.
The unit of mass $\mu$ is $1$ $\GeV$.
Only the infrared finite part is shown. 
The results are in $\GeV^{-2}$.
Unless indicated our relative error is below $10^{-7}$.}
\label{tabV131aBMR}
\end{table}
%--
\vspace{2.5cm}
%--
\begin{table}[ht]\centering
\setlength{\arraycolsep}{\tabcolsep}
\renewcommand\arraystretch{1.2}
\begin{tabular}{|c|c|c|c|l|l|}
\hline
Process &
$s$ & $M$, $m$ & $m_1$, $m_2$
& $\Reb\,V_{\acan{a}}^0$ 
& $\Imb\,V_{\acan{a}}^0$ \\
\hline
\hline
$ Z^* \rightarrow e^+ e^- $ &
$[ 500\,\GeV ]^2$ & $m_e$, $m_e$ & $m_b$, $m_b$
& $ -4.7574(4)\times 10^{-2} $ 
& $ 2.55694(6)\times 10^{-2} $ \\
\hline
$ $ &
$[ 500\,\GeV ]^2$ & $m_e$, $m_e$ & $m_t$, $m_t$
& $ -1.147833(1)\times 10^{-2}$ 
& $ 4.303633(8)\times 10^{-3}$ \\
\hline
\hline
$ Z \rightarrow e^+ e^- $ &
$\mz^2$ & $m_e$, $m_e$ & $m_b$, $m_b$
& $ -7.91577(3)\times 10^{-2} $ 
& $ 4.74278(5)\times 10^{-2} $ \\
\hline
$ $ &
$\mz^2$ & $m_e$, $m_e$ & $m_t$, $m_t$
& $ -0.2243587$ 
& $ 9.752214(1)\times 10^{-2}$ \\
\hline
\hline
$ Z^* \rightarrow e^+ e^- $ &
$100\,m_e^2$ & $m_e$, $m_e$ & $m_b$, $m_b$
& $ -5.665645 \times 10^6 $
& $ 9.745030 \times 10^5 $ \\
\hline
$ $ &
$100\,m_e^2$ & $m_e$, $m_e$ & $m_t$, $m_t$
& $ 1.987627\times 10^5$ 
& $ 4.155691\times 10^5$ \\
\hline
\hline
$ W \rightarrow e^+ \nu_e $ &
$\mw^2$ & $m_e$, $m_{\nu}$ & $m_t$, $m_t$
& $ -0.7168993 $ 
& $ 0.1227244 $ \\
\hline
\hline
$ W^* \rightarrow t \bar{b} $ &
$[ 200\,\GeV ]^2$ & $m_t$, $m_b$ & $m_b$, $m_b$
& $ -7.158(2)\times 10^{-3}$ 
& $ 3.3638(2)\times 10^{-2}$ \\
\hline
\hline
$ t \rightarrow W^+ b $ &
$m_t^2$ & $m_b$, $\mw$ & $\mw$, $\mw$
& $-2.709610(3) \times 10^{-2}$ 
& $3.182718(2) \times 10^{-2}$ \\
\hline
\end{tabular}
\vspace{0.3cm}
\caption[]{
{\bf Diagram ${\bf V_{\acan{a}}}$}
The setup refers to the diagram of \fig{fig131}
The neutrino mass has been arbitrarily set to $m_{\nu}=0.1\,\,\eV$
The unit of mass $\mu$ is $1$ $\GeV$
Only the infrared finite part is shown.
The results are in $\GeV^{-2}$
Unless indicated our relative error is below $10^{-7}$.}
\label{tabV131a}
\end{table}
%--
\begin{table}[ht]\centering
\setlength{\arraycolsep}{\tabcolsep}
\renewcommand\arraystretch{1.2}
\begin{tabular}{|l|c|l|l|}
\hline 
    & $s$          & $\Reb\,V_{\acan{b}}^0$ & $\Imb\,V_{\acan{b}}^0$ \\
\hline 
Our & $[ 500\,\GeV ]^2$   & $-7.102762\times 10^{-2}$  & $1.914262\times 10^{-2}$ \\
BMR &                 & $-7.102885\times 10^{-2}$  & $1.914295\times 10^{-2}$ \\
\hline 
Our & $M_Z^2$         & $-1.590738$                   & $0.4462335$ \\
BMR &                 & $-1.590738$                   & $0.4462335$ \\
\hline 
Our & $100\,m_e^2$    & $-7.073558\times 10^7$        & $4.1480(2)\times 10^7$ \\
BMR &                 & $-7.073558\times 10^7$         & $4.148499\times 10^7$ \\
\hline 
Our & $4.0001\,m_e^2$ & $-7.264768(2)\times 10^{10}$ & $4.364353\times 10^{11}$ \\
BMR &                 & $-7.264767\times 10^{10}$    & $4.364353\times 10^{11}$ \\
\hline 
Our & $      m_e^2$   & $1.000079\times 10^9$          & $0$ \\
BMR &                 & not available                 & $0$ \\
\hline 
Our & $-100\,m_e^2$   & $6.714222\times 10^7$          & $0$ \\
BMR &                 & $6.714222\times 10^7$          & $0$ \\
\hline 
\end{tabular}
\vspace{0.3cm}
\caption[]{
{\bf Diagram ${\bf V_{\acan{b}}}$}.
Comparison with the results of~\cite{Bonciani:2003te} (BMR). 
The setup for the diagram of \fig{fig131} is: $m_1 = m = M = m_e$, $m_2 = 0$.
The unit of mass $\mu$ is $1$ $\GeV$.
Only the infrared finite part is shown. 
The results are in $\GeV^{-2}$.
Unless indicated our relative error is below $10^{-7}$.}
\label{tabV131bBMR}
\end{table}
%--
\begin{table}[ht]\centering
\setlength{\arraycolsep}{\tabcolsep}
\renewcommand\arraystretch{1.2}
\begin{tabular}{|c|c|c|c|l|l|}
\hline
Process &
$s$ & $M$, $m$ & $m_1$, $m_2$
& $\Reb\,V_{\acan{b}}^0$ 
& $\Imb\,V_{\acan{b}}^0$ \\
\hline
\hline
$ Z^* \rightarrow e^+ e^- $ &
$[ 500\,\GeV ]^2$ & $m_e$, $m_e$ & $m_e$, $\mz$
& $ -8.675940\times 10^{-3} $ 
& $ 4.466520\times 10^{-3} $ \\
\hline
$  $ &
$[ 500\,\GeV ]^2$ & $m_e$, $m_e$ & $m_{\nu}$, $\mw$
& $ -8.462338\times 10^{-3} $ 
& $ 4.491430\times 10^{-3} $ \\
\hline
\hline
$ Z \rightarrow e^+ e^- $ &
$\mz^2$ & $m_e$, $m_e$ & $m_e$, $\mz$
& $ -0.1618863 $ 
& $ 7.892145\times 10^{-2} $ \\
\hline
$  $ &
$\mz^2$ & $m_e$, $m_e$ & $m_{\nu}$, $\mw$
& $ -0.1551424 $ 
& $ 7.718541\times 10^{-2} $ \\
\hline
\hline
$ Z^* \rightarrow e^+ e^- $ &
$100\,m_e^2$ & $m_e$, $m_e$ & $m_e$, $\mz$
& $ -5.936619\times 10^{5} $ 
& $ -9.349849\times 10^{4} $ \\
\hline
$  $ &
$100\,m_e^2$ & $m_e$, $m_e$ & $m_{\nu}$, $\mw$
& $ -7.285072\times 10^{5} $ 
& $ -1.167232\times 10^{5} $ \\
\hline
\hline
$ W^* \rightarrow t \bar{b} $ &
$[ 200\,\GeV ]^2$ & $m_t$, $m_b$ & $m_b$, $\mw$
& $ -2.383592\times 10^{-2} $ 
& $ 5.899083\times 10^{-2} $ \\
\hline
$  $ &
$[ 200\,\GeV ]^2$ & $m_t$, $m_b$ & $m_t$, $\mh$
& $ -3.364485\times 10^{-2} $ 
& $ 7.477385\times 10^{-2} $ \\
\hline
\hline
$ H^* \rightarrow W^+ W^- $ &
$[ 200\,\GeV ]^2$ & $\mw$, $\mw$ & $\mw$, $\mh$
& $ -7.102816\times 10^{-3} $ 
& $ 3.303713\times 10^{-2} $ \\
\hline
\end{tabular}
\vspace{0.3cm}
\caption[]{
{\bf Diagram ${\bf V_{\acan{b}}}$}.
The setup refers to the diagram of \fig{fig131}.
The neutrino mass has been arbitrarily set to $m_{\nu} = 0.1\,\,\eV$.
The unit of mass $\mu$ is $1$ $\GeV$.
Only the infrared finite part is shown. 
The results are in $\GeV^{-2}$.
Unless indicated our relative error is below $10^{-7}$.}
\label{tabV131b}
\end{table}
%--
\begin{table}[ht]\centering
\setlength{\arraycolsep}{\tabcolsep}
\renewcommand\arraystretch{1.2}
\begin{tabular}{|l|c|l|l|}
\hline 
    & $s$          & $\Reb\,V_{\adan{a}}^0$ & $\Imb\,V_{\adan{a}}^0$ \\
\hline 
Our & $[ 500\,\GeV ]^2$ & $2.0928(5)\times 10^3$      & $-1.95(3)\times 10^2$ \\
BMR &                  & $2.092863\times 10^3$       & $-1.941897\times 10^2$ \\
\hline 
Our & $M_Z^2$       & $5.659(1)\times 10^4$       & $-5.84(3)\times 10^3$ \\
BMR &               & $5.658944\times 10^4$       & $-5.837323\times 10^3$ \\
\hline 
Our & $100\,m_e^2$  & $6.28283(2)\times 10^{12}$  & $-1.70618(2)\times 10^{12}$ \\
BMR &               & $6.282796\times 10^{12}$    & $-1.706180\times 10^{12}$ \\
\hline 
Our & $4.0001\,m_e^2$ & $-1.447472\times 10^{20}$ & $1.6161(6)\times 10^{21}$ \\
BMR &                 & $-1.447472\times 10^{20}$ & $1.616432\times 10^{21}$ \\
\hline 
Our & $      m_e^2$ & $-3.11237240\times 10^{14}$ & $0$ \\
BMR &               & not available            & $0$ \\
\hline 
Our & $-100\,m_e^2$ & $-6.544718\times 10^{12}$   & $0$ \\
BMR &               & $-6.544718\times 10^{12}$   & $0$ \\
\hline 
\end{tabular}
\vspace{0.3cm}
\caption[]{
{\bf Diagram ${\bf V_{\adan{a}}}$}.
Comparison with the results of~\cite{Bonciani:2003te} (BMR). 
The setup refers to the diagram of\fig{fig141}: $m_1 = m_2 = m = M = m_e$.
The unit of mass $\mu$ is $1$ $\GeV$.
Only the infrared finite part is shown. 
The results are in $\GeV^{-4}$.
Unless indicated our relative error is below $10^{-7}$.}
\label{tabV141aBMR}
\end{table}
%--
\begin{table}[ht]\centering
\setlength{\arraycolsep}{\tabcolsep}
\renewcommand\arraystretch{1.2}
\begin{tabular}{|c|c|c|c|l|l|}
\hline
Process &
$s$ & $M$, $m$ & $m_1$, $m_2$
& $\Reb\,V_{\adan{a}}^0$ 
& $\Imb\,V_{\adan{a}}^0$ \\
\hline
\hline
$ Z^* \rightarrow e^+ e^- $ &
$[ 500\,\GeV ]^2$ & $m_e$, $m_e$ & $m_b$, $m_b$
& $ 91.02032$ %ok
& $ 1.5233(1)\times 10^{-6} $ \\ %ok
\hline
$  $ &
$[ 500\,\GeV ]^2$ & $m_e$, $m_e$ & $m_t$, $m_t$
& $ -24.61950 $ %ok
& $ -9.6641(1)\times 10^{-10} $ \\ %ok
\hline
\hline
$ Z \rightarrow e^+ e^- $ &
$\mz^2$ & $m_e$, $m_e$ & $m_b$, $m_b$
& $ 2.736577\times 10^3 $ %ok
& $ 4.4153(1)\times 10^{-5} $ \\ %ok
\hline
$  $ &
$\mz^2$ & $m_e$, $m_e$ & $m_t$, $m_t$
& $ -7.401994\times 10^2 $ %ok
& $ -2.114765\times 10^{-6} $ \\ %ok
\hline
\hline
$ Z^* \rightarrow e^+ e^- $ &
$100\,m_e^2$ & $m_e$, $m_e$ & $m_b$, $m_b$
& $ 8.061021\times 10^{11} $ %ok
& $ 3.308893\times 10^{10} $ \\ %ok
\hline
$  $ &
$100\,m_e^2$ & $m_e$, $m_e$ & $m_t$, $m_t$
& $ -2.392478\times 10^{11} $ %ok
& $ -4.086825\times 10^{10} $ \\ %ok
\hline
\hline
$ W^* \rightarrow t \bar{b} $ &
$[ 200\,\GeV ]^2$ & $m_t$, $m_b$ & $m_b$, $m_b$
& $ -6.5246(4)\times 10^{-5} $ %ok
& $ 3.5182(7)\times 10^{-5} $ \\ %ok
\hline
$  $ &
$[ 200\,\GeV ]^2$ & $m_t$, $m_b$ & $m_t$, $m_t$
& $ -6.132929\times 10^{-5} $ %ok
& $ -6.549233\times 10^{-6} $ \\ %ok
\hline
\hline
$ H^* \rightarrow W^+ W^- $ &
$[ 200\,\GeV ]^2$ & $\mw$, $\mw$ & $\mw$, $\mw$
& $ -1.493808(2)\times 10^{-7} $ %ok
& $ -3.513014(2)\times 10^{-7} $ \\ %ok
\hline
\hline
$ t \rightarrow W^+ b $ &
$m_t^2$ & $\mw$, $m_b$ & $m_b$, $m_b$
& $ -3.5328(2)\times 10^{-5} $ %ok
& $ 1.5688(4)\times 10^{-5} $ \\ %ok
\hline
$ $ &
$m_t^2$ & $\mw$, $m_b$ & $\mw$, $\mw$
& $ -2.341537\times 10^{-5} $ %ok
& $ -2.486382(2)\times 10^{-7} $ \\ %ok
\hline
\end{tabular}
\vspace{0.3cm}
\caption[]{
{\bf Diagram ${\bf V_{\adan{a}}}$}.
The setup refers to the diagram of \fig{fig141}.
The unit of mass $\mu$ is $1$ $\GeV$.
Only the infrared finite part is shown. 
The results are in $\GeV^{-4}$.
Unless indicated our relative error is below $10^{-7}$.}
\label{tabV141a}
\end{table}
%--
\begin{table}[ht]\centering
\setlength{\arraycolsep}{\tabcolsep}
\renewcommand\arraystretch{1.2}
\begin{tabular}{|c|c|c|c|l|l|}
\hline
Process &
$s$ & $M$, $m$ & $m_1$, $m_2$
& $\Reb\,V_{\adan{b}}^0$ 
& $\Imb\,V_{\adan{b}}^0$ \\
\hline
\hline
$ Z^* \rightarrow e^+ e^- $ &
$[ 500\,\GeV ]^2$ & $m_e$, $m_e$ & $m_e$, $\mz$
& $ -25.86641 $ 
& $ 2.008353\times 10^{-8} $ \\
\hline
$  $ &
$[ 500\,\GeV ]^2$ & $m_e$, $m_e$ & $m_{\nu}$, $\mw$
& $ -29.73137 $ 
& $ 2.488703\times 10^{-8} $ \\
\hline
\hline
$ Z \rightarrow e^+ e^- $ &
$\mz^2$ & $m_e$, $m_e$ & $m_e$, $\mz$
& $ -7.776883\times 10^{2} $ 
& $ 2.316778\times 10^{-6} $ \\
\hline
$  $ &
$\mz^2$ & $m_e$, $m_e$ & $m_{\nu}$, $\mw$
& $ -8.938906\times 10^{2} $ 
& $ 2.416842\times 10^{-6} $ \\
\hline
\hline
$ Z^* \rightarrow e^+ e^- $ &
$100\,m_e^2$ & $m_e$, $m_e$ & $m_e$, $\mz$
& $ -2.171299\times 10^{11} $ 
& $ 8.580100\times 10^{9} $ \\
\hline
$  $ &
$100\,m_e^2$ & $m_e$, $m_e$ & $m_{\nu}$, $\mw$
& $ -2.520681\times 10^{11} $ 
& $ 6.108271\times 10^{9} $ \\
\hline
\hline
$ W^* \rightarrow t \bar{b} $ &
$[ 200\,\GeV ]^2$ & $m_t$, $m_b$ & $m_b$, $\mw$
& $ -2.248708\times 10^{-7} $ 
& $ 1.048569\times 10^{-6} $ \\
\hline
$  $ &
$[ 200\,\GeV ]^2$ & $m_t$, $m_b$ & $m_t$, $\mh$
& $ -4.763299\times 10^{-7} $ 
& $ 1.548612\times 10^{-6} $ \\
\hline
\hline
$ H^* \rightarrow W^+ W^- $ &
$[ 200\,\GeV ]^2$ & $\mw$, $\mw$ & $\mw$, $\mh$
& $ 1.109640\times 10^{-7} $ 
& $ 4.589289\times 10^{-7} $ \\
\hline
\hline
$ t \rightarrow W^+ b $ &
$m_t^2$ & $\mw$, $m_b$ & $m_e$, $m_{\nu}$
& $ 3.567051\times 10^{-7} $ 
& $ -2.586087\times 10^{-7} $ \\
\hline
$ $ &
$m_t^2$ & $\mw$, $m_b$ & $\mw$, $\mz$
& $ -1.412470\times 10^{-7} $ 
& $ 3.224342\times 10^{-7} $ \\
\hline
\end{tabular}
\vspace{0.3cm}
\caption[]{
{\bf Diagram ${\bf V_{\adan{b}}}$}.
The setup refers to the diagram of \fig{fig141}.
The neutrino mass has been arbitrarily set to $m_{\nu}=0.1\,\,\eV$.
The unit of mass $\mu$ is $1$ $\GeV$.
Only the infrared finite part is shown. 
The results are in $\GeV^{-4}$.
Unless indicated our relative error is below $10^{-7}$.}
\label{tabV141b}
\end{table}
%--
\begin{table}[ht]\centering
\setlength{\arraycolsep}{\tabcolsep}
\renewcommand\arraystretch{1.2}
\begin{tabular}{|l|c|l|l|}
\hline 
    & $s$          & $\Reb\,V_{\adan{c}}^0$ & $\Imb\,V_{\adan{c}}^0$ \\
\hline 
Our & $[ 500\,\GeV ]^2$ & $-1.819980\times 10^5$     & $4.273610\times 10^4$ \\
BMR &                  & $-1.820011\times 10^5$     & $4.273685\times 10^4$ \\
\hline 
Our & $M_Z^2$          & $-4.194142\times 10^6$     & $1.086325\times 10^6$ \\
BMR &                  & $-4.194142\times 10^6$     & $1.086325\times 10^6$ \\
\hline 
Our & $100\,m_e^2$     & $-9.234450\times 10^{13}$  & $8.726096\times 10^{13}$ \\
BMR &                  & $-9.234450\times 10^{13}$  & $8.726096\times 10^{13}$ \\
\hline 
Our & $4.0001\,m_e^2$  & $1.449091\times 10^{20}$ & $-1.754699\times 10^{21}$ \\
BMR &                  & $1.449091\times 10^{20}$ & $-1.754699\times 10^{21}$ \\
\hline 
Our & $      m_e^2$    & $1.377881\times 10^{15}$   & $0$ \\
BMR &                  & not available             & $0$ \\
\hline 
Our & $-100\,m_e^2$    & $1.038162\times 10^{14}$   & $0$ \\
BMR &                  & $1.038162\times 10^{14}$   & $0$ \\
\hline 
\end{tabular}
\vspace{0.3cm}
\caption[]{
{\bf Diagram ${\bf V_{\adan{c}}}$}.
Comparison with the results of~\cite{Bonciani:2003te} (BMR). 
The setup referring to the diagram of \fig{fig141} is: $m = M = m_e$.
The unit of mass $\mu$ is $1$ $\GeV$.
Only the infrared finite part is shown. 
The results are in $\GeV^{-4}$.
Our relative error is everywhere below $10^{-7}$.}
\label{tabV141cBMR}
\end{table}
%--
\begin{table}[ht]\centering
\setlength{\arraycolsep}{\tabcolsep}
\renewcommand\arraystretch{1.2}
\begin{tabular}{|c|c|c|c|l|l|}
\hline
Process &
$s$ & $M$, $m$ & $m_1$, $m_2$, $m_3$
& $\Reb\,V_{\bcan{a}}^0$ 
& $\Imb\,V_{\bcan{a}}^0$ \\
\hline
\hline
$ Z \rightarrow e^+ e^- $ &
$[ 500\,\GeV ]^2$ & $m_e$, $m_e$ & $m_e$, $m_e$, $\mz$
& $ -8.6239(5)\times 10^{-9} $
%& $ -8.634(5)\times 10^{-9} $    % suave result without expansion
& $ 4.86523(5)\times 10^{-8} $ \\
\hline
$  $ &
$[ 500\,\GeV ]^2$ & $m_e$, $m_e$ & $\mw$, $\mw$, $m_{\nu}$
& $ -1.706(4)\times 10^{-8} $
%& $ -1.74(3)\times 10^{-8} $    % suave result without expansion
& $ 7.109(3)\times 10^{-8} $ \\
\hline
\hline
$ Z \rightarrow e^+ e^- $ &
$\mz^2$ & $m_e$, $m_e$ & $m_e$, $m_e$, $\mz$
& $ 5.11313(2)\times 10^{-6} $
%& $ 5.1285(7)\times 10^{-6} $    % suave result without expansion
& $ 5.05866(2)\times 10^{-6} $ \\
%& $ 5.0484(3)\times 10^{-6} $ \\ % suave result without expansion
\hline
$  $ &
$\mz^2$ & $m_e$, $m_e$ & $\mw$, $\mw$, $m_{\nu}$
& $ 4.552503(7)\times 10^{-6} $ 
& $ -1.269944(6)\times 10^{-6} $ \\
\hline
\hline
$ Z^* \rightarrow e^+ e^- $ &
$100\,m_e^2$ & $m_e$, $m_e$ & $m_e$, $m_e$, $\mz$
& $ -6.700055\times 10^{3} $ 
& $ 1.226283\times 10^{3} $ \\
\hline
$  $ &
$100\,m_e^2$ & $m_e$, $m_e$ & $\mw$, $\mw$, $m_{\nu}$
& $ -86.21982 $ 
& $ -11.84965 $ \\
\hline
\hline
$ W^* \rightarrow t \bar{b} $ &
$[ 200\,\GeV ]^2$ & $m_t$, $m_b$ & $m_b$, $m_t$, $\mz$
& $ 1.0060(2)\times 10^{-6} $ 
& $ -6.2131(5)\times 10^{-7} $ \\
\hline
$  $ &
$[ 200\,\GeV ]^2$ & $m_t$, $m_b$ & $\mw$, $\mh$, $m_t$
& $ 1.894547(1)\times 10^{-7} $ 
& $ -3.017159\times 10^{-7} $ \\
\hline
\hline
$ H^* \rightarrow W^+ W^- $ &
$[ 200\,\GeV ]^2$ & $\mw$, $\mw$ & $m_t$, $m_t$, $m_b$
& $ 4.216966(3)\times 10^{-8} $ 
& $ -1.225831\times 10^{-7} $ \\
\hline
\hline
$ t \rightarrow W^+ b $ &
$m_t^2$ & $\mw$, $m_b$ & $\mw$, $m_b$, $m_t$
& $ 3.4389(8)\times 10^{-7} $ 
& $ 6.082(2)\times 10^{-8} $ \\
\hline
$ $ &
$m_t^2$ & $\mw$, $m_b$ & $m_b$, $\mw$, $\mz$
& $ 6.3629(3)\times 10^{-7} $ 
& $ 3.0576(8)\times 10^{-7} $ \\
\hline
\end{tabular}
\vspace{0.3cm}
\caption[]{
{\bf Diagram ${\bf V_{\bcan{a}}}$}.
The setup refers to the diagram of \fig{fig231}.
The neutrino mass has been arbitrarily set to $m_{\nu}=0.1\,\,\eV$.
The unit of mass $\mu$ is $1$ $\GeV$.
Only the infrared finite part is shown. 
The results are in $\GeV^{-4}$.
Unless indicated our relative error is below $10^{-7}$.}
\label{tabV231a}
\end{table}
%--
\begin{table}[ht]\centering
\setlength{\arraycolsep}{\tabcolsep}
\renewcommand\arraystretch{1.2}
\begin{tabular}{|l|c|l|l|}
\hline 
    & $s$             & $\Reb\,V_{\bcan{b}}^0$ & $\Imb\,V_{\bcan{b}}^0$ \\
\hline 
Our & $[ 500\,\GeV ]^2$ & $-2.6228(2)\times 10^{-4}$ & $8.7973(7)\times 10^{-5}$ \\
BMR &                  & $-2.622920\times 10^{-4}$  & $8.797302\times 10^{-5}$ \\
\hline 
Our & $M_Z^2$         & $-1.64386(8)\times 10^{-3}$ & $5.8676(7)\times 10^{-4}$ \\
BMR &                 & $-1.6438612\times 10^{-3}$  & $5.867706\times 10^{-4}$ \\
\hline 
Our & $100\,m_e^2$    & $-3.8356(4)\times 10^{12}$  & $7.7519(4)\times 10^{12}$ \\
BMR &                 & $-3.835560\times 10^{12}$   & $7.751602\times 10^{12}$ \\
\hline 
Our & $4.0001\,m_e^2$ & $1.844(3)\times 10^{20}$    & unstable \\
BMR &                 & $1.842500\times 10^{20}$    & $5.160917\times 10^{19}$ \\
\hline 
Our & $      m_e^2$   & $-1.177792\times 10^{15}$   & $0$ \\
BMR &                 & not available              & $0$ \\
\hline 
Our & $-100\,m_e^2$   & $-5.682692\times 10^{12}$   & $0$ \\
BMR &                 & $-5.682692\times 10^{12}$   & $0$ \\
\hline 
\end{tabular}
\vspace{0.3cm}
\caption[]{
{\bf Diagram ${\bf V_{\bcan{b}}}$}.
Comparison with the results of~\cite{Bonciani:2003te} (BMR). 
The setup referring to the diagram of \fig{fig231} is: $m = M = m_e$.
The unit of mass $\mu$ is $1$ $\GeV$.
Only the infrared finite part is shown. 
The results are in $\GeV^{-4}$.
Unless indicated our relative error is below $10^{-7}$.}
\label{tabV231bBMR}
\end{table}
%--
\begin{table}[ht]\centering
\setlength{\arraycolsep}{\tabcolsep}
\renewcommand\arraystretch{1.2}
\begin{tabular}{|c|c|c|c|l|l|}
\hline
Process &
$s$ & $M$, $m$ & $m_1$, $m_2$, $m_3$
& $\Reb\,V_{\bcan{c}}^0$ 
& $\Imb\,V_{\bcan{c}}^0$ \\
\hline
\hline
$ Z \rightarrow e^+ e^- $ &
$[ 500\,\GeV ]^2$ & $m_e$, $m_e$ & $m_e$, $\mz$, $m_e$
& $ -2.1336(8)\times 10^{-6} $
& $ 6.87(2)\times 10^{-7} $ \\
\hline
\hline
$ Z \rightarrow e^+ e^- $ &
$\mz^2$ & $m_e$, $m_e$ & $m_e$, $\mz$, $m_e$
& $ -5.0064(3)\times 10^{-5} $
& $ 1.862(2)\times 10^{-6} $ \\
\hline
$  $ &
$m_e^2$ & $\mz$, $m_e$ & $m_t$, $m_t$, $m_t$
& $ -2.656960\times 10^{-7} $ 
& $ 2.424(5)\times 10^{-20} $ \\
\hline
$  $ &
$m_e^2$ & $\mz$, $m_e$ & $m_b$, $m_b$, $m_b$
& $ 6.8986(3)\times 10^{-6} $ 
& $ -6.3053(4)\times 10^{-6} $ \\
\hline
\hline
$ Z^* \rightarrow e^+ e^- $ &
$100\,m_e^2$ & $m_e$, $m_e$ & $m_e$, $\mz$, $m_e$
& $ -7.19834(2)\times 10^{3} $ 
& $ 3.73760(1)\times 10^{3} $ \\
\hline
\hline
$ W^* \rightarrow t \bar{b} $ &
$[ 200\,\GeV ]^2$ & $m_t$, $m_b$ & $m_t$, $\mz$, $m_t$
& $ 1.43046(3)\times 10^{-7} $ 
& $ -2.256417(8)\times 10^{-7} $ \\
\hline
$  $ &
$[ 200\,\GeV ]^2$ & $m_t$, $m_b$ & $\mw$, $m_b$, $\mw$
& $ 5.962(5)\times 10^{-7} $ 
& $ 6.528(3)\times 10^{-7} $ \\
\hline
\hline
$ H^* \rightarrow W^+ W^- $ &
$[ 200\,\GeV ]^2$ & $\mw$, $\mw$ & $m_t$, $m_b$, $m_t$
& $ 3.62055(7)\times 10^{-8} $ 
& $ -1.075947(8)\times 10^{-7} $ \\
\hline
\hline
$ t \rightarrow W^+ b $ &
$m_t^2$ & $\mw$, $m_b$ & $m_t$, $m_b$, $m_t$
& $ 1.22769(4)\times 10^{-7} $ 
& $ -1.084614(6)\times 10^{-7} $ \\
\hline
$ $ &
$m_t^2$ & $\mw$, $m_b$ & $\mw$, $\mh$, $\mw$
& $ 1.6920(4)\times 10^{-7} $ 
& $ -1.49596(9)\times 10^{-7} $ \\
\hline
\end{tabular}
\vspace{0.3cm}
\caption[]{
{\bf Diagram ${\bf V_{\bcan{c}}}$}.
The setup refers to the diagram of \fig{fig231}.
The unit of mass $\mu$ is $1$ $\GeV$.
Only the infrared finite part is shown. 
The results are in $\GeV^{-4}$.
Unless indicated our relative error is below $10^{-7}$.}
\label{tabV231c}
\end{table}
%--
\begin{table}[ht]\centering
\setlength{\arraycolsep}{\tabcolsep}
\renewcommand\arraystretch{1.2}
\begin{tabular}{|l|c|l|l|}
\hline 
    & $s$             & $\Reb\,V_{\bcan{d}}^0$ & $\Imb\,V_{\bcan{d}}^0$ \\
\hline 
Our & $[ 500\,\GeV ]^2$ & $1.58634(5)\times 10^5$     & $-3.868(8)\times 10^4$ \\
BMR &                  & $1.586339\times 10^5$        & $-3.878825\times 10^4$ \\
\hline 
Our & $M_Z^2$          & $3.61488(4)\times 10^6$      & $-9.75(2)\times 10^5$ \\
BMR &                  & $3.614876\times 10^6$        & $-9.774586\times 10^5$ \\
\hline 
Our & $100\,m_e^2$     & $6.7595(2)\times 10^{13}$    & $-7.0650(4)\times 10^{13}$ \\
BMR &                  & $6.759415\times 10^{13}$     & $-7.065317\times 10^{13}$ \\
\hline 
Our & $4.0001\,m_e^2$  & $-1.680564(3)\times 10^{17}$ & $1.1570(6)\times 10^{17}$ \\
BMR &                  & $-1.680565\times 10^{17}$    & $1.156966\times 10^{17}$ \\
\hline 
Our & $      m_e^2$    & $-7.970954\times 10^{14}$    & $0$ \\
BMR &                  & not available                & $0$ \\
\hline 
Our & $-100\,m_e^2$    & $-7.782868\times 10^{13}$    & $0$ \\
BMR &                  & $-7.782868\times 10^{13}$    & $0$ \\
\hline 
\end{tabular}
\vspace{0.3cm}
\caption[]{
{\bf Diagram ${\bf V_{\bcan{d}}}$}.
Comparison with the results of~\cite{Bonciani:2003te} (BMR). 
The setup referring to the diagram of \fig{fig231} is: $m = M = m_e$.
The unit of mass $\mu$ is $1$ $\GeV$.
Only the infrared finite part is shown. 
The results are in $\GeV^{-4}$.
Unless indicated our relative error is below $10^{-7}$.}
\label{tabV231dBMR}
\end{table}
%--
%\begin{table}[ht]\centering
%\setlength{\arraycolsep}{\tabcolsep}
%\renewcommand\arraystretch{1.2}
%\begin{tabular}{|c|c|c|c|l|l|}
%\hline
%Process &
%$s$ & $M$, $m$ & $\Reb\,V_{\bbb}^0$ & $\Imb\,V_{\bbb}^0$ \\
%\hline
%\hline
%$ W^* \rightarrow t \bar{b} $ & 
%$[ 500\,\GeV ]^2$ & $m_t$, $m_b$ 
%& $ \times 10^{-6} $ 
%& $ \times 10^{-7} $ \\
%\hline
%$  $ &
%$[ 200\,\GeV ]^2$ & $m_t$, $m_b$ 
%& $ \times 10^{-7} $ 
%& $ \times 10^{-7} $ \\
%\hline
%\hline
%$ t \rightarrow W^+ b $ &
%$[ 500\,\GeV ]^2$ & $\mw$, $m_b$ 
%& $ \times 10^{-7} $ 
%& $ \times 10^{-8} $ \\
%\hline
%$ $ &
%$m_t^2$ & $\mw$, $m_b$ 
%& $ \times 10^{-7} $ 
%& $ \times 10^{-7} $ \\
%\hline
%\end{tabular}
%\caption[]{
%{\bf Diagram ${\bf V_{\bbb}}$}.
%The setup refers to \fig{fig231}.
%The unit of mass $\mu$ is $1$ $GeV$.
%Only the infrared finite part is shown. 
%The results are in $GeV^{-4}$.
%Unless indicated our relative error is below $10^{-7}$.}
%\label{tabV222}
%\end{table}
%--
\begin{table}[ht]\centering
\setlength{\arraycolsep}{\tabcolsep}
\renewcommand\arraystretch{1.2}
\begin{tabular}{|l|c|l|l|}
\hline 
    & $s$             & $\Reb\,V_{\bbb}^0$ & $\Imb\,V_{\bbb}^0$ \\
\hline 
Our & $500^2GeV^2$    & $-1.342815\times 10^{-6}$ & $1.052948\times 10^{-6}$ \\
BMR &                 & $-1.342815\times 10^{-6}$ & $1.052948\times 10^{-6}$ \\
\hline 
Our & $M_Z^2$         & $-4.562983\times 10^{-4}$ & $5.478876\times 10^{-4}$ \\
BMR &                 & $-4.552983\times 10^{-4}$ & $5.478876\times 10^{-4}$ \\
\hline 
Our & $100\,m_e^2$    & $2.801(1)\times 10^{11}$ & $-2.0846(8)\times 10^{12}$ \\
BMR &                 & $2.801721\times 10^{11}$ & $-2.084294\times 10^{12}$ \\
\hline 
Our & $4.01\,m_e^2$   & $-1.87(1)\times 10^{16}$  & $2.185(32)\times 10^{16}$ \\
BMR &                 & $-1.866708\times 10^{16}$   & $2.152803\times 10^{16}$ \\
\hline 
Our & $      m_e^2$   & $1.424912\times 10^{14}$ & $0$ \\
BMR &                 & not available & $0$ \\
\hline 
Our & $-100\,m_e^2$   & $1.21504(6)\times 10^{12}$ & $0$ \\
BMR &                 & $1.214934\times 10^{12}$ & $0$ \\
\hline 
\end{tabular}
\vspace{0.3cm}
\caption[]{
{\bf Diagram ${\bf V_{\bbb}}$}.
Comparison with the results of~\cite{Bonciani:2003te} (BMR). 
The setup referring to the diagram of \fig{fig222} is: $m = M = m_e$.
The unit of mass $\mu$ is $1$ $\GeV$.
Only the infrared finite part is shown. 
The results are in $\GeV^{-4}$.
We have found numerical instabilities for values of $s$ too close to the 
normal threshold, %s = 4\,m^2_e$.
Unless indicated our relative error is below $10^{-7}$.}
\label{tabV222BMR}
\end{table}
%--
\begin{table}[ht]\centering
\setlength{\arraycolsep}{\tabcolsep}
\renewcommand\arraystretch{1.2}
\begin{tabular}{|l|c|l|l|}
\hline 
 {} & {$\sqrt{s}\,\,[\GeV]$} & {$\Reb\,V_{\bcan{0}}\quad[\GeV^{-4}]$} & 
{$\Imb\,V_{\bcan{0}}\quad[\GeV^{-4}]$} \\
\hline 
 { Our} & $ 400$  & {$5.1343(1)\times 10^{-8}$} & {$1.94009(8)\times 10^{-8}$} \\
     { DK}  &         &      $5.13445  \times 10^{-8}$  &      $1.94008\times 10^{-8}$  \\
\hline 
 { Our} & $ 300$  & {$5.68801\times 10^{-8}$} & {$-1.61218\times 10^{-8}$} \\
     { DK}  &         &      $5.68801\times 10^{-8}$    &      $-1.61218\times 10^{-8}$  \\
\hline 
 { Our} & $ 200$  & {$9.36340\times 10^{-8}$} & {$-2.84232\times 10^{-8}$} \\
     { DK}  &         &      $9.36340\times 10^{-8}$    &      $-2.84232\times 10^{-8}$  \\
\hline 
 { Our} & $ 100$  & {$2.94726\times 10^{-7}$} & {$-9.74218\times 10^{-8}$} \\
     { DK}  &         &      $2.94726\times 10^{-7}$    &      $-9.74218\times 10^{-8}$  \\
\hline 
 {} & {$\sqrt{-t}\,\,[\GeV]$} & {$\Reb\,V_{\bcan{0}}\quad[\GeV^{-4}]$} & 
{$\Imb\,V_{\bcan{0}}\quad[\GeV^{-4}]$} \\
\hline 
 { Our} & $ 100$  & {$-2.85709\times 10^{-7}$} & {$0$} \\
     { DK}  &         &      $-2.85709\times 10^{-7}$    &      $0$  \\
\hline 
 { Our} & $ 200$  & {$-7.61695\times 10^{-8}$} & {$0$} \\
     { DK}  &         &      $-7.61695\times 10^{-8}$    &      $0$  \\
\hline 
 { Our} & $ 300$  & {$-3.29938\times 10^{-8}$} & {$0$} \\
     { DK}  &         &      $-3.29938\times 10^{-8}$    &      $0$  \\
\hline 
 { Our} & $ 400$  & {$-1.74228\times 10^{-8}$} & {$0$} \\
     { DK}  &         &      $-1.74228\times 10^{-8}$    &      $0$  \\
\hline
\end{tabular}
\vspace{0.3cm}
\caption[]{Comparison with the results of~\cite{Davydychev:2003mv} (DK) in the
setup of \eqn{DK}. Only the infrared finite part is shown. Unless indicated
our relative error is below $10^{-5}$.}
\label{tabDK}
\end{table}
%--
\begin{table}[ht]\centering
\setlength{\arraycolsep}{\tabcolsep}
\renewcommand\arraystretch{1.2}
\begin{tabular}{|l|c|l|l|}
\hline 
& {$s\,\,[\GeV^2]$} & {$\Reb\,V_{\bcan{0}}\quad[\GeV^{-4}]$} & 
{$\Imb\,V_{\bcan{0}}\quad[\GeV^{-4}]$} \\
\hline 
 { Our} & $10.2$ & {$16.346(5)$} & {$-18.059(4)$} \\
     { DS}  &        &      $16.3459$    &      $-18.0590$    \\
\hline 
 { Our} & $ 9.2$ & {$19.928(6)$} & {$-22.175(4)$} \\
     { DS}  &        &      $19.9189$    &      $-22.1755$    \\
\hline 
 { Our} & $ 8.2$ & {$24.898(8)$} & {$-27.980(4)$} \\
     { DS}  &        &      $24.9015$    &      $-27.9753$    \\
\hline 
 { Our} & $ 7.2$ & {$32.185(8)$} & {$-36.547(8)$} \\
     { DS}  &        &      $32.1805$    &      $-36.5550$    \\
\hline 
 { Our} & $ 6.2$ & {$43.51(1)$}  & {$-50.114(8)$} \\
     { DS}  &        &      $43.4927$    &      $-50.1010$    \\
\hline 
 { Our} & $ 5.2$ & {$62.62(2)$}  & {$-73.51(2)$} \\
     { DS}  &        &      $62.6575$    &      $-73.5359$   \\
\hline 
 { Our} & $ 4.2$ & {$99.58(3)$}  & {$-120.09(2)$} \\
     { DS}  &        &      $99.6039$    &      $-120.086$    \\
\hline 
 { Our} & $ 3.2$ & {$188.04(5)$} & {$-237.07(5)$} \\
     { DS}  &        &      $188.017$    &      $-237.028$    \\
\hline
\end{tabular}
\vspace{0.3cm}
\caption[]{Comparison with the results of~\cite{Davydychev:2002hy} (DK) in the
setup of \eqn{DS}. Only the infrared finite part is shown.}
\label{tabDS}
\end{table}
%--
\begin{table}[ht]\centering
\setlength{\arraycolsep}{\tabcolsep}
\renewcommand\arraystretch{1.2}
\begin{tabular}{|c|l|l|}
\hline
Tensor coeff. & Real part & Imaginary part \\
\hline
\hline
$V_{\bcan{c}}^0$
& $ 6.8986(3)\times 10^{-6} $ 
& $ -6.3053(4)\times 10^{-6} $ \\
\hline
\hline
$V_{\bcan{c}|11}^0$
& $ -4.210(2)\times 10^{-6} $ 
& $ 3.423(3)\times 10^{-6} $ \\
\hline
$V_{\bcan{c}|12}^0$
& $ -2.9453(4)\times 10^{-6} $ 
& $ 2.4067(6)\times 10^{-6} $ \\
\hline
\hline
$V_{\bcan{c}|11|11}^0$
& $ 3.043(2)\times 10^{-6} $ 
& $ -2.372(3)\times 10^{-6} $ \\
\hline
$V_{\bcan{c}|11|12}^0$
& $ 2.1699(3)\times 10^{-6} $ 
& $ -1.7005(4)\times 10^{-6} $ \\
\hline
$V_{\bcan{c}|12|12}^0$
& $ 1.7392(2)\times 10^{-6} $ 
& $ -1.5014(2)\times 10^{-6} $ \\
\hline
\end{tabular}
\vspace{0.3cm}
\caption[]{
{\bf Tensor Diagrams}. 
Results for the tensor coefficients of rank $0,1$ and $2$ for the diagram 
$V_{\bcan{c}}$, see \eqn{tensS}, with setup $m_1 = m_2 = m_3 = m_b$, 
$M = \mz$, $m = m_e$ and $s= \mzs$ (see \fig{fig231}).
The unit of mass $\mu$ is $1$ $\GeV$.
Only the infrared finite part is shown. 
The results are in $\GeV^{-4}$.}
\label{tabTensors}
\end{table}
%-
\clearpage
%--

%--
\clearpage
%--
\end{document}